\documentclass[fleqn,usenatbib]{mnras}

\usepackage[T1]{fontenc}
\usepackage{ae,aecompl}

\usepackage{graphicx}   
\usepackage{amsmath}    
\usepackage{amssymb}    
\usepackage{hyperref}
\usepackage{subcaption}

\title{Galaxy Pairs in the Sloan Digital Sky Survey XV: Properties of Ionised Outflows}

\author[Matzko et al.]{
William Matzko$^{1}$\thanks{E-mail: wmatzko@gmu.edu},
Shobita Satyapal$^{1}$,
Sara L. Ellison$^{2}$,
Remington O. Sexton$^{1,3}$,
Nathan J. Secrest$^{3}$
\newauthor
Gabriela Canalizo$^{4}$,
\ Laura Blecha$^{5}$,
David R. Patton$^{6}$,
Jillian M. Scudder$^{7}$
\\
$^{1}$Department of Physics and Astronomy, George Mason University, MS3F3, 4400 University Drive, Fairfax, VA 22030, USA\\
$^{2}$Department of Physics and Astronomy, University of Victoria, Victoria, BC V8P 1A1, Canada\\
$^{3}$U.S. Naval Observatory, 3450 Massachusetts Ave NW, Washington, DC 20392-5420, USA\\
$^{4}$Department of Physics and Astronomy, University of California, Riverside, 900 University Avenue, Riverside CA, 92521, USA\\
$^{5}$Department of Physics, University of Florida, P.O. Box 118440, Gainesville, FL 32611-8440, USA\\
$^{6}$Department of Physics and Astronomy, Trent University, 1600 West Bank Drive, Peterborough, ON K9L 0G2, Canada\\
$^{7}$Department of Physics and Astronomy, Oberlin College, Oberlin, OH 44074, USA
}

\date{Accepted XXX. Received YYY; in original form ZZZ}

\pubyear{2022}

\usepackage{newtxtext,newtxmath}
\begin{document}
\label{firstpage}
\pagerange{\pageref{firstpage}--\pageref{lastpage}}
\maketitle

\begin{abstract}

Powerful outflows are thought to play a critical role in galaxy evolution and black hole growth. We present the first large-scale systematic study of ionised outflows in paired galaxies and post-mergers compared to a robust control sample of isolated galaxies. We isolate the impact of the merger environment to determine if outflow properties depend on merger stage. Our sample contains $\sim$4,000 paired galaxies and $\sim$250 post-mergers in the local universe ($0.02 \leq z \leq 0.2$) from the SDSS DR 7 matched in stellar mass, redshift, local density of galaxies, and [OIII]~$\lambda$5007 luminosity to a control sample of isolated galaxies. By fitting the [OIII]~$\lambda$5007 line, we find ionised outflows in $\sim$15 per cent of our entire sample. Outflows are much rarer in star-forming galaxies compared to AGN, and outflow incidence and velocity increase with [OIII]~$\lambda$5007 luminosity. Outflow incidence is significantly elevated in the optical+mid-infrared selected AGN compared to purely optical AGN; over 60 per cent show outflows at the highest luminosities ($L_{\mathrm{[OIII]~\lambda5007}}$ $\gtrsim$ 10$^{42}$~erg~s$^{-1}$), suggesting mid-infrared AGN selection favours galaxies with powerful outflows, at least for higher [OIII]~$\lambda$5007 luminosities. However, we find no statistically significant difference in outflow incidence, velocity, and luminosity in mergers compared to isolated galaxies, and there is no dependence on merger stage. Therefore, while interactions are predicted to drive gas inflows and subsequently trigger nuclear star formation and accretion activity, when the power source of the outflow is controlled for, the merging environment has no further impact on the large-scale ionised outflows as traced by [OIII]~$\lambda5007$.

\end{abstract}




\section{Introduction} \label{sec: Introduction}

Hierarchical models of the Universe and observations suggest that mergers play a critical role in galaxy evolution. Mergers are thought to drive morphological transformations \citep[e.g.][]{1972ApJ...178..623T, 1982ApJ...252..455S, 2004AJ....128.2098R}, trigger star formation \citep[e.g.][]{2008AJ....135.1877E,2010AJ....139.1857W,2011ApJ...728..119W, 2012MNRAS.426..549S,2012ApJ...745...94L,2011MNRAS.412..591P,2013MNRAS.433L..59P}, and trigger AGN \citep[e.g.][]{2001ApJ...555..719C,2008ApJS..175..390H,2011MNRAS.418.2043E,2014AJ....148..137L,2014MNRAS.441.1297S,2017MNRAS.464.3882W,2018PASJ...70S..37G,2018MNRAS.478.3056B}, although the merger-AGN connection is a topic of vigorous debate, particularly at higher redshifts \citep[][]{2011ApJ...726...57C,2012ApJ...744..148K,2012ApJ...761...75S,2014MNRAS.439.3342V,2017MNRAS.466..812V,2015MNRAS.451.2517S,2015A&A...573A..85R,2016MNRAS.458.2391B,2016ApJ...830..156M,2020ApJ...904..107S}. Similarly, the connection between mergers and star formation at higher redshifts is also debated  \citep[][]{2011MNRAS.410L..42K,2012ApJ...760...72X,2013MNRAS.429L..40K,2014A&A...562A...1P,2017MNRAS.465.1934F,2019A&A...631A..51P}. Despite these debates, the currently accepted paradigm of galaxy evolution begins with young, blue disk galaxies merging and ends with the formation of large `red and dead' elliptical galaxies, where some form of feedback from the enhanced nuclear activity is required to quench star formation in the host galaxies. 
\par
An outflow is one form of feedback that is typically invoked as the primary mechanism that suppresses star formation in such mergers. However, it is not completely clear if these outflows are sufficient to cause quenching \citep[e.g.][]{2005MNRAS.361..776S,2006ApJS..163....1H,2018A&A...616A.171P, 2019MNRAS.483.4586F, 2021MNRAS.505L..46E}, or if they result in an elevation in the star formation activity \citep[e.g.][]{2013ApJ...772..112S,2013MNRAS.433.3079Z,2015A&A...582A..63C}.  Nonetheless, outflows have long been known to be present in purely star-forming, isolated galaxies at low redshifts \citep[e.g.][]{1990ApJS...74..833H, 2005ApJS..160...87R} and more moderate ($0.5 \lesssim z \lesssim 1$) redshifts \citep[e.g.][]{2011MNRAS.412.1559N, 2018ApJ...864L...1G}. For recent reviews, see \citet[][]{2018Galax...6..138R} and \citet[][]{2020A&ARv..28....2V}. In star-forming galaxies, there are clear indications that the incidence and velocity of the outflows are correlated with the stellar mass, central concentration, star formation rate, and star formation surface density, with a clear increase in the incidence in the most active star forming galaxies \citep[e.g.][]{2005ApJS..160..115R, 2019ApJ...873..122D, 2019ApJ...886...11L}, with the latter two examining high redshift samples ($1 \lesssim z \lesssim 4$). Indeed, powerful outflows are found to be ubiquitous in high redshift star-forming galaxies, where the star formation rates and surface densities are high \citep[][]{2009ApJ...692..187W, 2010ApJ...719.1503R, 2011ApJ...733..101G, 2019ApJ...875...21F, 2019MNRAS.487..381S}. Outflows have also long been known to be present in isolated AGN, with clear indications that their incidence and velocities are correlated with AGN luminosity and perhaps Eddington ratio \citep[e.g.][]{2016ApJ...817..108W, 2020A&A...633A.134L, 2020MNRAS.492.4680W, 2021MNRAS.503.5134A}. Radio properties (e.g. radio luminosity) of galaxies can also play a role in outflow incidence and properties \citep[e.g.][]{2014MNRAS.442..784Z,2019A&A...631A.132M, 2020A&A...644A..54S}; however, we do not explore the radio properties of the galaxies in our sample in the present work. 
\par
It is well-known that nuclear star formation and accretion activity are both enhanced in mergers and show a clear dependence on merger stage \citep[e.g.][]{2011MNRAS.418.2043E,2013MNRAS.435.3627E,  2014MNRAS.441.1297S}. It is thus expected that outflows would be prevalent in mergers, and that their incidence and properties might show a dependence on merger stage. Indeed, many studies have detected outflows in interacting galaxies and mergers \citep[e.g.][]{2005ApJS..160...87R, 2005ApJ...632..751R, 2012ApJS..203....3S, 2012MNRAS.424..416W, 2013ApJ...776...27V, 2018MNRAS.480.3993B, 2018ApJ...864L...1G, 2018A&A...616A.171P, 2020A&A...635A..47H, 2021MNRAS.502.3618G}. However, these studies have by and large been carried out on small samples of galaxies, and often target `extreme' objects, such as ultra-luminous infrared galaxies (ULIRGs) or dust reddened quasars \citep[e.g.][and references therein]{2011ApJ...733L..16S, 2014A&A...562A..21C, 2017ApJ...850...40R, 2020A&A...633A.134L, 2020A&ARv..28....2V, 2021MNRAS.505.5753F}, which are not representative of the general galaxy population. In addition to these studies, follow-up observations of the several hundred galaxies with double-peaked narrow emission line profiles in optical surveys have revealed that the vast majority are associated with gas outflows in mergers \citep[e.g.][]{2013ApJ...769...95B, 2015ApJ...813..103M, 2018ApJ...867...66C, 2018MNRAS.473.2160N} in both low and moderate ($z \lesssim 1.6$) redshifts, again suggesting a close tie between galaxy-galaxy interactions and the presence of outflows.
There have also been attempts to quantify the relative role played by star formation and accretion activity on driving the outflows, with indications that the outflow incidence and velocities are higher in galaxies harbouring AGN  \citep[e.g.][]{2016MNRAS.456.1195H, 2017A&A...606A..36C,2019MNRAS.489.4016S, 2021MNRAS.503.5134A}, even at redshifts out to $z\sim 3.8$ \citep[][]{2019ApJ...886...11L}.
\par
While it is clear that outflows are associated with galaxy mergers, it is unclear how the merger environment itself impacts the incidence and properties of these outflows.  In order to determine the effect of the merger on the outflows, a systematic study of outflows in mergers compared to a matched control sample of isolated galaxies must be conducted. However, there has thus far been no large-scale systematic study that examines the properties of outflows in galaxy mergers compared with isolated galaxies to determine if the merger environment enhances or suppresses the outflowing material, and to explore how the presence and properties of outflows depend on merger stage. Most importantly, since there have not yet been any outflow studies in which galaxy mergers have been compared to a matched control sample of isolated galaxies, the exact effect of mergers on outflows is not yet known, and the dependence of outflow properties on merger stage is completely unexplored. Since mergers induce gas inflows \citep[e.g.][]{1991ApJ...370L..65B, 2018MNRAS.479.3952B}, it might be natural to expect that the outflow incidence and properties are suppressed in merging galaxies compared to isolated galaxies. Alternatively, the gas inflows and subsequent increase in the central gas densities might cause the outflows to sweep up more material, thereby increasing the outflow luminosity and resulting in an enhancement of the outflow incidence and energetics in mergers. Mergers also cause a dilution of the central gas phase metallicity \citep[e.g.][]{2010ApJ...721L..48K,2013MNRAS.435.3627E, 2019MNRAS.482L..55T}. Together with the enhanced nuclear activity, mergers may disturb the interstellar medium (ISM) in some way that enhances the outflows. Given that galaxy mergers are thought to play a major role in a galaxy's morphology, star formation rate, nuclear accretion rate, metallicity, and gas content, exploring the incidence and properties of outflowing gas in mergers is of fundamental importance to our understanding of the critical role galaxy mergers play in galaxy evolution.

In this paper, we carry out the first large-scale systematic study of outflows in merging environments compared to non-merging environments. This paper builds upon a large body of work on galaxy pairs and post-mergers, drawn from the Sloan Digital Sky Survey (SDSS), that uses robust control samples of isolated galaxies to quantify the impact of mergers on galaxy and AGN properties \citep[][]{2008AJ....135.1877E, 2010MNRAS.407.1514E, 2011MNRAS.412..591P, 2011MNRAS.418.2043E, 2012MNRAS.426..549S, 2013MNRAS.433L..59P, 2013MNRAS.430.3128E, 2013MNRAS.435.3627E, 2014MNRAS.441.1297S, 2015MNRAS.449.3719S, 2015MNRAS.451L..35E, 2016MNRAS.461.2589P}. Here, we utilize the [OIII]~$\lambda$5007 emission line to trace ionised gas in pairs, post-mergers, and isolated galaxies, with the goal of isolating the impact of the merging environment on outflow properties. We examine $\sim$4,000 paired galaxies and $\sim$250 post-mergers, consisting of AGN and star-forming (SF) galaxies, matched to a control sample of $\sim$12,000 galaxies. In Section \ref{sec: Sample Selection}, we provide details on the construction of the pair, post-merger and control samples, and discuss our fitting procedure and outflow selection criteria in Section \ref{sec:Spectral Fitting and Outflow Criteria}. In Section \ref{sec: General Outflow Characteristics in Sample}, we discuss the bulk outflow characteristics of our sample, while in Section \ref{sec: Outflow Characteristics as a Function of Merger Stage} we examine outflow properties as a function of projected physical separation ($r_p$). In Section \ref{sec: Discussion}, we discuss the implications of our results, and Section \ref{sec: Conclusion} contains a results summary and directions of future work. Throughout this paper, we assume a flat $\Lambda$CDM cosmology with $H_0=70$~km~s$^{-1}$~Mpc$^{-1}$, $\Omega_m = 0.27$, and $\Omega_\Lambda = 0.73$.

\section{Sample Selection}\label{sec: Sample Selection}

\subsection{Galaxy Pair and Post-Merger Samples}

The galaxy merger sample used in this work has been described in depth in previous papers in this series and is drawn specifically from the sample described in \citet[][]{2014MNRAS.441.1297S}. We refer the reader to this previous work for full details on the sample, and the careful considerations applied to remove any selection effects. In brief, the pairs sample consists of close spectroscopic galaxy pairs,  and visually classified post-mergers from the  Sloan Digital Sky Survey Data Release 7 \citep[SDSS DR7,][]{2009ApJS..182..543A} Main Galaxy Sample. The extinction-corrected r-band Petrosian magnitudes are between $14.0 \leq m_r \leq 17.77$, the redshift range is $0.02 \leq z \leq 0.2$, and all objects are spectroscopically classified as a galaxy. Projected physical separations are required to be $r_p \leq 80 h_{70}^{-1}$ kpc and the relative velocity must be $\Delta V \leq 300$~km~s$^{-1}$. Further, the stellar mass ratios between the pairs must be $0.25 \leq M_1/M_2 \leq 4$ in order to select major mergers. It should be noted that not every galaxy pair is visibly interacting; it is possible some of these galaxies may never merge or are non-interacting projected pairs. These galaxies are referred to as `paired galaxies'. 

The visually classified post-merger sample is initially drawn from the Galaxy Zoo \citep[][]{2008MNRAS.389.1179L} catalogue presented by \citet[][]{2010MNRAS.401.1043D}.  \citet[][]{2013MNRAS.435.3627E} imposed further restrictions on the post-merger sample by removing `normal' irregular galaxies and pairs that have not fully coalesced into an appropriate post-merger state, among other requirements. This sample of visually classified post-merger galaxies is referred to as the post-merger sample.

For convenience, we will refer to the paired galaxies, which represent the `early' stages of galaxy interactions (0-80 kpc), and the post-merger sample, which represents the `late' stages of galaxy interactions (i.e. post-coalescence), collectively as the `merger' sample. 

Finally, the merger sample is matched to the final public all-sky WISE source catalogue, \footnote{\url{https://wise2.ipac.caltech.edu/docs/release/allsky/}} where a merger is matched if the positions agree to within six arcsec. Since we employ mid-infrared AGN classification in one of our sub-samples (as discussed in the paragraph below), we require that all objects be detected with a signal-to-noise greater than $5\sigma$ in the 3.4 $\mu$m $W1$ and 4.6 $\mu$m $W2$ bands. Our initial merger sample thus consists of 321 post-mergers and 5026 paired galaxies, with $\sim$ 55 per cent of galaxies meeting the WISE detection criteria.

The galaxies in our sample are classified as AGN or SF galaxies based on optical narrow emission line ratio Baldwin-Phillips-Terlevich (BPT) diagnostics \citep[][]{1981PASP...93....5B} and mid-infrared colour selection. For optical selection, we use the traditional K03 \citep[][]{2003MNRAS.346.1055K} and K01 \citep[][]{2001ApJS..132...37K} selection criteria to classify galaxies as K01 AGN, K03 AGN, and K03 SF using extinction-corrected fluxes provided by \citet[][]{2012MNRAS.423.2690S}. K01 AGN are taken to be `pure' AGN, whereas K03 AGN are dominated by AGN activity, but starbursts may contribute to the ionised emission. K03 SF galaxies (hereafter SF galaxies) are dominated by stellar activity, but may have a small contribution \citep[up to 3 per cent,][]{2006MNRAS.371..972S} from AGN. However, see \citet[][]{2021ApJ...922..156A} for an alternative view on the `mixing sequence' picture. We remove possible AGN contamination from our SF sample by requiring our SF galaxies to have a WISE colour cut of $W1 - W2 < 0.5$ since optical flags might not detect some AGN, especially in late-stage mergers when AGN are highly obscured by gas and dust \citep[][]{2014MNRAS.441.1297S, 2019MNRAS.487.2491E}. Since mid-infrared selected AGN are more common in mergers \citep{2014MNRAS.441.1297S, 2018MNRAS.478.3056B}, we also sub-divide the merger sample into `WISE+BPT' AGN to determine if mid-infrared selection has an impact on outflow properties. Here, we require both an optical AGN (either K01 or K03) and an infrared AGN detection ($W1 - W2 > 0.5$). While the WISE colour cut we impose here is more relaxed than the common $W1 - W2 > 0.8$ cut, it is not significantly less reliable and provides a more complete sample of AGN at $z \lesssim 0.5$ \citep[][]{2018MNRAS.478.3056B}. We require the WISE AGN to be BPT AGN so we can control for the power source of the outflow; since we impose an [OIII]~$\lambda$5007 luminosity matching requirement between mergers and controls (see below paragraph), we want to ensure the [OIII]~$\lambda$5007 is tracing the AGN luminosity. Galaxy properties such as the stellar mass and total star formation rate (SFR) are taken from the \citet[][]{2014ApJS..210....3M} and \citet[][]{2004MNRAS.351.1151B} catalogues. We note that while the SFR for SF galaxies is computed by modelling the H$\alpha$ emission line, the SFR for AGN is computed using D4000. 

\subsection{Matched Control Sample}

The merger sample is compared to a set of physically similar isolated galaxies, drawn from the same database as the mergers, so that each merger has a matched set of controls associated with them. We consider a galaxy to be isolated if the projected distance $r_p$ to its closest spectroscopic companion is greater than 200 kpc. As discussed in \citet[][]{2009MNRAS.397..748P} and following \citet[][]{2010MNRAS.407.1514E,2012MNRAS.426..549S,2013MNRAS.430.3128E}, controls are matched simultaneously in redshift, mass, and local density of galaxies, all of which may show a dependence on outflow incidence and properties. Since the [OIII]~$\lambda$5007 luminosity can affect outflow properties \citep[see][]{2017A&A...601A.143F}, we also match in this parameter. Further, matching in redshift and [OIII]~$\lambda$5007 luminosity ensures that our ability to detect an outflow will not be biased towards either the mergers or controls. SF galaxies are additionally matched in SFR. The local density of galaxies is defined as 
\begin{equation}
    \Sigma_n = \frac{n}{\pi d_n^2}
\end{equation}
\noindent where $d_n$ is the projected distance in Mpc to the $n$th nearest neighbour within $\pm1000$~km~s$^{-1}$. Normalized densities $\delta_n$ are computed relative to the median $\Sigma_n$ within a redshift slice of $\pm 0.01$. Following \citet[][]{2014MNRAS.441.1297S}, we take $n=5$. Intuitively, this local density of galaxies describes the large-scale neighbourhood of a given galaxy. In our case, we use the fifth nearest neighbour to characterise this density. As explained in \citet[][]{2009MNRAS.397..748P} and \citet[][]{2010MNRAS.407.1514E}, this parameter is sensitive to the SFR and gas content of a galaxy -- isolated galaxies can have larger amounts of cold gas and subsequently enhanced star formation compared to mergers. Hence, an unbiased control sample must account for this parameter.

The matching tolerance on redshift is $\pm 0.05$, while for mass and local galaxy density it is $\pm 0.1$ dex. The [OIII]~$\lambda$5007 luminosity tolerance is $\pm 10$ per cent. We attempt to match each object in our merger sample with three \textit{unique} controls. Mergers that could not be matched with at least three unique controls were dropped from the sample. Our matching criteria leaves us with 1906 SF mergers, 1375 K03 AGN mergers, 562 K01 AGN mergers, and 72 WISE+BPT AGN mergers, including paired galaxies and post-mergers. 

We note that while we do employ WISE colour cuts on our control sample when appropriate (e.g. when making the control sample for WISE+BPT AGN and filtering AGN from our SF control sample), the WISE controls do not necessarily have the same 5$\sigma$ detection in the first three WISE bands as the mergers do.

\begin{figure*}
    \centering
    
    \begin{subfigure}[b]{\paperwidth}
    \includegraphics[width=0.85\textwidth,height=0.23\textheight]{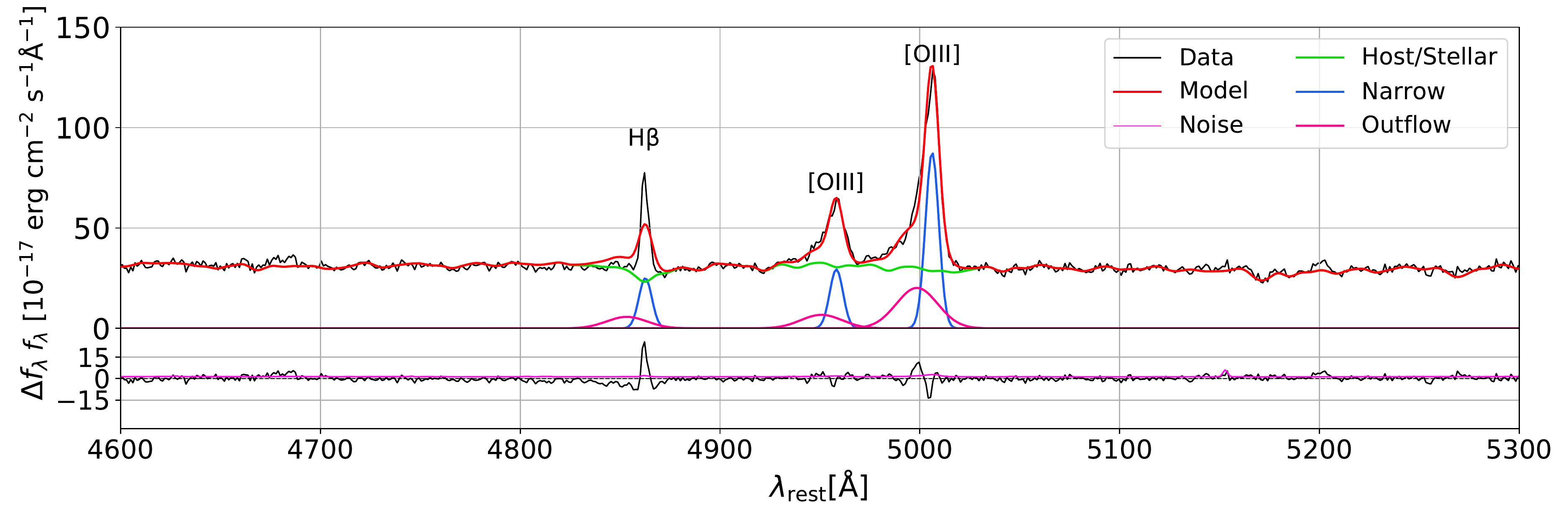}
    \end{subfigure}

    \begin{subfigure}[b]{\paperwidth}
    \includegraphics[width=0.85\textwidth,height=0.23\textheight]{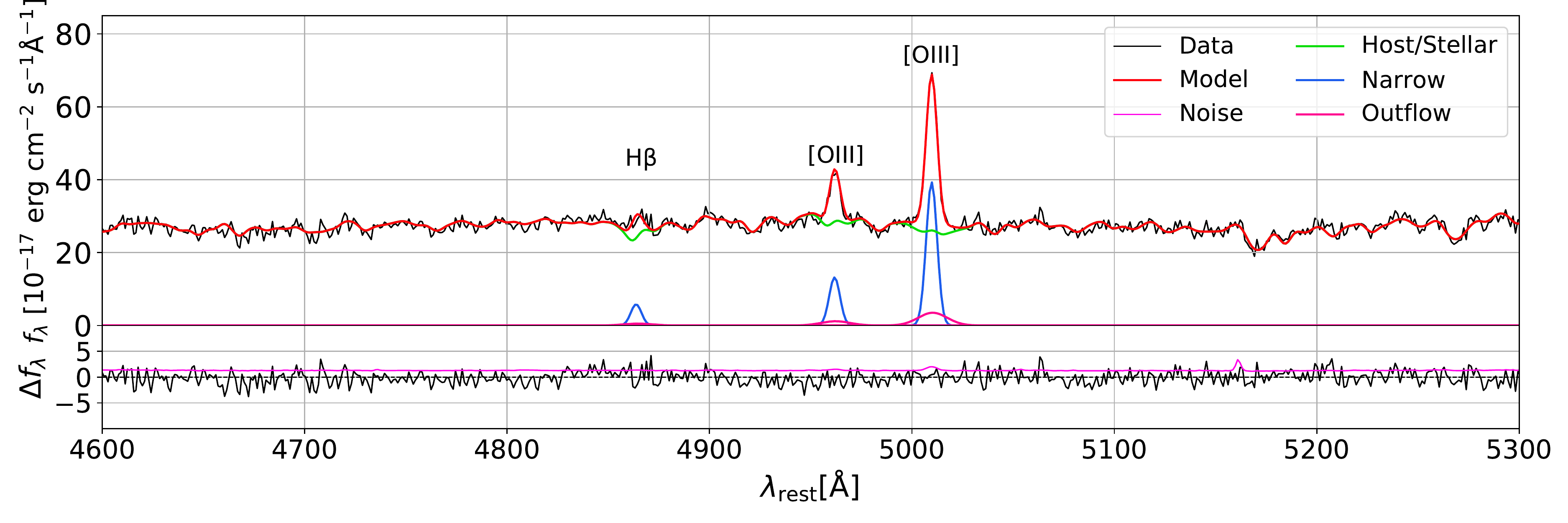}
    \end{subfigure}

    \begin{subfigure}[b]{\paperwidth}
    \includegraphics[width=0.85\textwidth,height=0.23\textheight]{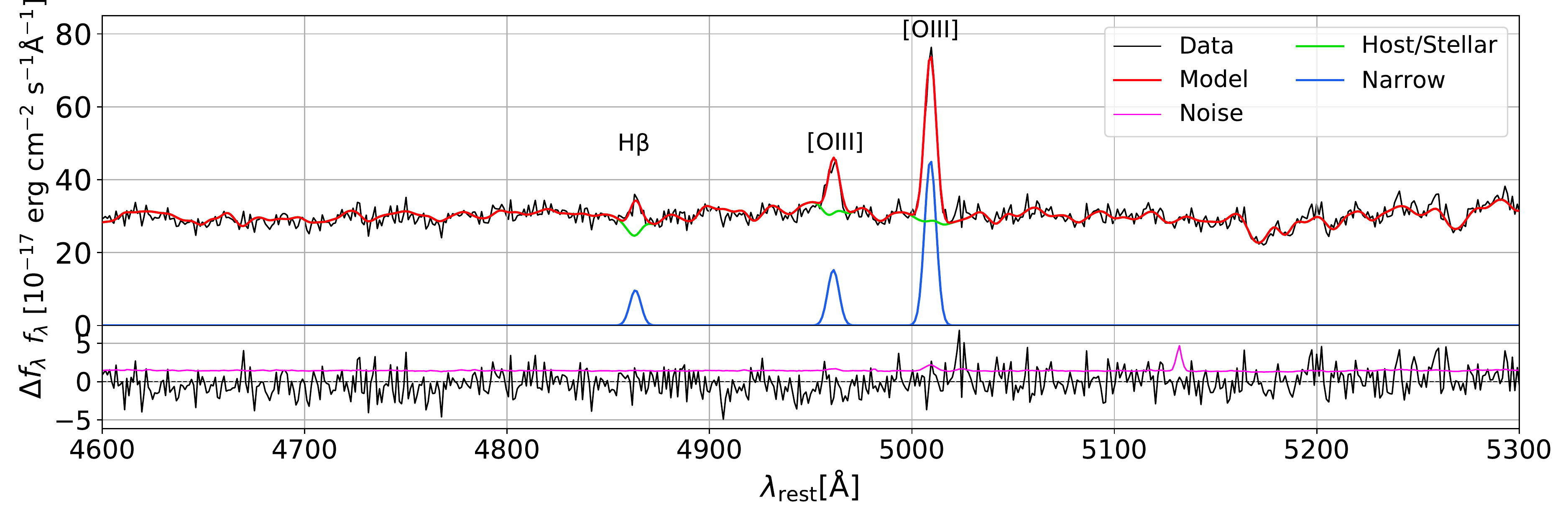}
    \end{subfigure}
    
    \begin{subfigure}[b]{\paperwidth}
    \includegraphics[width=0.85\textwidth,height=0.23\textheight]{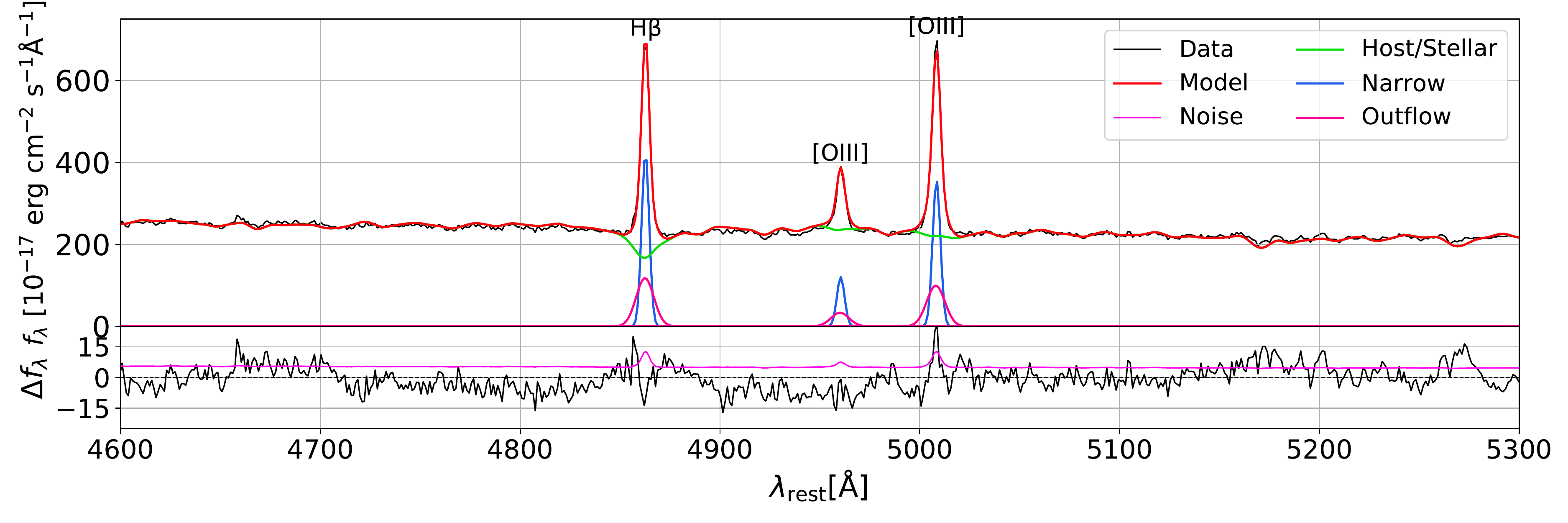}
    \end{subfigure}
    
\caption{Example spectra fits. Subsequent identifiers are the DR7 SpecObjIDs. \textit{Top Panel}: AGN SDSS 456217686624436224 exhibits a reasonably strong outflow detected both in the initial testing and with \textsc{emcee}. The bad fit to the H$\beta$ emission line is caused by a poor stellar template match to the continuum and does not impact our results. \textit{Second to Top Panel}: AGN SDSS 407803917544456192 was fit for an outflow with \textsc{emcee}, but the fit was not significant enough to be classified as an outflow. \textit{Second to Bottom Panel}: AGN SDSS 160665218618228736 failed the initial testing, hence \textsc{emcee} did not fit it for an outflow. \textit{Bottom Panel}: SDSS 631860897941291008 shows a reasonably strong outflow in a SF galaxy.}
\label{fig:OutflowFits}
\end{figure*}

\section{Spectral Fitting Procedures and Outflow Criteria}\label{sec:Spectral Fitting and Outflow Criteria}

\subsection{Spectral Fitting Software}\label{subsec: Spectral Fitting Software} 

We use the [OIII]~$\lambda$5007 emission line to trace outflows in all mergers and controls. This line is one of the strongest features in the optical spectrum of all emission line galaxies, and is located in a wavelength region free from strong stellar absorption features. Since it is produced in the lower density narrow line region, any asymmetry and broadening of the profile can often be attributed to large-scale winds. For these reasons, it is commonly used in the literature to search for and characterize outflows \citep[e.g.][]{2014MNRAS.441.3306H, 2016ApJ...828...97B, 2016A&A...588A..41C, 2016A&A...592A.148K, 2016MNRAS.459.3144Z, 2020MNRAS.492.4680W}. 
Spectral fitting is done with the open-source Python code Bayesian AGN Decomposition Analysis for SDSS Spectra (\textsc{badass}) version 7.6.5.\footnote{https://github.com/remingtonsexton/BADASS3} A full description of the code can be found in \citet[][]{2021MNRAS.500.2871S}. Here, we just give a brief summary of the fitting techniques and input parameters. \textsc{badass} works by modelling various spectral components (e.g. power-law continuum, emission lines, etc.) and sequentially subtracting them from the spectra to be fit independently. Once these components are subtracted off, only the underlying stellar continuum remains. The stellar continuum may be fit with either empirical stellar templates or a single stellar population model, depending on the quality (signal-to-noise ratio, S/N) of the spectrum. Emission lines may be fit with a variety of line profiles, but here we use a Gaussian profile that is fit for amplitude, FWHM, and velocity offset. Initial fits to the spectra are obtained using the \textsc{slsqp} algorithm from the standard Python library \textsc{scipy.optimize.minimize}. The parameter fits from this algorithm are then used as the initial guesses for a Markov Chain Monte Carlo (MCMC) algorithm, implemented by the \textsc{emcee} package \citep[][]{2013PASP..125..306F}, to estimate robust parameter uncertainties. 

Outflow presence may be (and, in this work, is) tested for during the initial fit by fitting both a single Gaussian and a double Gaussian model to the [OIII]~$\lambda$5007 emission line. The fit is repeated for a set number of iterations (50 in our case) with slightly different initial guesses, which provides a more reliable estimation of parameter values than a single fit. The median value for the fitted parameters is used to assess the possible presence of outflows by comparing the fitted parameters (amplitude, FWHM, and velocity offset) of the `core' and outflowing line components. If the uncertainty overlap between the core and outflowing component for one of these parameters is within a certain user-defined $\sigma$ value (e.g. 1$\sigma$, 2$\sigma$), then \textsc{badass} assumes there is no outflow. In addition to this uncertainty check, a statistical F-test is performed to compare the single and double Gaussian models. This test effectively calculates how much better the double Gaussian fit is and if using it is justified. The resulting F-statistic is converted to a p-value, and if it is less than a user-defined value then \textsc{badass} assumes there is no outflow. If the preliminary outflow tests are passed, the final fit using \textsc{emcee} will include an outflow component in the model. 

The AGN we are fitting are type II AGN, so we disable unnecessary fitting components (namely AGN power-law, broad lines, and Fe II contamination). For the preliminary outflow testing, we check for overlapping uncertainties in the FWHM and velocity offset parameters, and apply the F-test as described above. Although the purpose of the preliminary outflow testing is to discard spectra that clearly do not have outflows, there is a danger that a weak outflow will be overlooked during the initial fit. To guard against that, we apply generous cutoffs for the outflow tests. Specifically, we only require the core component and outflow component FWHM and velocity offset parameters to be different at a $0.5\sigma$ level, and the F-test p-value be greater than 0.70. Fits with outflows that do not meet the initial criteria for detection will not significantly improve in the final fit using MCMC, and will hence not be reliably detected. False-positives detected by \textsc{emcee} will be flagged when the final fits are analysed (see Section \ref{subsec: Outflow Selection Criteria}). The SF spectra are fit in the same way as the AGN spectra. However, we do not require outflows to be blue-shifted in the SF galaxies because outflows in SF galaxies have a much more symmetrical line profile even in the presence of outflows and are thus much harder to detect (see Sections \ref{subsec: Outflow Selection Criteria} and \ref{sec: General Outflow Characteristics in Sample}). 

For both the AGN and SF spectra, we run the MCMC sampler for a minimum of 5,000 iterations and a maximum of 25,000 iterations using 20 walkers. We use the `median' convergence option in the autocorrelation analysis which uses the median autocorrelation time of all free parameters to assess if the MCMC chains have converged. Effectively, if enough parameters have converged, the MCMC analysis will stop before it reaches 25,000 iterations. Once the parameters have converged, the MCMC will continue sampling for 5,000 iterations. These iterations are used to generate the posterior distribution and subsequently the best-fit parameter values and their uncertainties. 

It should be noted that \textsc{badass} will not fit spectra that are of insufficient quality, as determined by the number of `good' channels. A good channel is defined by having flux and flux errors greater than zero at each pixel, and not being flagged by the SDSS for having bad pixels. If the fraction of good channels in a spectrum is less than 60 per cent, then the spectrum will not be fit. This will cause additional mergers to be dropped from the final analysis, since a merger that was matched with three controls may not have three controls that are of sufficient quality.


\begin{figure}
    \centering
    \begin{subfigure}[b]{\columnwidth}
    
    \includegraphics[width=\columnwidth]{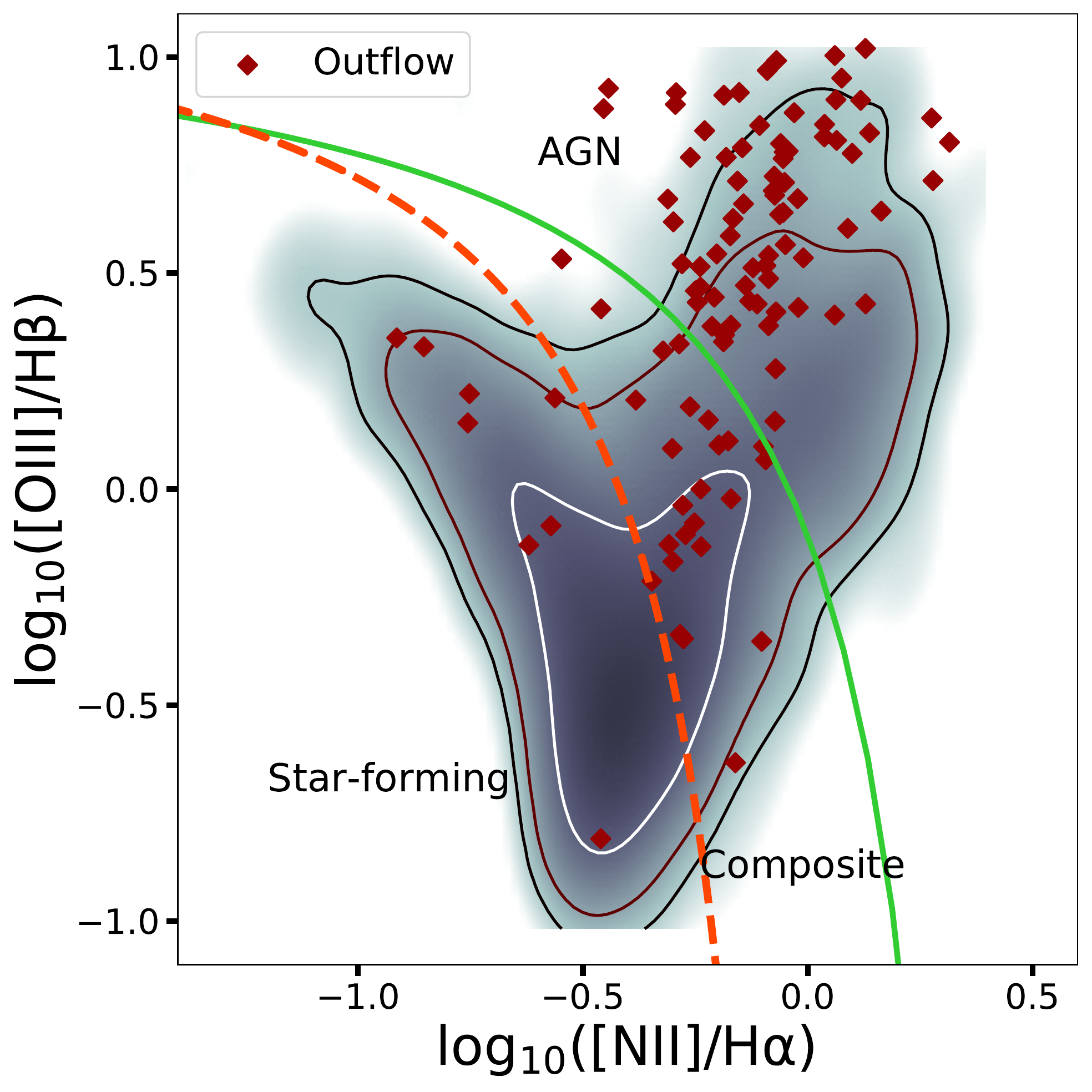}
    
    \end{subfigure}

\begin{subfigure}[b]{\columnwidth}

    \includegraphics[width=\columnwidth]{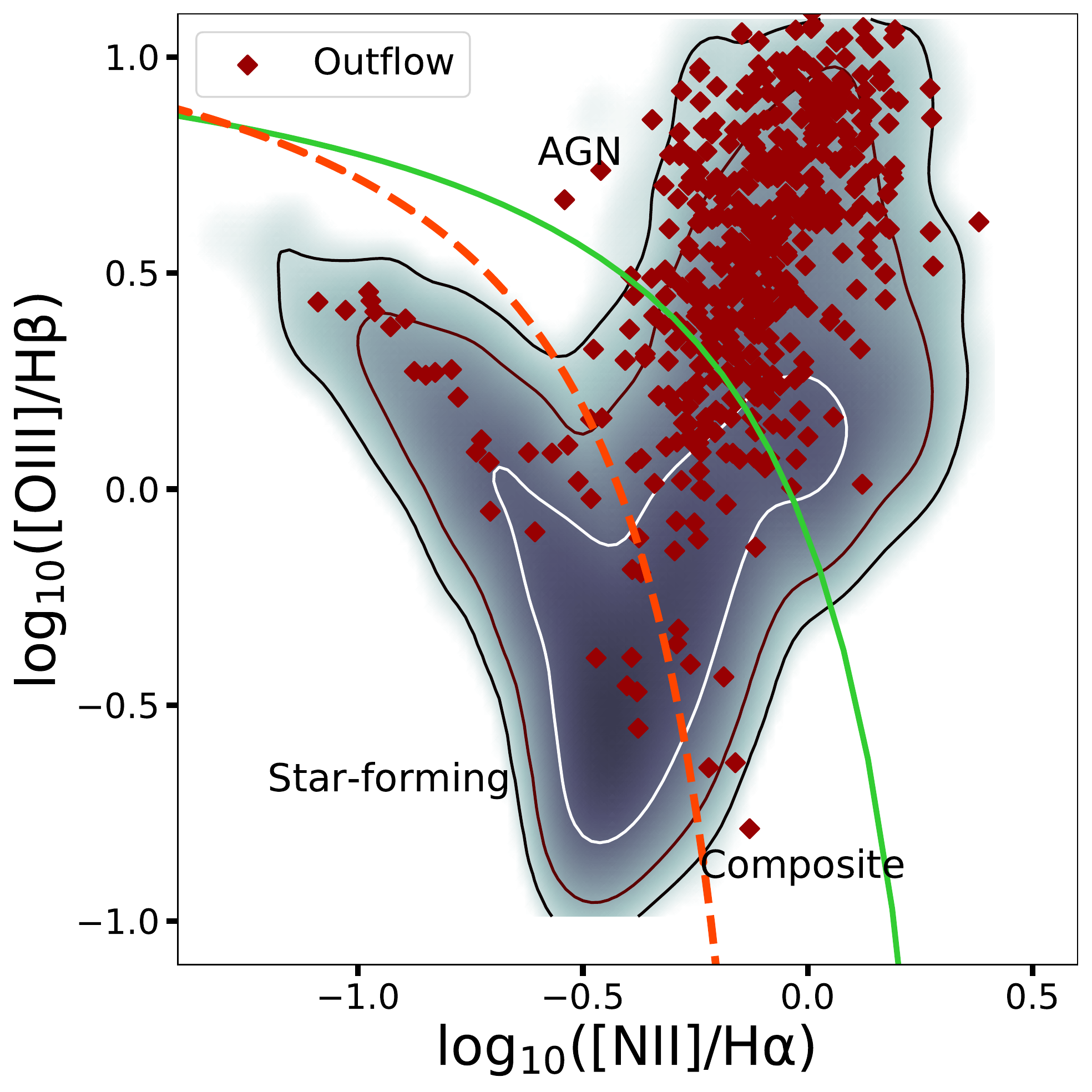}

    \end{subfigure}
\caption{BPT diagrams of the entire merger (top panel) and control (bottom panel) samples. Darker colours indicate denser regions populated by objects with no outflows. Contours enclose the regions containing 99.5 per cent, 95 per cent and 68 per cent of the objects without outflows. The orange dashed line denotes the K03 SF curve, below which galaxies are dominated by star formation. The solid green line denotes the K01 AGN curve, above which galaxies are dominated by AGN. Red diamonds indicate objects with outflows. }
\label{fig:BPT}
\end{figure}

\begin{figure}
    \centering
    \begin{subfigure}[b]{\columnwidth}
    
    \includegraphics[width=\columnwidth]{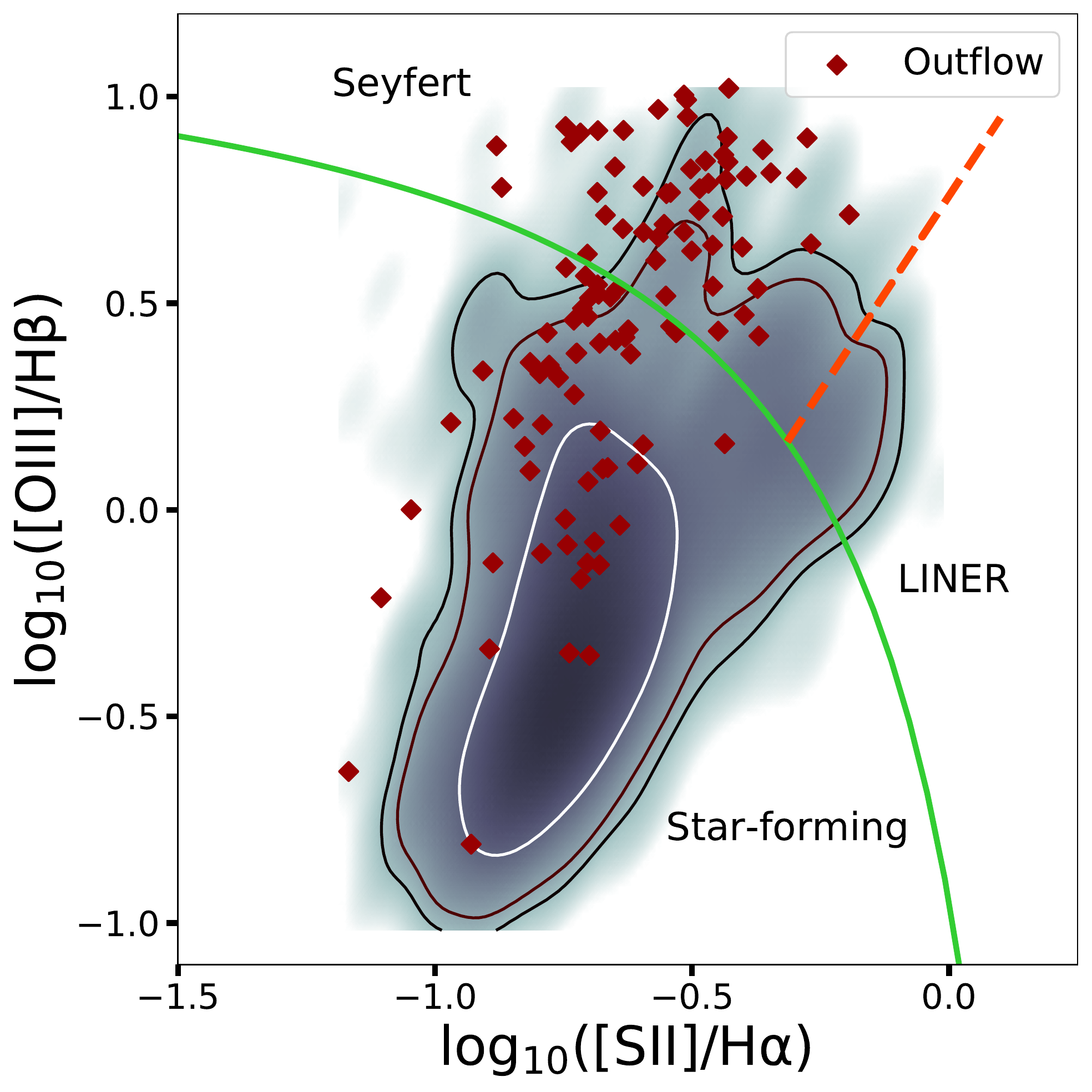}

    \end{subfigure}

\begin{subfigure}[b]{\columnwidth}
    \includegraphics[width=\columnwidth]{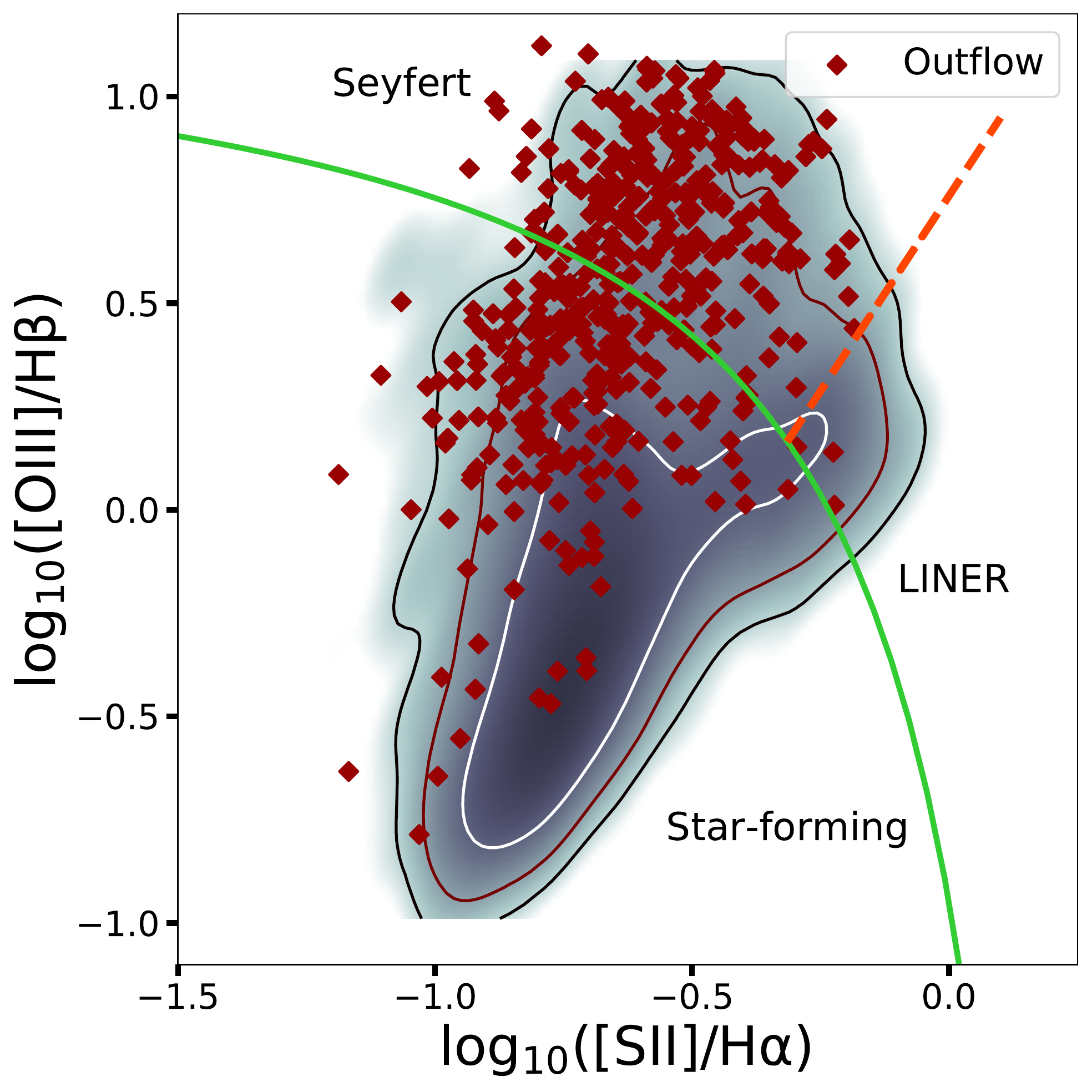}

    \end{subfigure}
\caption{Alternate BPT diagram of the entire merger (top panel) and control (bottom panel) samples. Darker colours indicate denser regions populated by objects with no outflows. Contours enclose the regions containing 99.5 per cent, 95 per cent and 68 per cent of the objects without outflows. The solid green line separates AGN from SF galaxies, while the orange dashed line separates Seyfert galaxies from LINERs. Red diamonds indicate objects with outflows. }
\label{fig:BPT2}
\end{figure}

\begin{figure}
    \centering
    \begin{subfigure}[b]{\columnwidth}
    
    \includegraphics[width=\columnwidth]{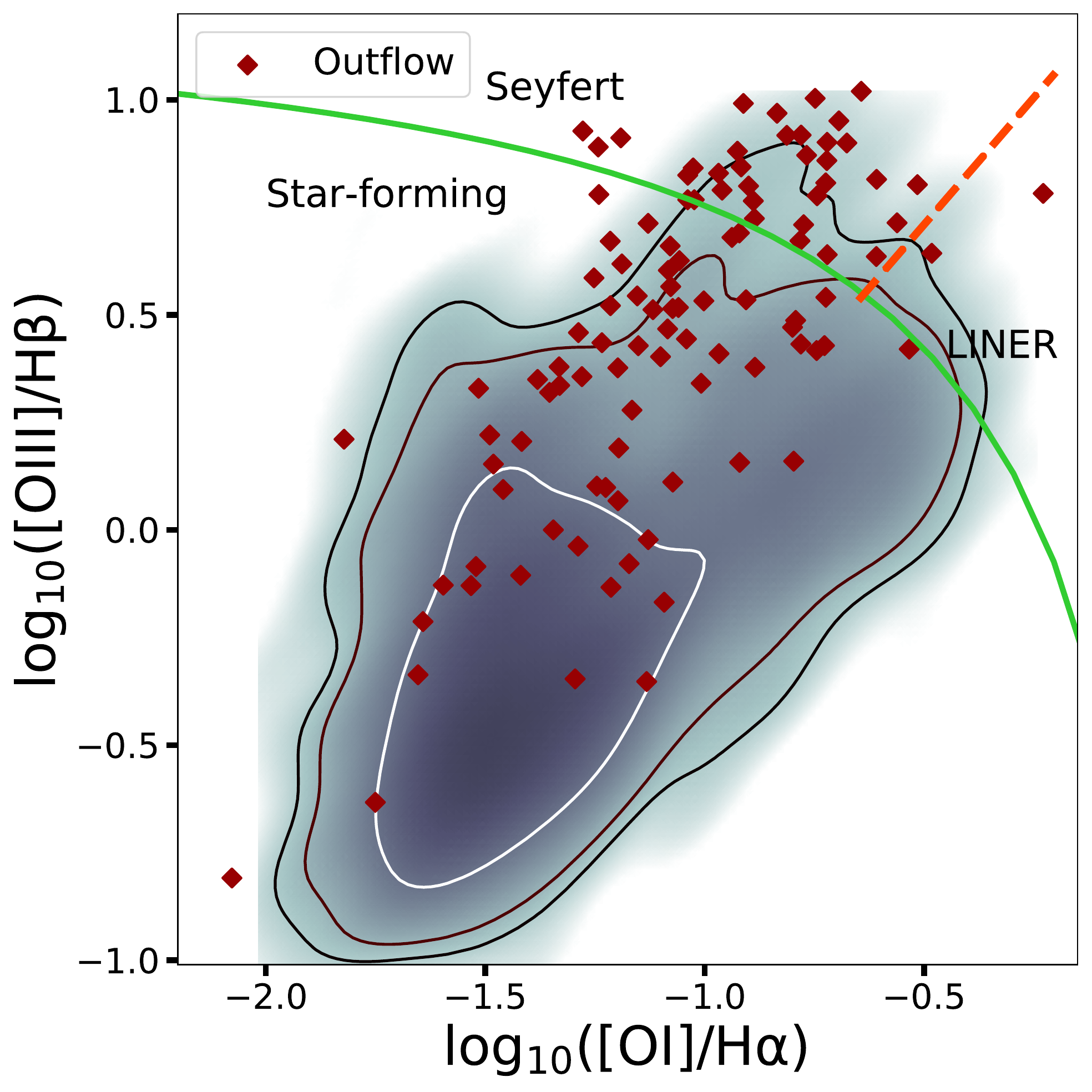}

    \end{subfigure}

\begin{subfigure}[b]{\columnwidth}
    \includegraphics[width=\columnwidth]{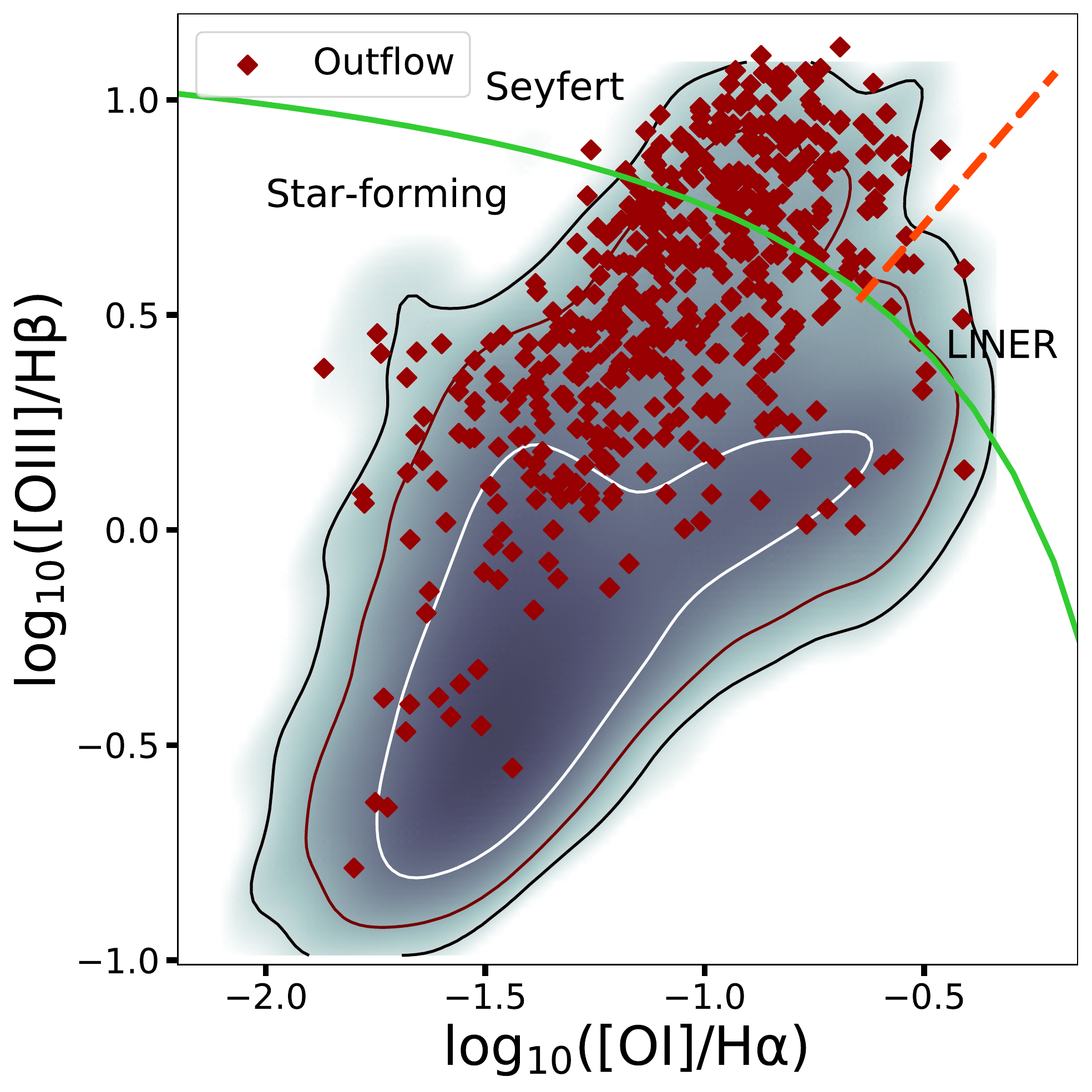}

    \end{subfigure}
\caption{Alternate BPT diagram of the entire merger (top panel) and control (bottom panel) samples. Darker colours indicate denser regions populated by objects with no outflows. Contours enclose the regions containing 99.5 per cent, 95 per cent and 68 per cent of the objects without outflows. The solid green line separates AGN from SF galaxies, while the orange dashed line separates Seyfert galaxies from LINERs. Red diamonds indicate objects with outflows. }
\label{fig:BPT3}
\end{figure}

\begin{figure}
    \centering
    \begin{subfigure}[b]{\columnwidth}
    
    \includegraphics[width=\columnwidth]{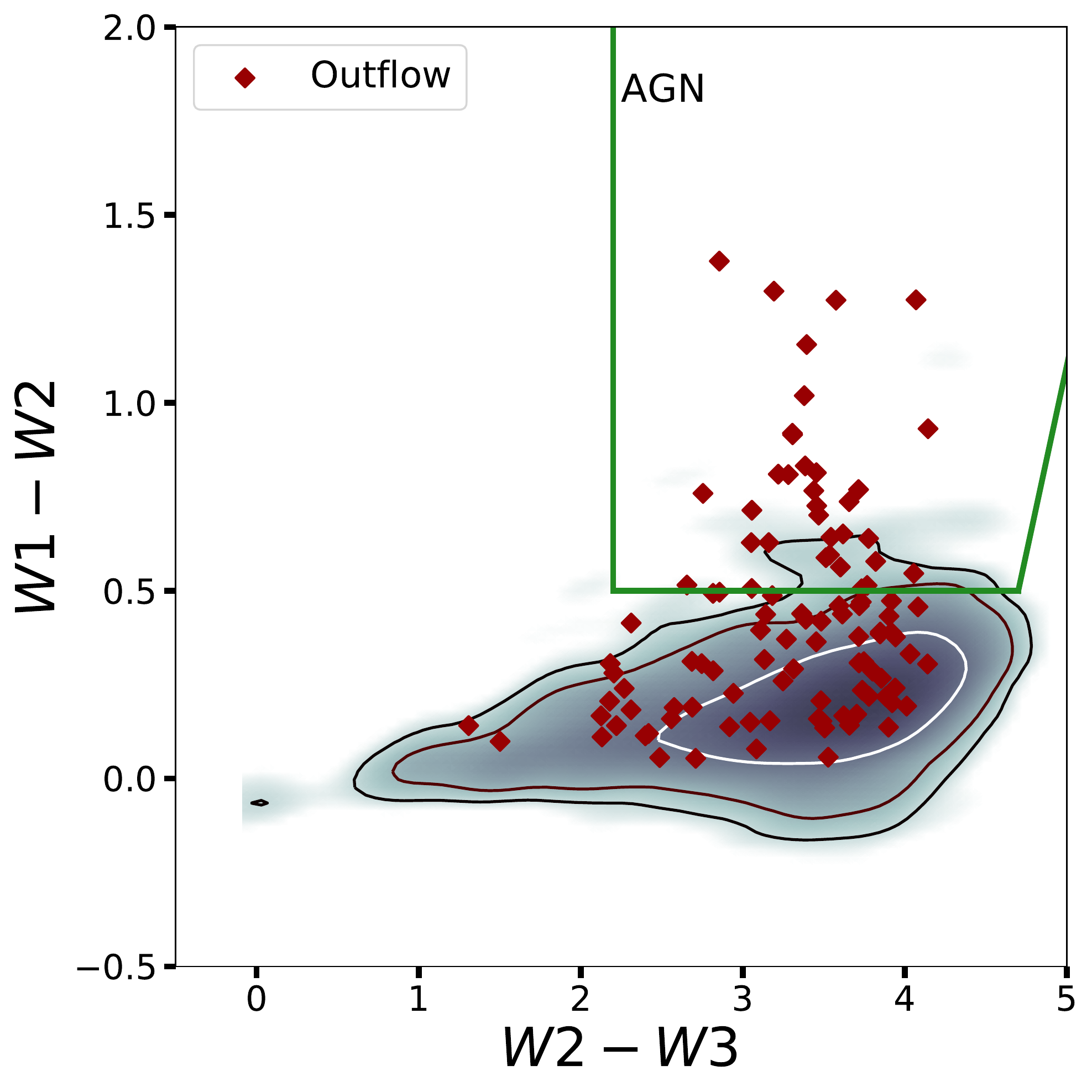}

    \end{subfigure}

\begin{subfigure}[b]{\columnwidth}

    \includegraphics[width=\columnwidth]{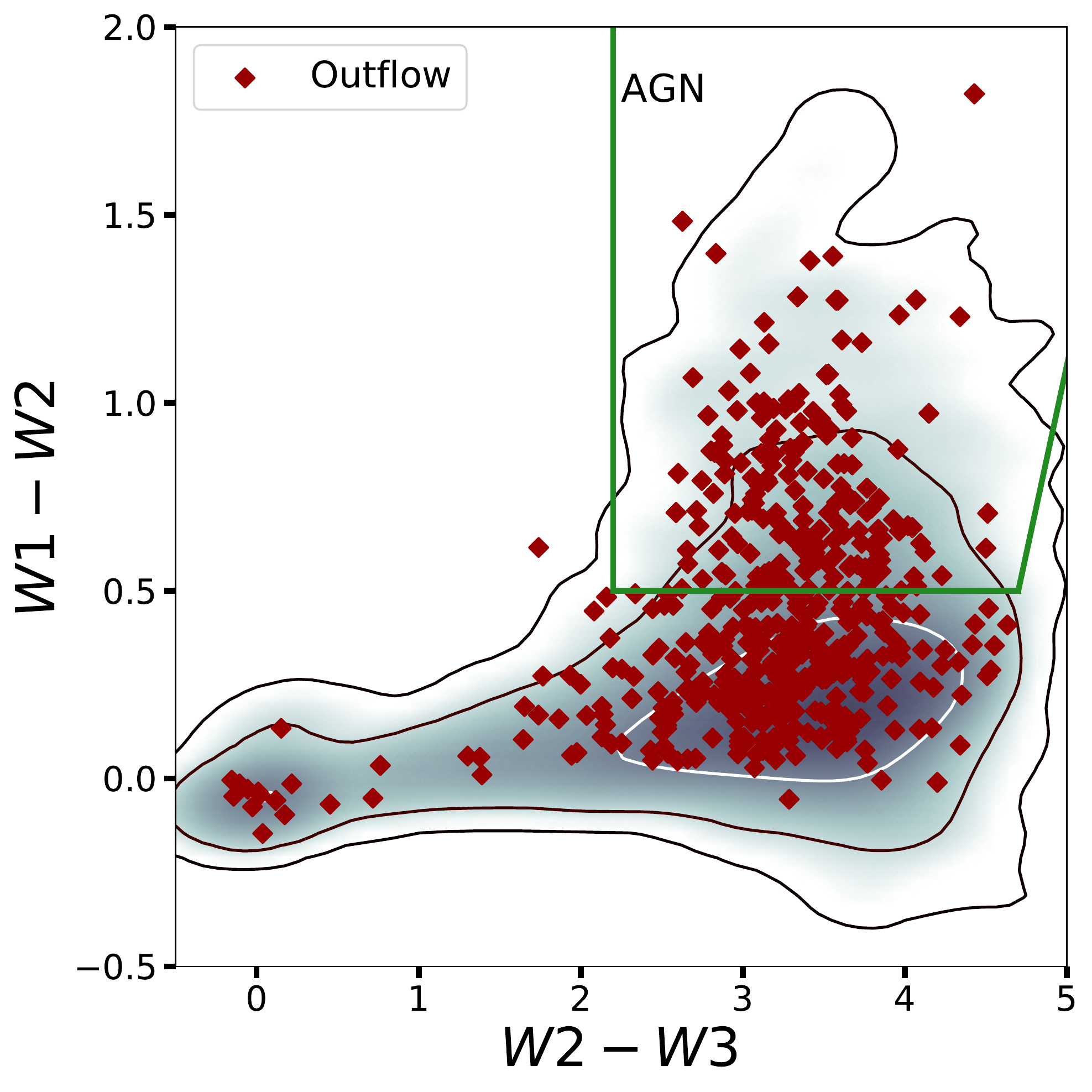}

    \end{subfigure}
\caption{WISE colour-colour diagrams of the entire merger (top panel) and control (bottom panel) sample. The green box displays the mid-infrared colour cut described in \citet[][]{2018MNRAS.478.3056B}, with WISE AGN populating the box's interior. Note our colour cut only requires $W1-W2 > 0.5$. Red diamonds indicate objects with outflows. Darker colours indicate denser regions populated by objects with no outflows. Contours enclose the regions containing 99.5 per cent, 95 per cent and 68 per cent of the objects without outflows.}
\label{fig:CC}
\end{figure}

\subsection{Outflow Selection Criteria}\label{subsec: Outflow Selection Criteria}

As discussed in Section \ref{subsec: Spectral Fitting Software}, we use a SLSQP bootstrap algorithm to perform an initial test for outflows, as well as a statistical F-test. For this initial test, we require that the outflowing component be broader than the core by only $0.5~\sigma$. Further, the p-value returned by the F-test must be greater than 0.70, indicating we are 70 per cent confident that our [OIII]~$\lambda$5007 line profile is best modelled by two Gaussian components instead of one. If a spectrum passes the initial outflow test, then an outflow component is included in the final MCMC model. Even if \textsc{badass} detects an outflow in the initial fit, the final fit given by \textsc{emcee} may not be reliable. To ensure outflows are robustly detected, the final fits must meet the following criteria. First, the outflow amplitude must be at least $1\sigma$ above the noise floor of the spectrum. Second, the FWHM of the outflow component must be larger than the FWHM of the `core' line within $1\sigma$ (in other words, the FWHM values between the line core and outflow component should not overlap within the uncertainties given by MCMC). Finally, for AGN only, the velocity offset of the outflow component must be blue-shifted with respect to the core line by at least $1\sigma$. This blue-shift requirement is consistent with numerous other studies that interpret broad, blue-wing components of emission lines as outflow signatures \citep[e.g.][]{2008MNRAS.387..639H,2013MNRAS.433..622M,2014MNRAS.442..784Z,2015A&A...580A.102C,2016MNRAS.456.1195H,2016A&A...585A.148B,2016A&A...592A.148K,2018ApJ...864L...1G,2021MNRAS.502.3618G}. Note that we do not impose a blue-shift requirement on the outflow component for SF galaxies. The vast majority of [OIII]~$\lambda$5007 line profiles in SF galaxies are highly symmetrical, but some show an underlying broad component or a slightly blue-shifted asymmetry in their line profile. We choose to be inclusive in our outflow definition for SF galaxies such that the profiles with an underlying broad component are flagged as outflows. The physical motivation for this decision is presented in Kadir et al. (in prep), where they employ the biconical outflow model of \citet[][]{2016ApJ...828...97B} and simulate the effects of extinction on the outflowing [OIII]~$\lambda$5007 line profile in an obscuring medium. These simulations show that when outflows arise from extended regions, as expected in SF galaxies, the observed line profile arising from the \textit{integrated} emission from the outflow is less affected by extinction and therefore displays a more symmetrical line profile compared to an outflow arising from a compact region of deeply embedded gas around an AGN. While we choose to relax the blue-shift requirement for SF galaxies for these reasons, we point out that the outflow incidence in SF galaxies is not significantly altered if the blue-shift requirement is imposed -- the outflow incidence decreases from $\sim$0.5 per cent to $\sim$0.4 per cent in both the merger and control samples -- leaving the results from this work unchanged. Similarly, if we relax the blue-shift requirement in our AGN to account for more uncommon outflow line profiles, such as red-shifted profiles or underlying broad profiles, the number of outflows in our AGN merger and control samples increases by 280. However, this increase does not alter our conclusions. We finally visually inspect the final fits and discard 11 lingering spurious outflow detections in the entire sample of mergers and controls. 

Of course, our ability to detect an outflow will depend on the S/N of the spectrum. The majority (>90 per cent) of spectra in our sample have a S/N greater than five in the [OIII]~$\lambda$5007 line, with only about two percent of the sample having a S/N less than three. Although we have no explicit S/N requirement when matching the mergers to controls, by matching in [OIII]~$\lambda$5007 luminosity and redshift we ensure that our ability to detect an outflow is not biased towards either the merger or its associated controls (assuming the aperture and exposure time is the same, or nearly so, for all targets). 

It should be noted that there is no single definition of an outflow when examining emission lines. While blue-shifted absorption lines provide an unambiguous signature of outflows \citep[][]{2013ApJ...776...27V}, the exact origin of the broad, blue-shifted, component of emission lines is more uncertain. As discussed in \citet[][]{2016A&A...588A..41C}, such a component may be the result of galaxy interactions or virialised motions within the galaxy. Further, outflows can produce a variety of emission line profiles; depending on outflow orientation and dust extinction, an outflow might create a non-Gaussian or red-shifted Gaussian line profile \citep[][]{2016ApJ...828...97B}. Outflows may also create a double-peaked [OIII]~$\lambda$5007 emission line profile \citep[e.g.][]{2018ApJ...867...66C, 2018MNRAS.473.2160N}. Dual AGN may also generate a double-peaked line profile \citep[e.g.][]{2011ApJ...739...44R}, but high resolution imaging and spectroscopy suggest that the majority of these double-peaked line profiles are associated with outflows \citep[e.g.][]{2011ApJ...735...48S, 2015ApJ...813..103M}. Regardless, our outflow selection criteria, as explained earlier in Section \ref{subsec: Outflow Selection Criteria}, is well-justified by the literature.

Similarly, there is no single definition of outflow velocity. A common measure of outflow velocity is $W_{80}$ \citep[e.g.][]{2014MNRAS.442..784Z}, which is defined as the flux enclosing 80 per cent of the emission line. For a Gaussian profile, there is a simple relation between $W_{80}$ and the FWHM of the line, namely $W_{80} = 1.088 \; \text{FWHM}$. However, there are other common measures of outflow velocity, the main difference being the velocity offset $v_{\text{off}}$ between the outflow and core emission line component is taken into account. While this definition is also used \citep[e.g.][]{2017ApJ...850..140T,2019ApJ...884...54M}, we prefer to use the $W_{80}$ metric because $v_{\text{off}}$ is sensitive to the amount of extinction in a galaxy \citep[][]{2016ApJ...828...97B}, and our mergers have a higher amount of extinction than our controls (see Section \ref{sec: Discussion}). 

Figure \ref{fig:OutflowFits} shows example fits to a variety of spectra. Various cases corresponding to different outflow test conditions described above are highlighted, including a strong outflow detection in an AGN, an outflow that was `found' during preliminary testing but was not significant when fit with \textsc{emcee}, no outflow detection at all, and an outflow detection in a SF galaxy. We note that the outflow in the AGNs are much more asymmetrical than it is in the SF galaxy. Indeed, this is a general characteristic with outflows in our sample -- SF galaxies, even in the presence of outflows, are much more symmetrical than their AGN counterparts (again, see Section \ref{sec: General Outflow Characteristics in Sample}).

\section{General Outflow Characteristics in the Sample}\label{sec: General Outflow Characteristics in Sample}

\begin{table*}
    \centering
    \caption{Bulk properties of each sub-sample. The average for each sub-sample property along with upper and lower errors are given. Outflow fraction errors were computed using binomial statistics with a 95 per cent confidence interval. A 2$\sigma$ mean standard error is used for all other errors. Note that the outflow luminosity listed here refers to the luminosity in the outflow component of the line profile, and we only require mergers to have one successfully analysed control.}

    \begin{tabular}{lccccccccc}
        \hline
         & & K03 SF & K03 SF & K01 AGN & K01 AGN & K03 AGN & K03 AGN  & WISE+BPT & WISE+BPT \\[-0.2cm]
         Property & & & & & & & & &\\[-0.25cm] 
         & & Merger & Control & Merger & Control & Merger & Control & Merger & Control  \\ 
         \hline
         Total objects &  & 1906 & 5396 & 562& 1610& 1375& 3922 &72 & 205\\[0.15cm]
         $\log_{10}$(L$_{\mathrm{[O\,\textsc{iii}]}\,\lambda5007}$) {(erg s$^{-1}$)} &  & $40.44_{0.03}^{0.03}$ & $40.44_{0.2}^{0.2}$& $41.09_{0.09}^{0.09}$ & $41.09_{0.05}^{0.05}$& $40.91_{0.06}^{0.06}$ & $40.90_{0.03}^{0.03}$ &$41.63_{0.13}^{0.13}$ &$41.63_{0.08}^{0.08}$ \\[0.15cm]
         Outflow fraction (\%) &  & $0.52_{0.27}^{0.44}$ & $0.56_{0.18}^{0.24}$ & $14.06_{2.80}^{3.20}$& $14.47_{1.68}^{1.81}$& $7.34_{1.32}^{1.51}$& $7.73_{0.82}^{0.88}$ &$38.89_{11.27}^{12.22}$ & $42.44_{6.90}^{7.10}$\\[0.15cm]
         $\log_{10}$(L$_{\mathrm{outflow}}$) {(erg s$^{-1}$)} &  & $40.06_{0.31}^{0.31}$ & $40.03_{0.18}^{0.18}$ & $40.71_{0.13}^{0.13}$& $40.72_{0.07}^{0.07}$& $40.66_{0.12}^{0.12}$& $40.69_{0.06}^{0.06}$ & $40.80_{0.18}^{0.18}$ & $40.87_{0.09}^{0.09}$\\[0.15cm]
         Outflow velocity {(km s$^{-1}$)} &  & $400_{250}^{250}$ & $320_{80}^{80}$ & $690_{70}^{70}$& $730_{40}^{40}$& $700_{70}^{70}$& $720_{30}^{30}$ &$710_{140}^{140}$ & $830_{80}^{80}$\\[0.15cm]
         $\log_{10}$(SFR) {(M$_{\star}$ yr$^{-1}$)} &  & $0.54_{0.02}^{0.02}$ & $0.53_{0.01}^{0.01}$ & $0.44_{0.28}^{0.28}$& $0.15_{0.04}^{0.04}$& $0.52_{0.10}^{0.10}$& $0.44_{0.02}^{0.02}$ & $0.94_{0.17}^{0.17}$ & $0.81_{0.09}^{0.09}$\\[0.15cm]
         \hline 
    \end{tabular}
    \label{Table:BulkProp}
\end{table*}

Since our sample allows us to explore the dependence of outflow incidence and properties with various galaxy properties in a large, statistically significant sample, we first present the general results from our analysis in our full sample. In Table \ref{Table:BulkProp}, we show selected properties of the full merger and control samples for each of the sub-classes. Quoted values are the average of the entire sub-sample that could be matched with three unique controls, at least one of which was successfully fitted by \textsc{badass}. Uncertainties on the outflow incidence fraction are given by binomial counting statistics with a two-sided 95 per cent confidence interval, in accordance with \citet[][]{1986ApJ...303..336G}. All other uncertainties in the table are given by the standard error of the mean, multiplied by 1.96 for the 95 per cent confidence interval. 

From Table \ref{Table:BulkProp}, we observe several clear results on the general characteristics of the sample. First, the fraction of SF galaxies (both mergers and controls) with outflows is significantly lower (between a factor of $\sim$14 for K03 AGN and a factor of $\sim$80 for WISE+BPT AGN) compared to all of the AGN sub-samples. This is also clearly seen in Figure \ref{fig:BPT}, where it is readily apparent that the outflow fraction is much higher in the AGN-dominated region of the BPT diagram. This is in qualitative agreement with a number of other studies that reveal that the outflow incidence is significantly lower in SF galaxies compared to AGNs \citep[e.g.][]{2016A&A...588A..41C,2017A&A...606A..36C,2019ApJ...884...54M}. Further, Figures \ref{fig:BPT2} and \ref{fig:BPT3} show alternate BPT diagrams, both of which indicate outflows are not common in low-ionization nuclear emission-line regions (LINERs). This is in contrast to what was found by \citet[][]{2022arXiv220105080H}, where they detected ionised outflows in $\sim$50 per cent of their LINER sample. However, even though our sample luminosities are comparable, their observations make use of higher resolution IFU data which is generally more sensitive to outflow detections than our lower resolution data (see Section \ref{sec: Discussion}). 

From Table \ref{Table:BulkProp}, we also see that the outflow fraction in optical+mid-infrared selected AGNs is roughly a factor of four times higher than the outflow fraction in optically selected AGNs, and almost two orders of magnitude higher than that found in SF galaxies. This is also clearly seen in Figure \ref{fig:CC}, which shows the WISE colour-colour diagram for our full sample. 

We further note that there is no statistically significant difference between the mergers and controls in any of our four samples (SF, K03 AGN, K01 AGN, and WISE+BPT AGN) in Table \ref{Table:BulkProp}. However, the four samples do show a statistically significant difference in their outflow incidence, with the more stringent AGN selection techniques displaying higher outflow incidences. Indeed, only $\sim$ 0.5 per cent of the SF galaxies have an ionized outflow. K03 AGN have a much higher outflow incidence near $\sim$ 7.5 per cent, while K01 AGN have an outflow incidence of $\sim$ 14 per cent. The WISE+BPT AGN have the largest outflow incidence by far at $\sim$ 40 per cent.

\begin{figure}
    \centering
    \begin{subfigure}[b]{\columnwidth}
    
    \includegraphics[width=\columnwidth]{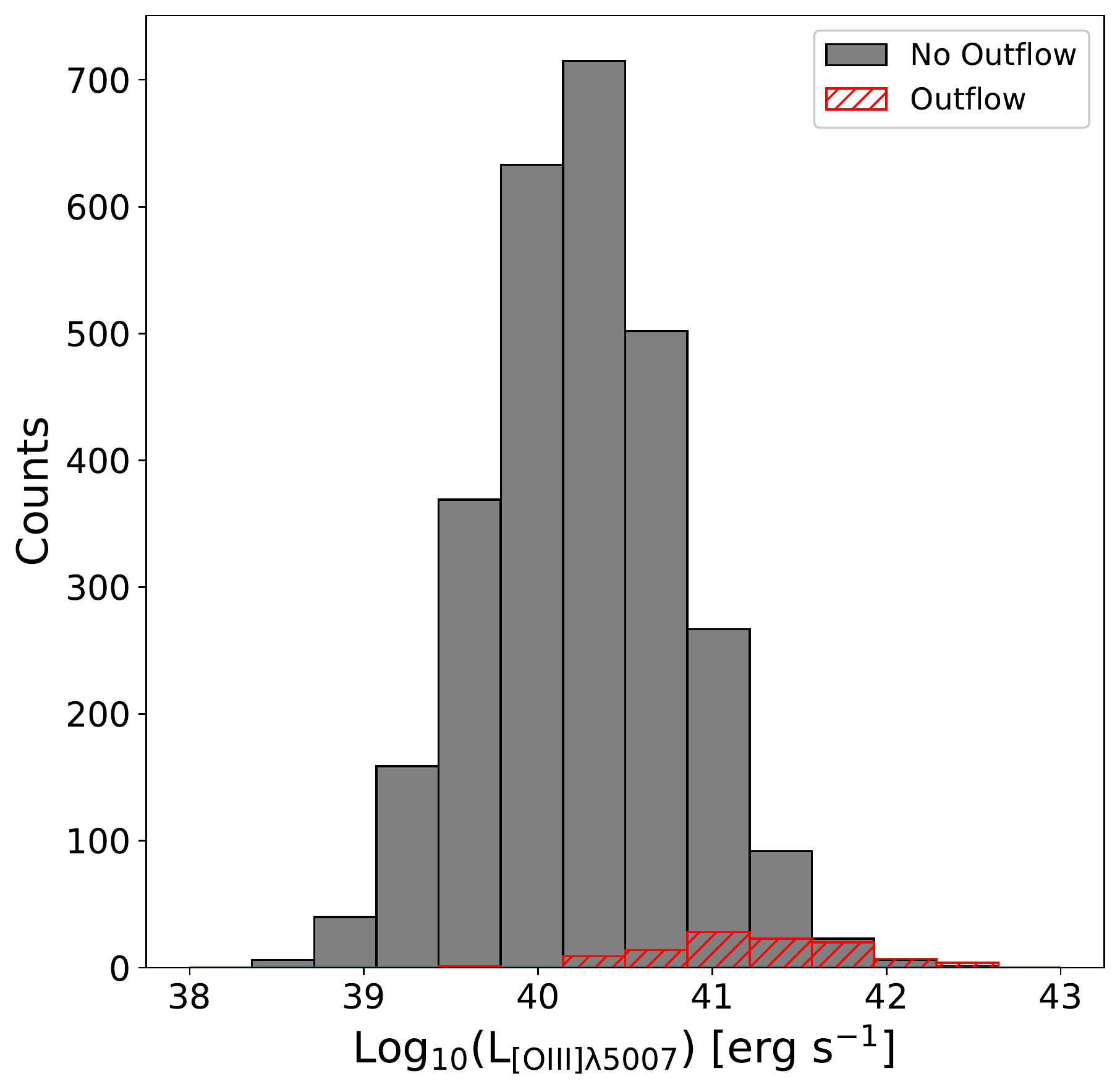}

    \end{subfigure}

\begin{subfigure}[b]{\columnwidth}
    \includegraphics[width=\columnwidth]{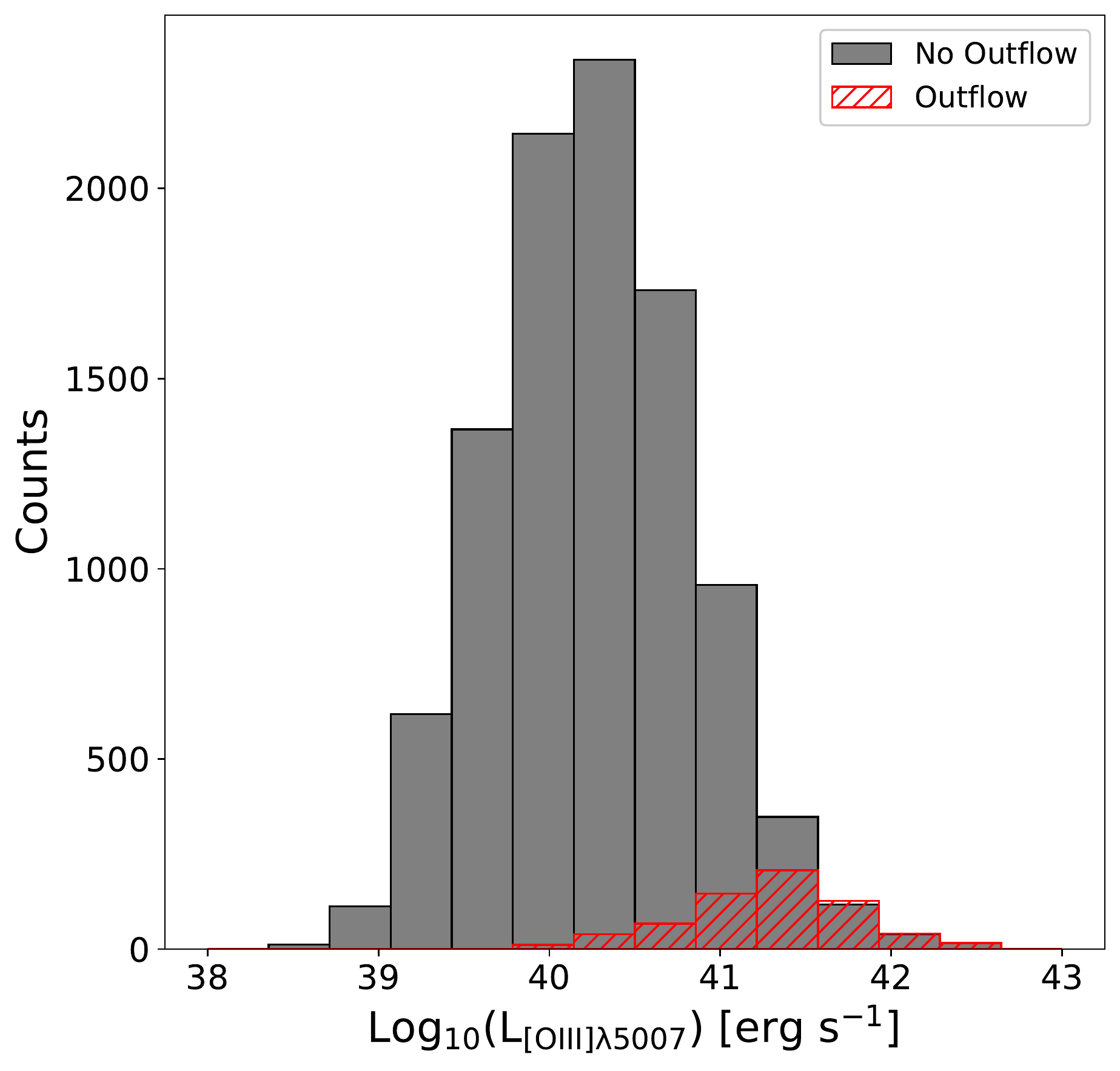}

    \end{subfigure}
\caption{Distribution of {[OIII]~$\lambda$5007} luminosity of the entire merger (top panel) and control (bottom panel) sample. The gray bars indicate galaxies with no significant outflows, while red bars indicate galaxies with significant outflows.}
\label{fig:LumHist}
\end{figure}

\begin{figure}
    \centering
    \begin{subfigure}[b]{\columnwidth}
    
    \includegraphics[width=\columnwidth]{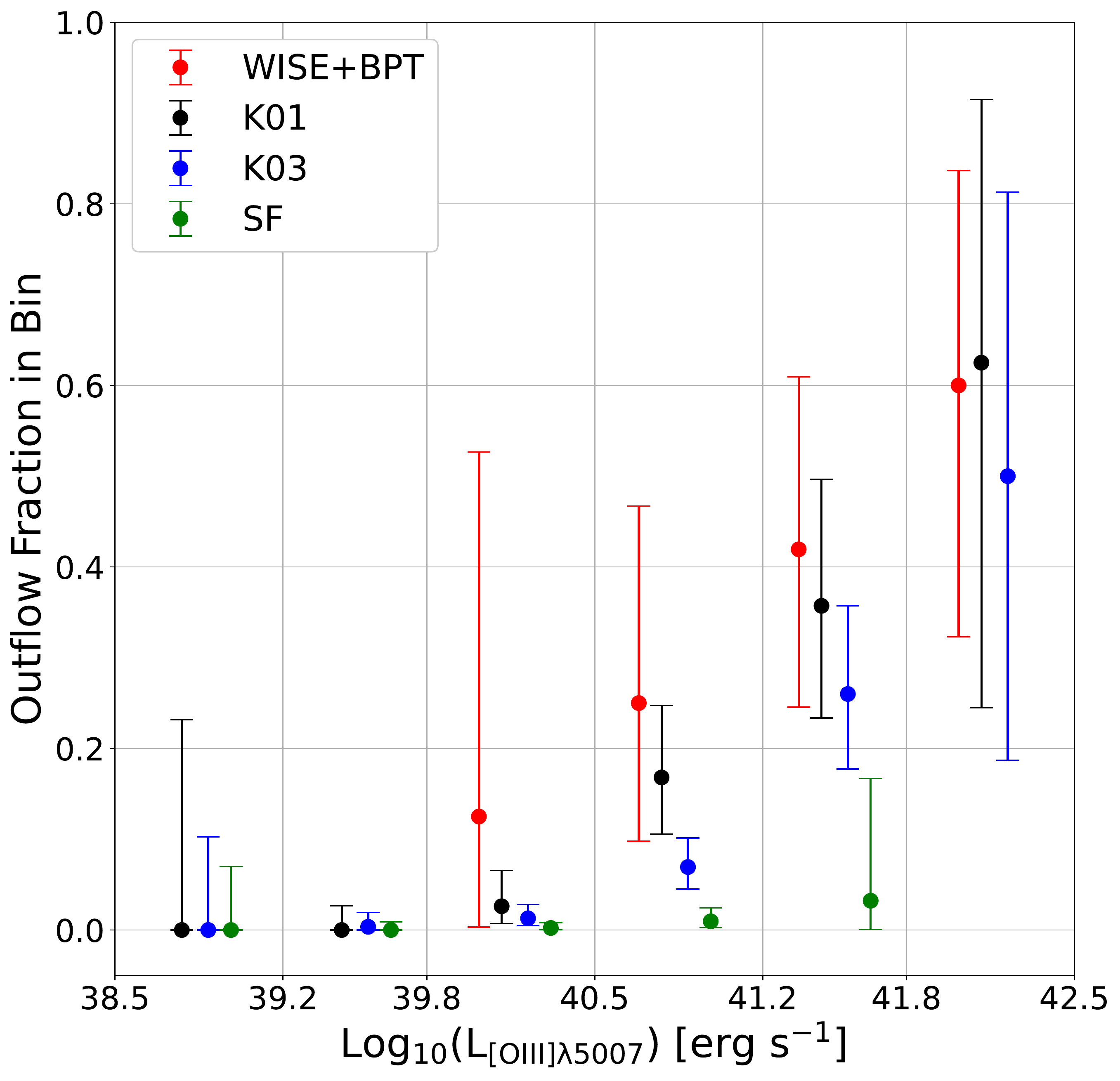}

    \end{subfigure}

\begin{subfigure}[b]{\columnwidth}
    \includegraphics[width=\columnwidth]{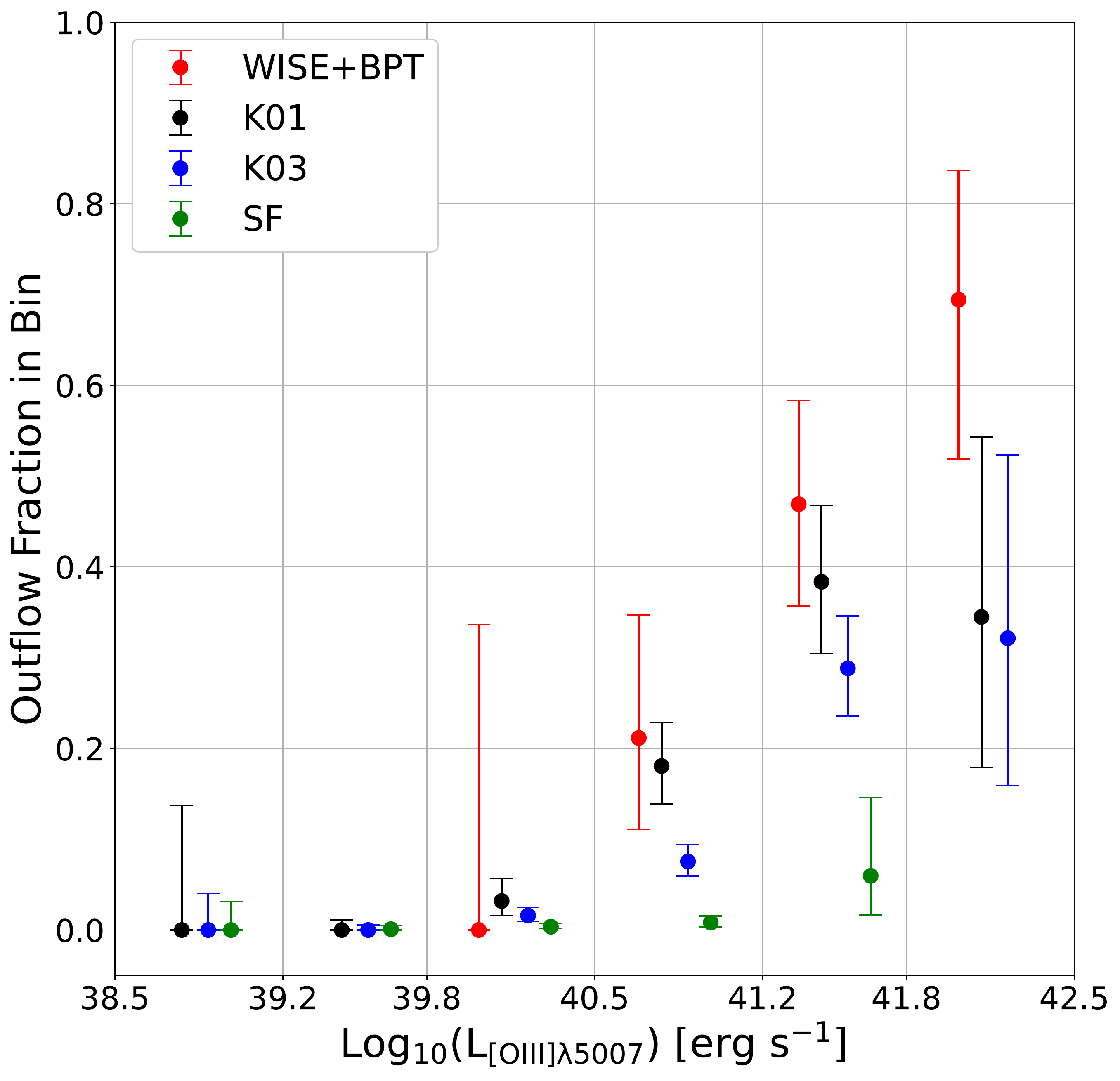}

    \end{subfigure}
\caption{Outflow incidence as a function of luminosity, separated into our various sub-samples for the entire merger (top panel) and control (bottom panel) sample. Errors are computed using binomial statistics with a two-sided 95 per cent confidence interval. Horizontal spacing between data points \textit{within} each bin is arbitrary and only serves to enhance readability.}
\label{fig:LumInc}
\end{figure}

At first glance, the excess of outflows in optical+mid-infrared selected AGN might simply seem to be a consequence of the variation in the luminosity across the sub-samples, with mid-infrared selected AGNs representing the more luminous objects in the sample. Intuitively, more luminous galaxies can, in principle, drive more luminous outflows which are easier to detect, a result that is indeed reported in a number of previous works \citep[][]{2014MNRAS.439.2701H,2016ApJ...817..108W,2017A&A...606A..36C,2021MNRAS.503.5134A}. The trend of outflow incidence increasing with [OIII]~$\lambda$5007 luminosity is also seen in the general properties of our samples, as shown in Figure \ref{fig:LumHist}. Here, objects with outflows are noticeably skewed towards higher [OIII]~$\lambda$5007 luminosity values. However, the increased outflow incidence in optical+mid-infrared selected AGN is not simply a consequence of variation in [OIII]~$\lambda$5007 luminosity across the sub-classes, as can be seen in Figure \ref{fig:LumInc} where we plot the mean outflow fraction as a function of the [OIII]~$\lambda$5007 luminosity for each of the various sub-classes. It is readily apparent that while the outflow incidence increases with increasing [OIII]~$\lambda$5007 luminosity, the outflow fraction is consistently significantly higher in AGN compared with SF galaxies at all [OIII]~$\lambda$5007 luminosities. Further, optical+mid-infrared selected AGN display elevated outflow fractions compared with optically selected AGN, particularly at the highest [OIII]~$\lambda$5007 luminosities. Our results therefore indicate that the incidence of outflows is higher in AGNs, a result that is independent of total luminosity or SFR. Additionally, our findings suggest that mid-infrared selection in particular may favour outflows, a result that is reported for the first time in this work (see Sections \ref{subsec: Outflow Incidence} and \ref{sec: Discussion} for further discussion of this result).

Finally, we also see from Table \ref{Table:BulkProp} that the SF galaxies typically have outflow velocities on the order of $\sim$300~km~s$^{-1}$ while AGN typically have outflow velocities on the order of $\sim$700~km~s$^{-1}$. This is again generally consistent with the literature \citep[e.g.][]{2002ApJ...570..588R,2005ApJS..160..115R,2005ASPC..331..313V,2014MNRAS.441.3306H,2020ApJ...905..166L,2021MNRAS.507.3985S}. The average outflow velocity in our SF control sample is lower than the average AGN outflow velocity by about four sigma. We additionally note that there is no significant difference between the outflow velocity \textit{among} our three AGN samples. 

While there is a statistically significant enhancement in the [OIII]~$\lambda$5007 outflow luminosities in the AGN compared to the SF galaxies, this enhancement appears only marginal.


\section{Outflow Characteristics as a Function of Merger Stage}\label{sec: Outflow Characteristics as a Function of Merger Stage}

In this section, we examine outflows as a function of merger stage.\footnote{Strictly speaking, we examine outflow properties as a function of the projected physical separation $r_p$. While later stage mergers are typically found at smaller values of $r_p$, paired galaxies can increase in $r_p$ after a close encounter. Hence, it is possible to find a `late stage' merger at higher $r_p$ values.} In each plot, we bin the data in increments of 20 kpc. While this choice of binning is somewhat arbitrary, this particular choice allows for a generally acceptable number of objects in each bin, while still probing an acceptable range of $r_p$. Note that in all figures with $r_p$ displayed on the horizontal axis, the first bin covering the -20 to 0 kpc range corresponds to the post-merger objects. As discussed in Section \ref{sec: Sample Selection}, our controls are matched in redshift, stellar mass, local density of galaxies, and [OIII]~$\lambda$5007 luminosity. SF galaxies are additionally matched in SFR. 

\subsection{Outflow Incidence}\label{subsec: Outflow Incidence}
\begin{figure*}
\centering
\begin{subfigure}{0.33\textwidth}
  \centering
  
  \includegraphics[width=1\linewidth]{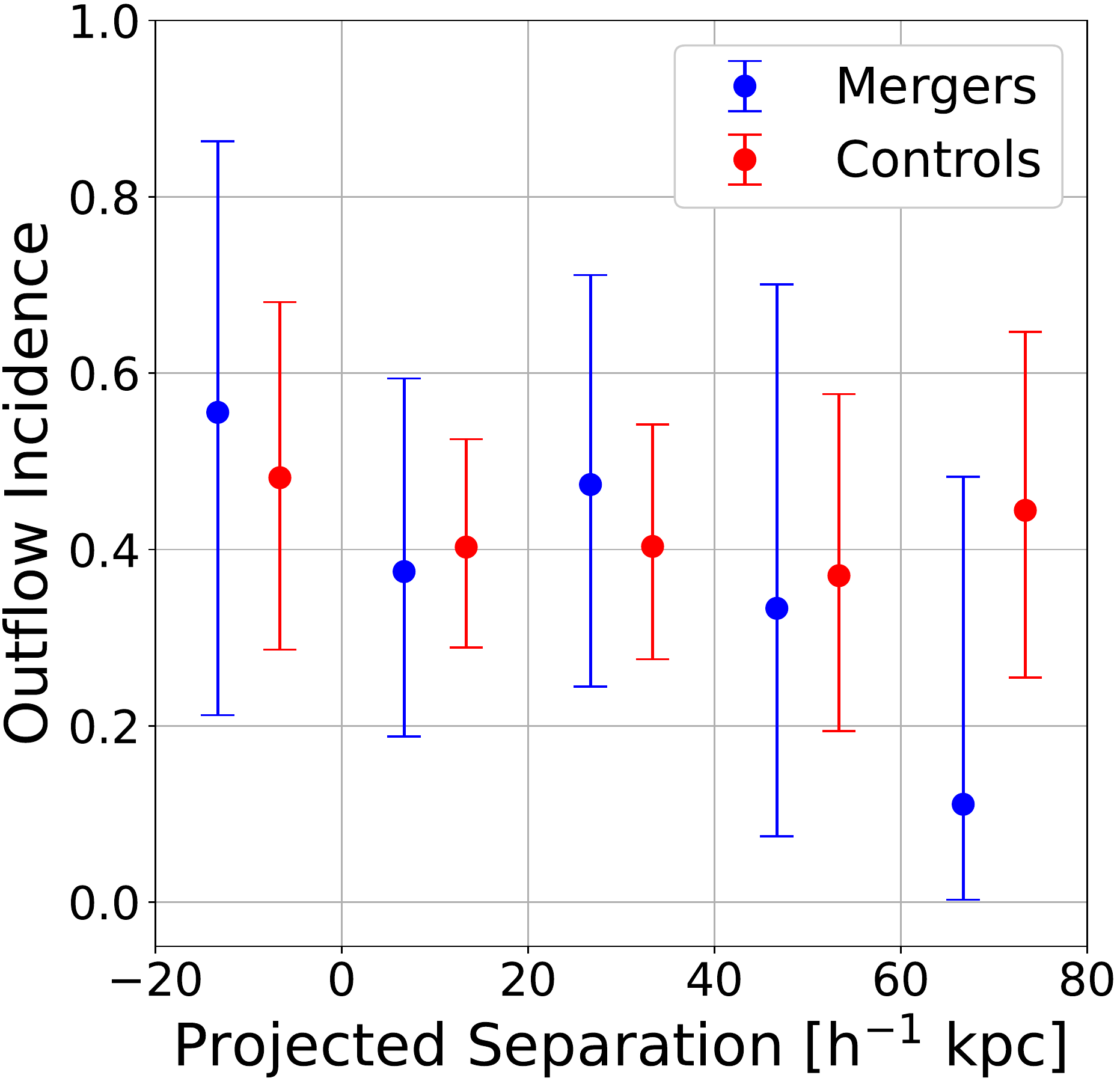}

\end{subfigure}%
\begin{subfigure}{0.33\textwidth}
  \centering
  
  \includegraphics[width=1\linewidth]{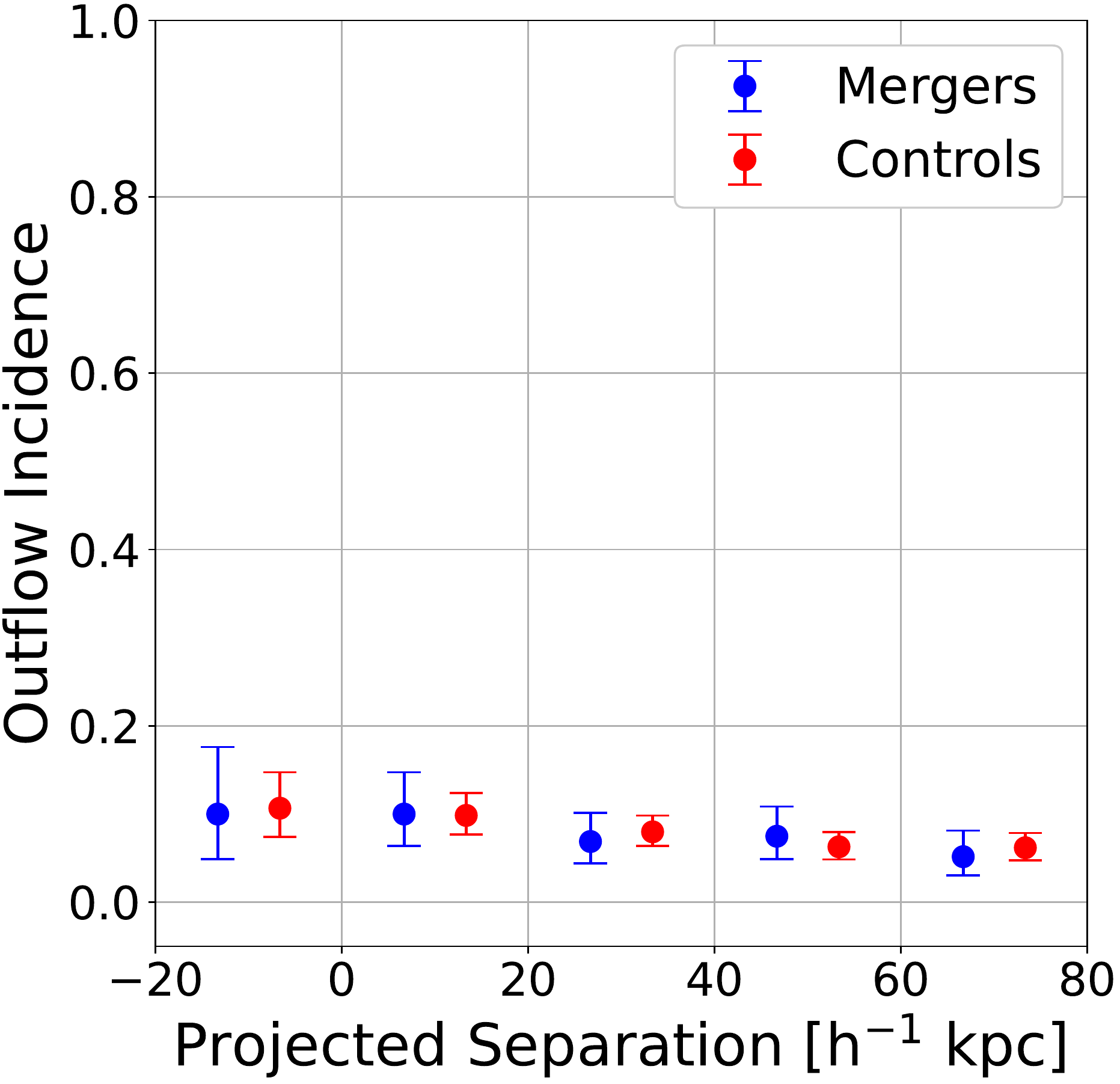}

\end{subfigure}
\begin{subfigure}{0.33\textwidth}
  \centering
  
  \includegraphics[width=1\linewidth]{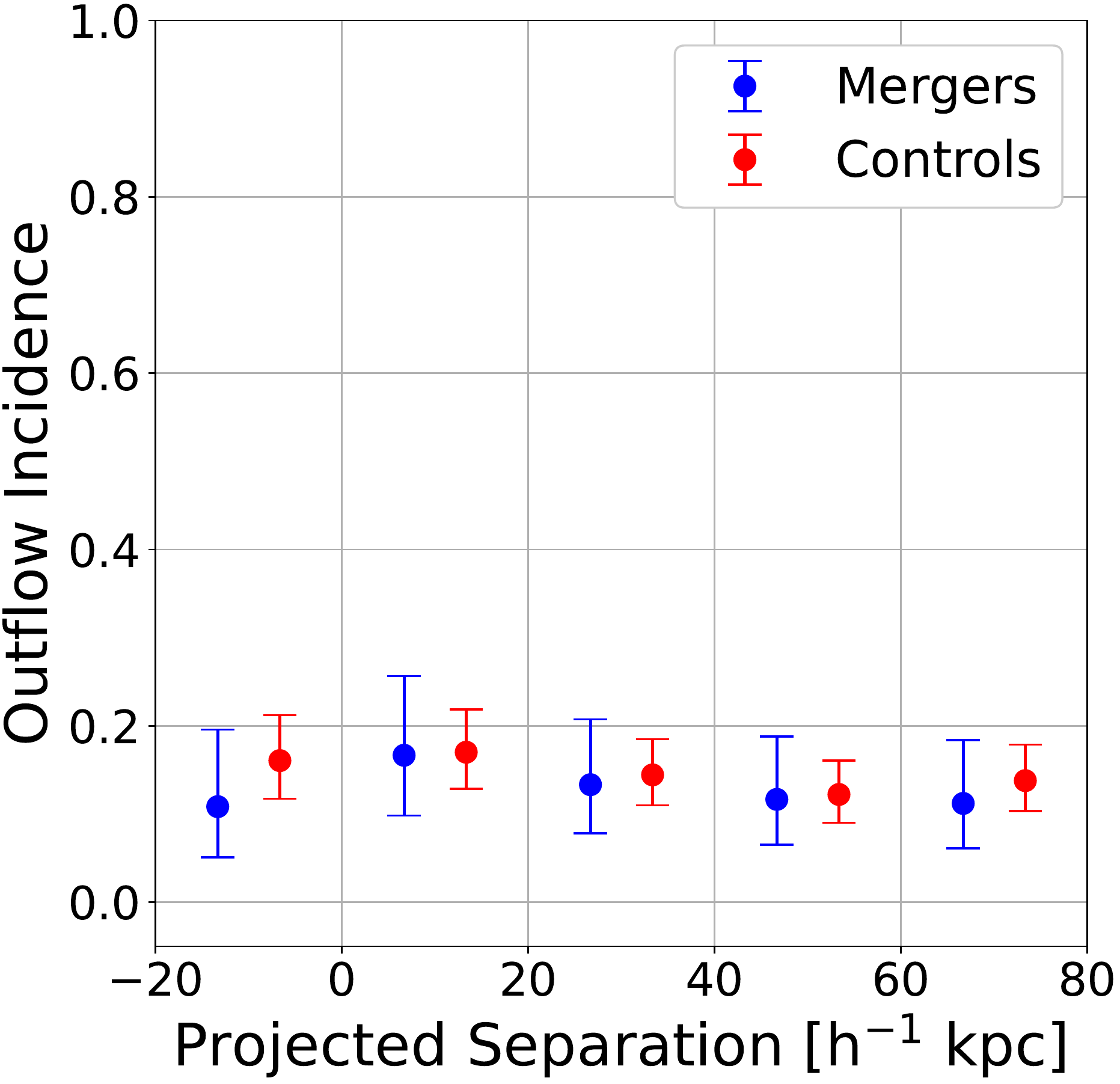}

\end{subfigure}
\caption{Average outflow incidence in each bin of projected physical separation $r_p$ for the WISE+BPT sample (left panel), K03 AGN (middle panel), and K01 AGN (right panel). Errors are computed using binomial statistics with a 95 per cent confidence interval. Horizontal spacing between data points \textit{within} each bin is arbitrary and only serves to enhance readability.}
\label{fig:Binned_AGN_Rate}
\end{figure*}

Figure \ref{fig:Binned_AGN_Rate} shows the average outflow incidence rate in each bin for both the mergers and controls as a function of $r_p$ in AGN. As in Table \ref{Table:BulkProp}, we compute error bars according to binomial statistics using a two-sided 95 per cent confidence interval -- the confidence interval covers 2.5 per cent to 97.5 per cent, or 95 per cent, of the uncertainty. Unlike in Table \ref{Table:BulkProp}, we require all mergers to have three controls successfully analysed by \textsc{badass} in order to be included in this Figure. In Figure \ref{fig:Binned_AGN_Rate}, we see that the outflow fraction in the mergers and their matched controls shows no statistically significant difference in any $r_p$ bin. 

The WISE+BPT pairs and their controls typically have an outflow incidence around 40 per cent in each bin, with each bin typically having around 10 total mergers. While the post-mergers and their controls have an outflow incidence around 50 per cent, the large error bars in each bin suggest there is no significant difference between the merger and control samples in outflow incidence as $r_p$ changes. 

The outflow incidence in the optical AGN is much lower, with the K03 AGN incidence being slightly below 10 per cent in each bin and the K01 AGN being slightly above 10 per cent in each bin. The K03 AGN typically have a few hundred mergers in each bin, while the K01 AGN typically have around 100 mergers in each bin. In both cases, there is no statistically significant difference between the mergers and controls in each bin.

\begin{figure}
    \centering
    \includegraphics[width=0.70\linewidth]{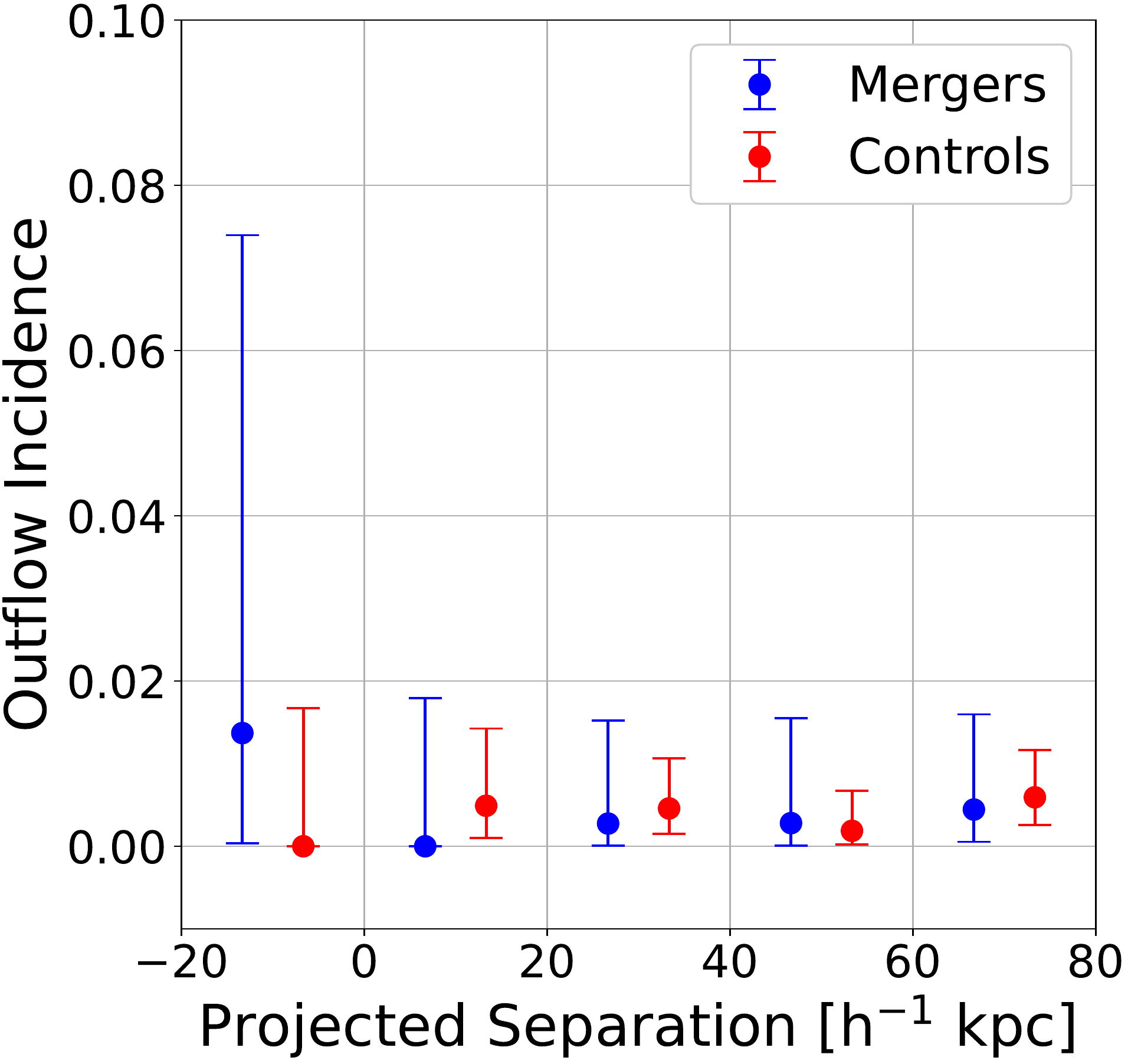}
    \caption{Average outflow incidence in each bin of projected physical separation $r_p$ for K03 star forming galaxies. Errors are computed using binomial statistics with a 95 per cent confidence interval. Horizontal spacing between data points \textit{within} each bin is arbitrary and only serves to enhance readability.}
    \label{fig:Binned_SF_Rate}
\end{figure}

Similarly, Figure \ref{fig:Binned_SF_Rate} shows the average outflow incidence in SF galaxies. The outflow incidence here is much less than the corresponding value for AGN, with a typical outflow incidence being only a fraction of a percent. While there is an apparent jump in outflow incidence in the post-merger bin, this is because there are only 73 total objects in the post-merger bin (with a single outflow detection), while the other bins typically have a couple of hundred mergers in them. 

Naively, it might be expected that the outflow incidence in both mergers and matched controls will increase at smaller pair separations, since AGN fraction and mid-infrared luminosity increase at smaller pair separations \citep[][]{2013MNRAS.430.3128E,2014MNRAS.441.1297S}. However, this is not the case in our sample because of the way our sample is constructed. Since we require that mergers and controls are matched in [OIII]~$\lambda$5007 luminosity, the most luminous mergers, which are preferentially found at the smallest pair separations, do not have enough corresponding control galaxies with comparable luminosities and are therefore omitted from our sample. As a result, no enhancement in outflow incidence at smaller $r_p$ is seen. 

Further, we note that the difference in the outflow occurrence rates between our samples is not \textit{solely} due to different luminosity ranges between the samples. As discussed in Section \ref{sec: General Outflow Characteristics in Sample}, Figure \ref{fig:LumInc} shows that even for the same [OIII]~$\lambda$5007 luminosity range, SF galaxies have fewer outflows compared to AGN by a factor of about eight. Additionally, the K01 and K03 AGN samples have comparable luminosities across all $r_p$ bins, yet outflows are more common in K01 AGN by as much as a factor of about two. While the WISE+BPT AGN do indeed have higher luminosities than the optical AGN in each $r_p$ bin, we again see in Figure \ref{fig:LumInc} that even in the same luminosity bin, outflows are more common in the WISE+BPT sample by as much as a factor of about two at the highest [OIII]~$\lambda$5007 luminosities. This suggests that not only is the presence of an AGN necessary to create and drive powerful outflows, but the specific AGN selection technique can play an important role in characterising the ionized outflows. We discuss the significance of the enhanced outflow rate in the WISE+BPT AGN in Section \ref{sec: Discussion}.

\subsection{Outflow Velocity}\label{subsec: Outflow Velocity}
 
\begin{figure*}
\centering
\begin{subfigure}{0.33\textwidth}
  \centering
  
  \includegraphics[width=1\linewidth]{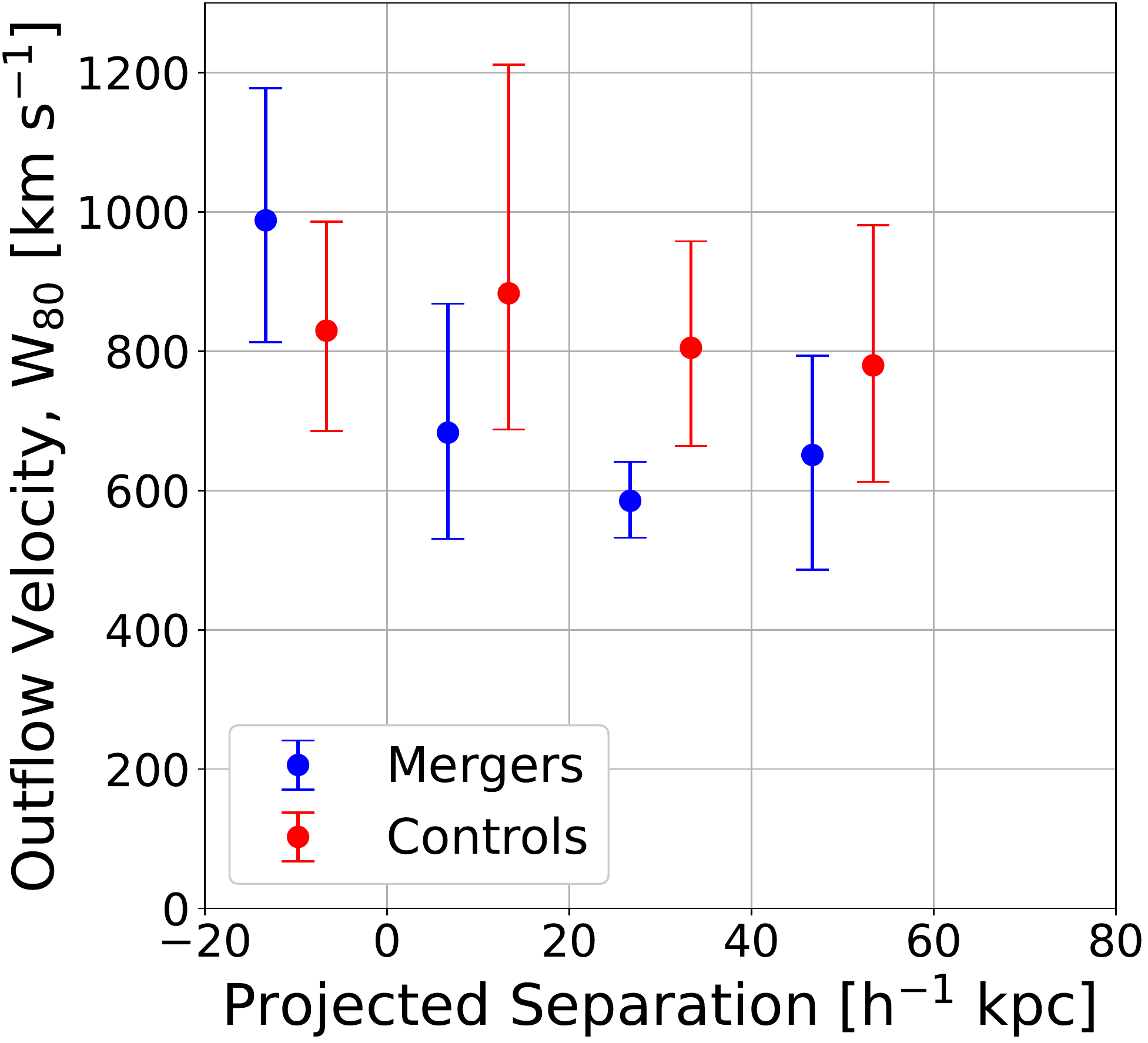}

\end{subfigure}%
\begin{subfigure}{0.33\textwidth}
  \centering
  
  \includegraphics[width=1\linewidth]{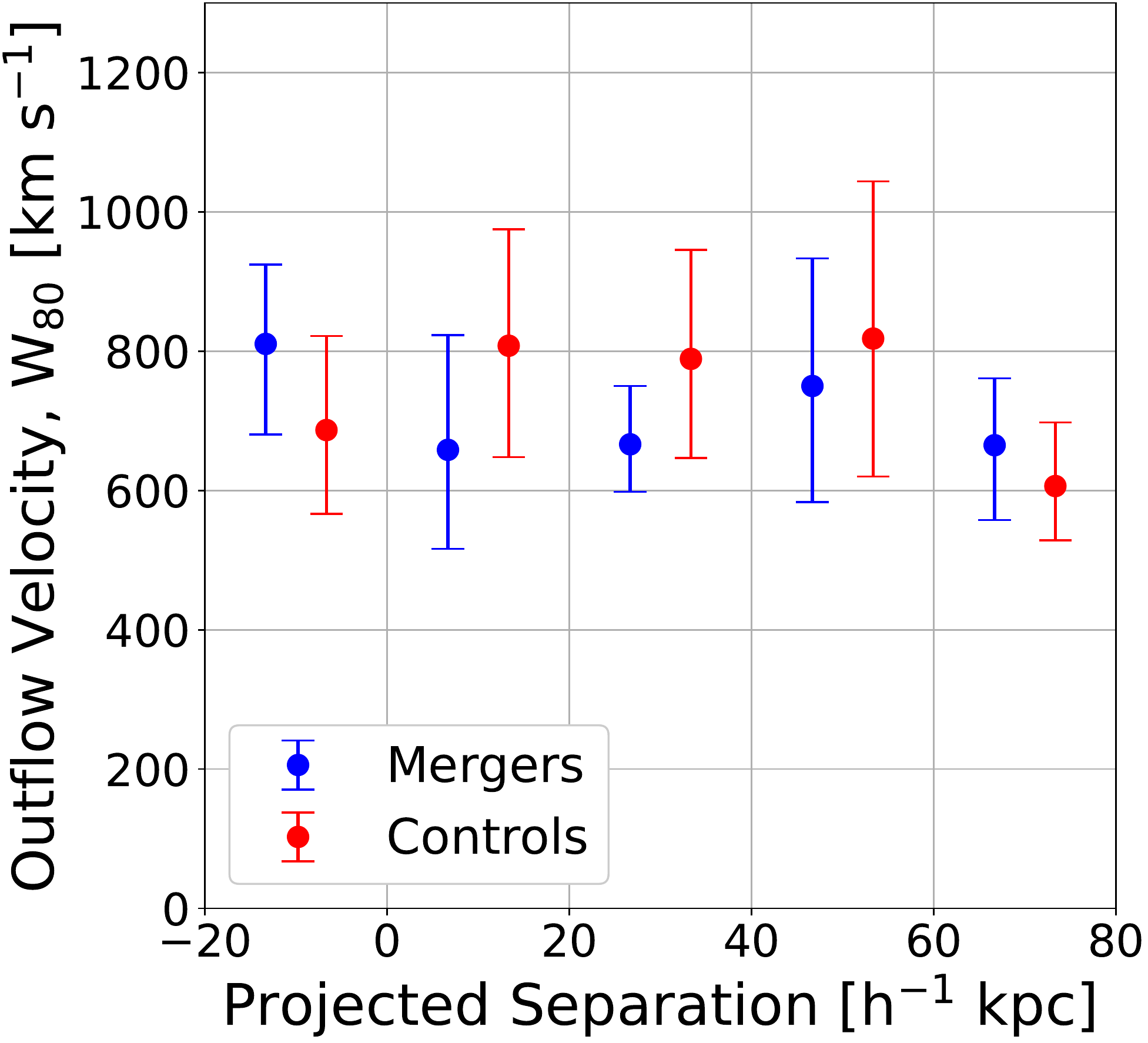}

\end{subfigure}
\begin{subfigure}{0.33\textwidth}
  \centering
  
  \includegraphics[width=1\linewidth]{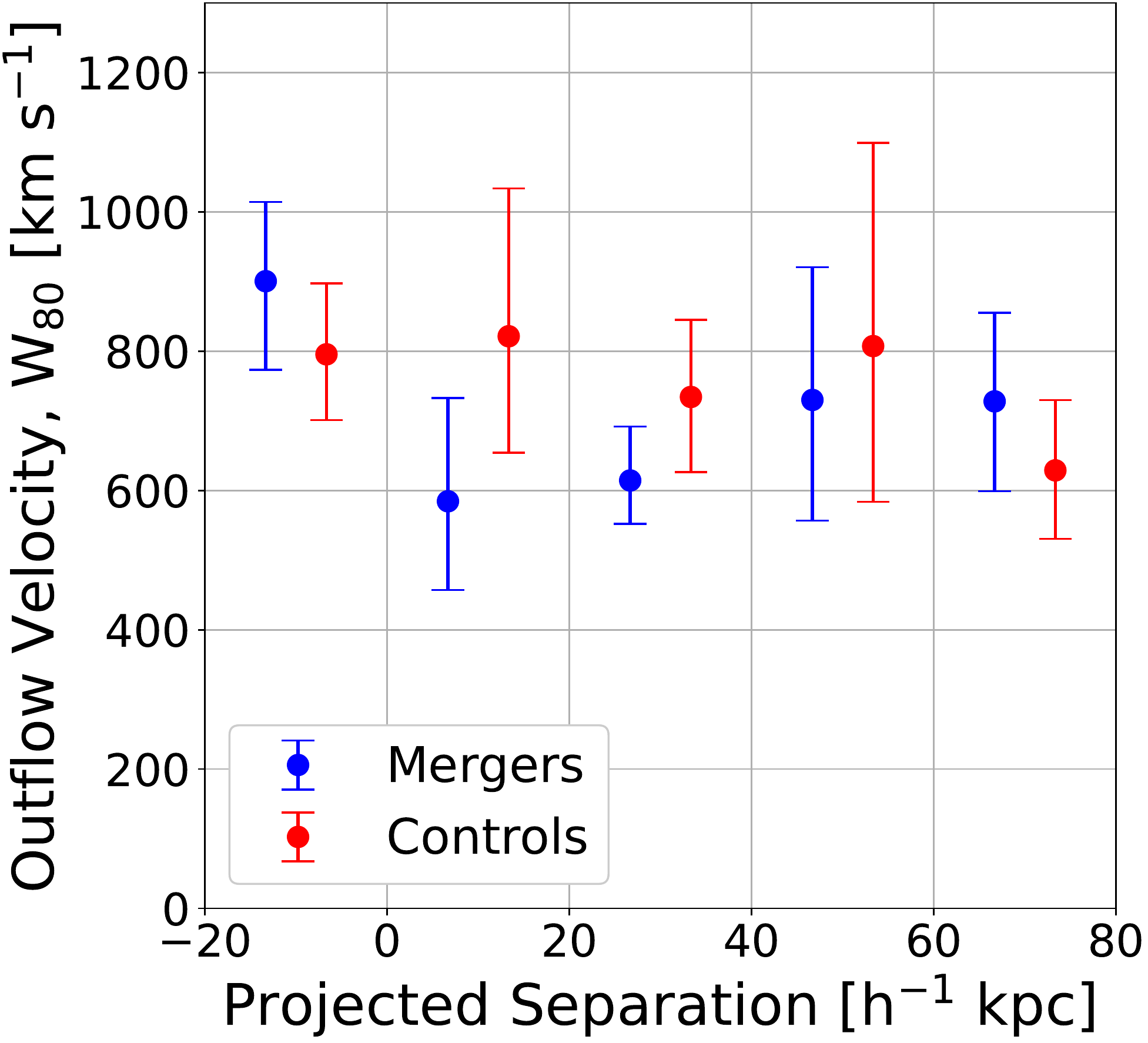}

\end{subfigure}
\caption{Average outflow velocity in each bin of projected physical separation $r_p$ for the WISE+BPT sample (left panel), K03 AGN sample (middle panel) and K01 AGN sample (right panel). Errors are computed by averaging the upper (lower) uncertainties associated with each data point in each bin to obtain the final upper (lower) uncertainty for the average in each bin. Each merger is only required to have a minimum of \textit{one} control with an outflow in order to be included in the plot. Horizontal spacing between data points \textit{within} each bin is arbitrary and only serves to enhance readability.}
\label{fig:Binned_AGN_Velocity}
\end{figure*}

We also explore the outflow velocity in AGN mergers compared to their matched controls as a function of $r_p$. As discussed in Section \ref{sec: General Outflow Characteristics in Sample}, we use $W_{80} = 1.088 \; \text{FWHM}$ to measure outflow velocity. However, we do not use the standard error of the mean to compute the uncertainties in this section. Each individual data point has its own upper and lower uncertainty value given by our final MCMC fit for each spectrum, and we attempt to take that uncertainty into account here. The uncertainties reported correspond to a 1$\sigma$ level, so we multiply them by 1.96 for consistency with previous sections.\footnote{The posterior distribution of the outflow FWHM parameter is generally Gaussian for objects that have an outflow.} Figure \ref{fig:Binned_AGN_Velocity} shows the average outflow velocity for each $r_p$ bin. The uncertainties on these data points are obtained by averaging the upper uncertainties together to obtain the average upper uncertainty, and similarly for the lower uncertainty. 

In order for a velocity to be included in the plot, we must have a merger with an outflow and at least one control with an outflow. For this reason, the 60-80 kpc bin in the WISE+BPT outflow velocity plot does not contain any data points. The third bin (20-40 kpc) in this figure hints at a possible suppression of outflow velocities at this merger stage. However, this dearth is marginal and disappears if we use error bars corresponding to the 99 per cent level. The K03 and K01 AGN outflow velocities do not show any significant differences as a function of $r_p$. We do note that the WISE+BPT and K01 AGN outflow velocities are slightly elevated in the post-merger sample; however, the error bars in each of the bins are too large to draw any conclusions here. We omit the corresponding plot of the outflow velocity as a function of $r_p$ for the SF sample because there are not enough outflows in this sample to make any meaningful remarks.

\begin{figure}
    \centering
    \begin{subfigure}[b]{\columnwidth}
    
    \includegraphics[width=\columnwidth]{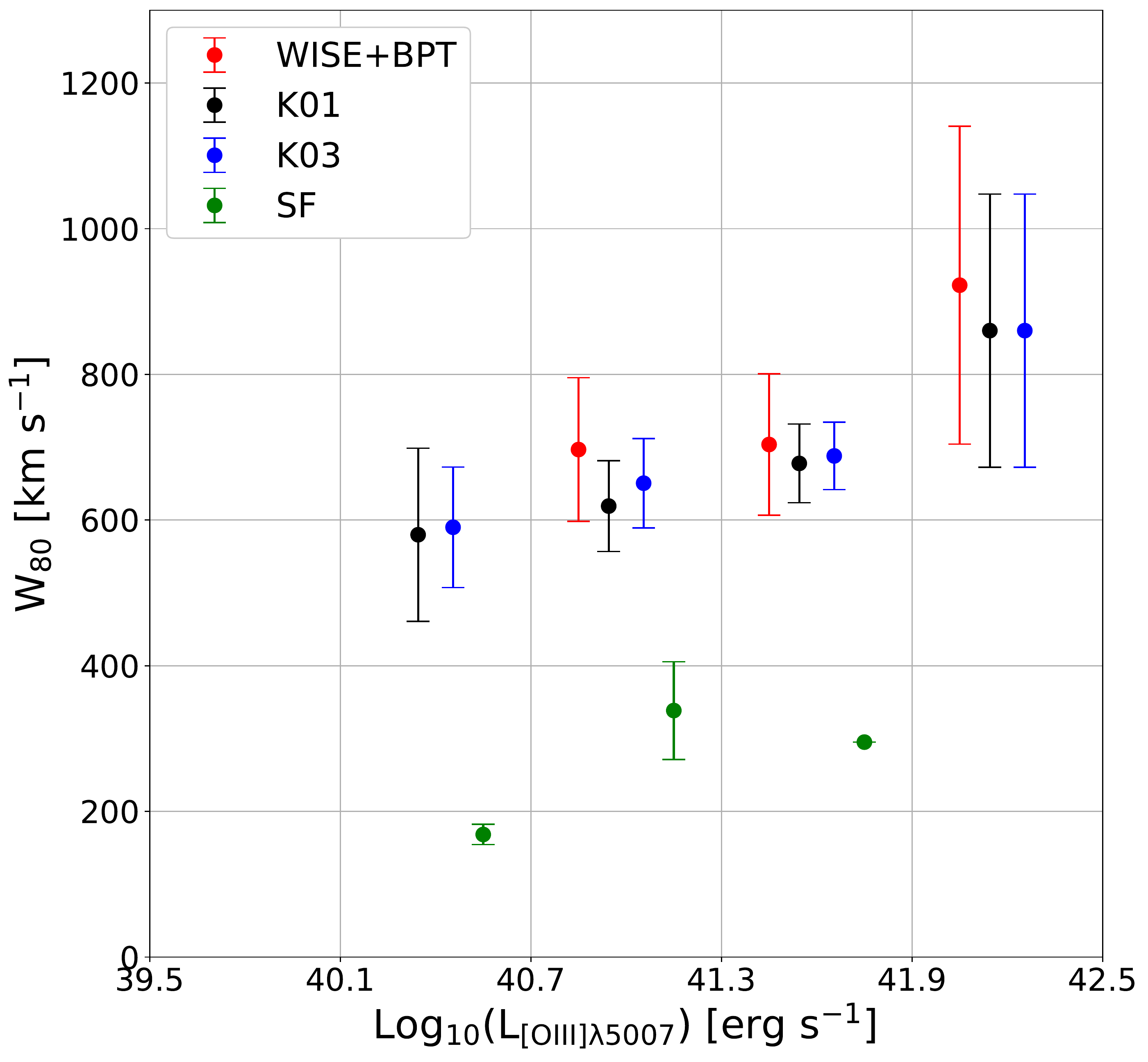}

    \end{subfigure}

\begin{subfigure}[b]{\columnwidth}
    \includegraphics[width=\columnwidth]{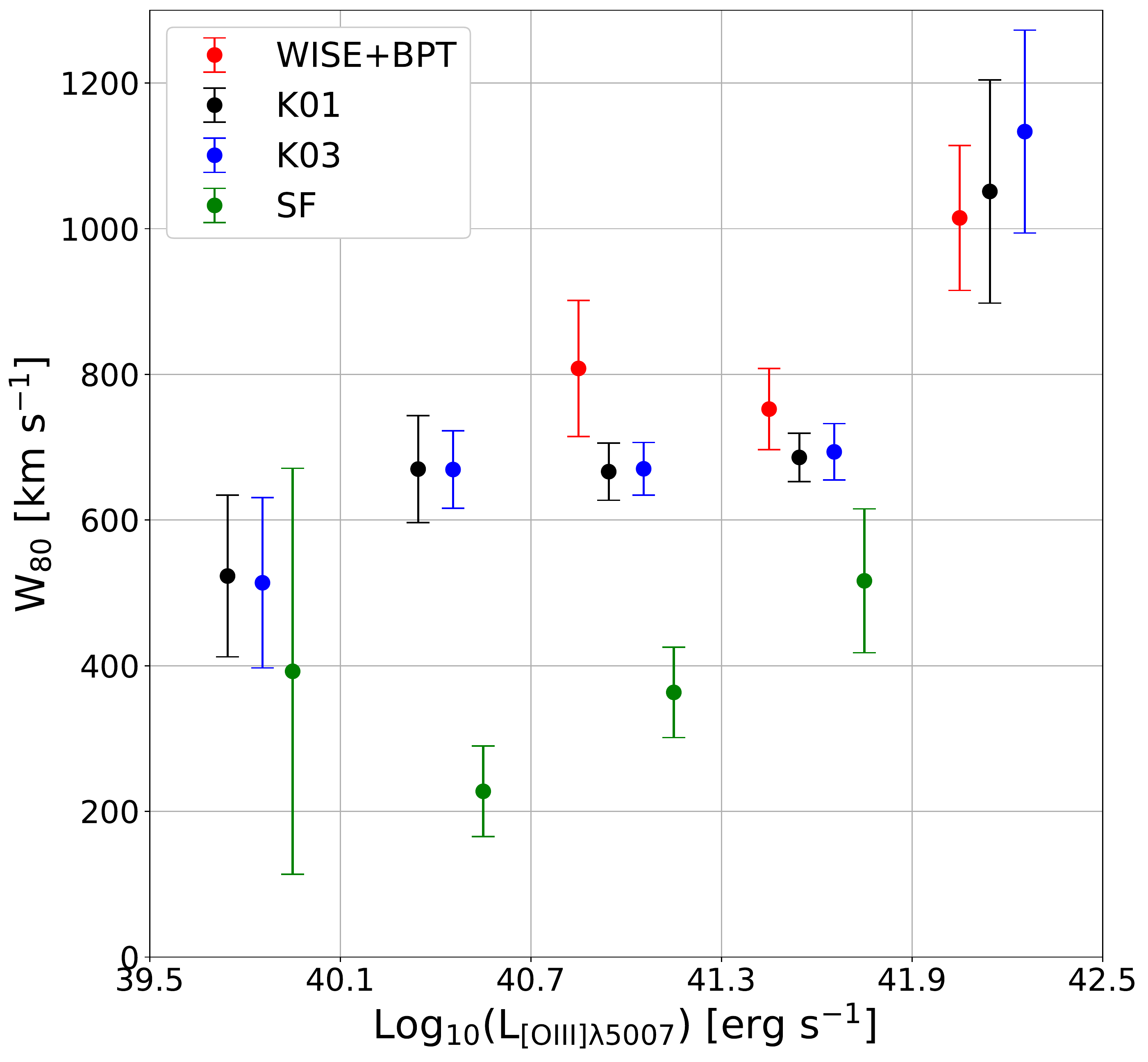}

    \end{subfigure}
\caption{Average outflow velocity in each bin of [OIII]~$\lambda$5007 luminosity for the merger (top panel) and control (bottom panel), separated into our various sub-samples. Errors are computed using a 2$\sigma$ standard error of the mean. Horizontal spacing between data points \textit{within} each bin is arbitrary and only serves to enhance readability. The SF data point in the fourth bin of the merger plot has zero error because it is the only object in that bin.}
\label{fig:VelLum}
\end{figure}

Figure \ref{fig:VelLum} shows the outflow velocities as a function of [OIII]~$\lambda$5007 luminosity for each sub-sample. First, we note that the SF sub-sample has systematically lower outflow velocities than the AGN samples at all luminosities. Further, as the [OIII]~$\lambda$5007 luminosity increases, the outflow velocity in the SF sub-sample steadily increases. However, we only observe a significant increase in outflow velocity in the highest luminosity bin for the AGN sub-samples. Evidently, the presence of an AGN is important in driving high-velocity outflows. 

\subsection{Outflow Luminosity}\label{subsec: Outflow Luminosity}
\begin{figure*}
\centering
\begin{subfigure}{0.33\textwidth}
  \centering
  
  \includegraphics[width=1\linewidth]{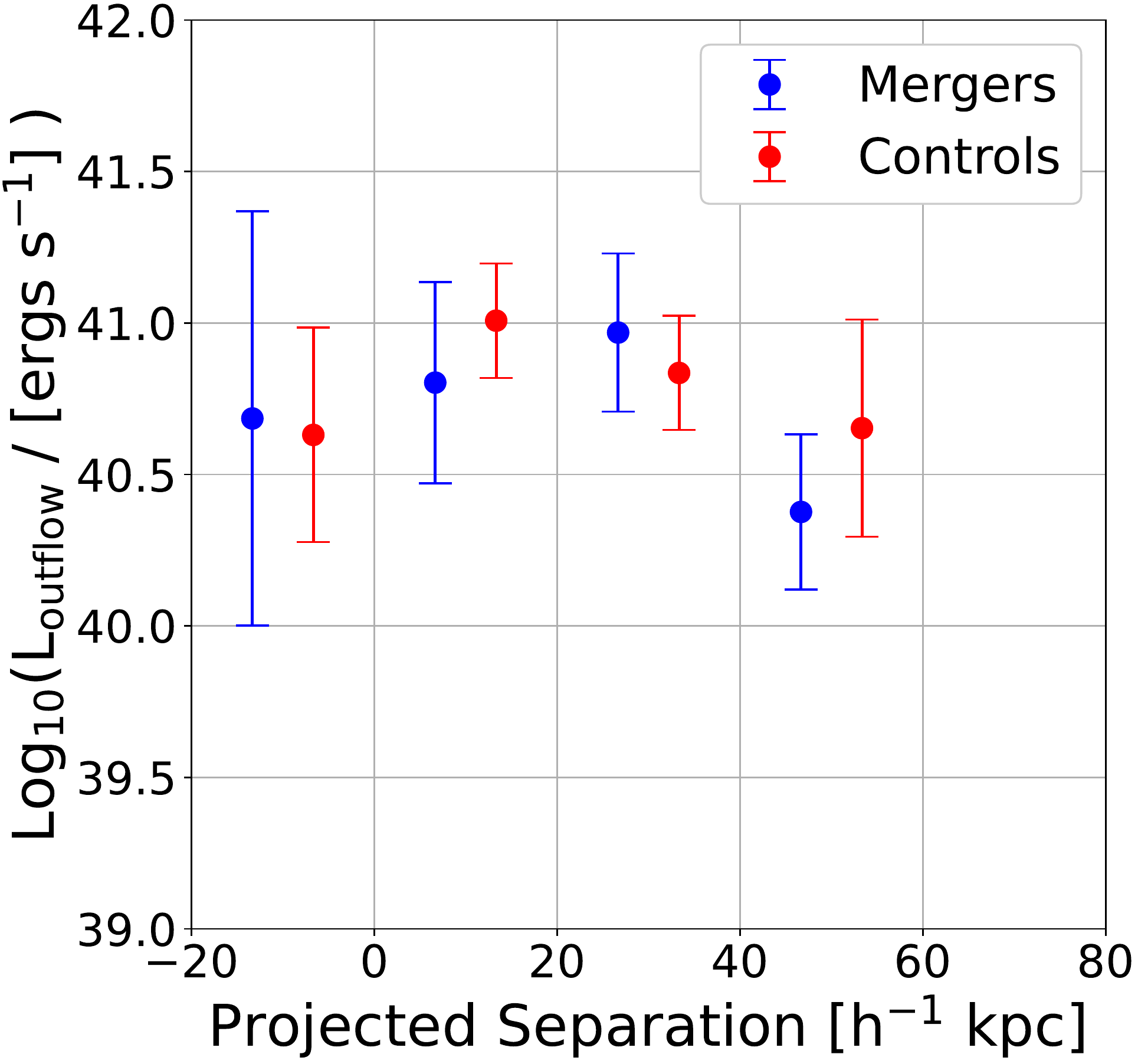}

\end{subfigure}%
\begin{subfigure}{0.33\textwidth}
  \centering
  
  \includegraphics[width=1\linewidth]{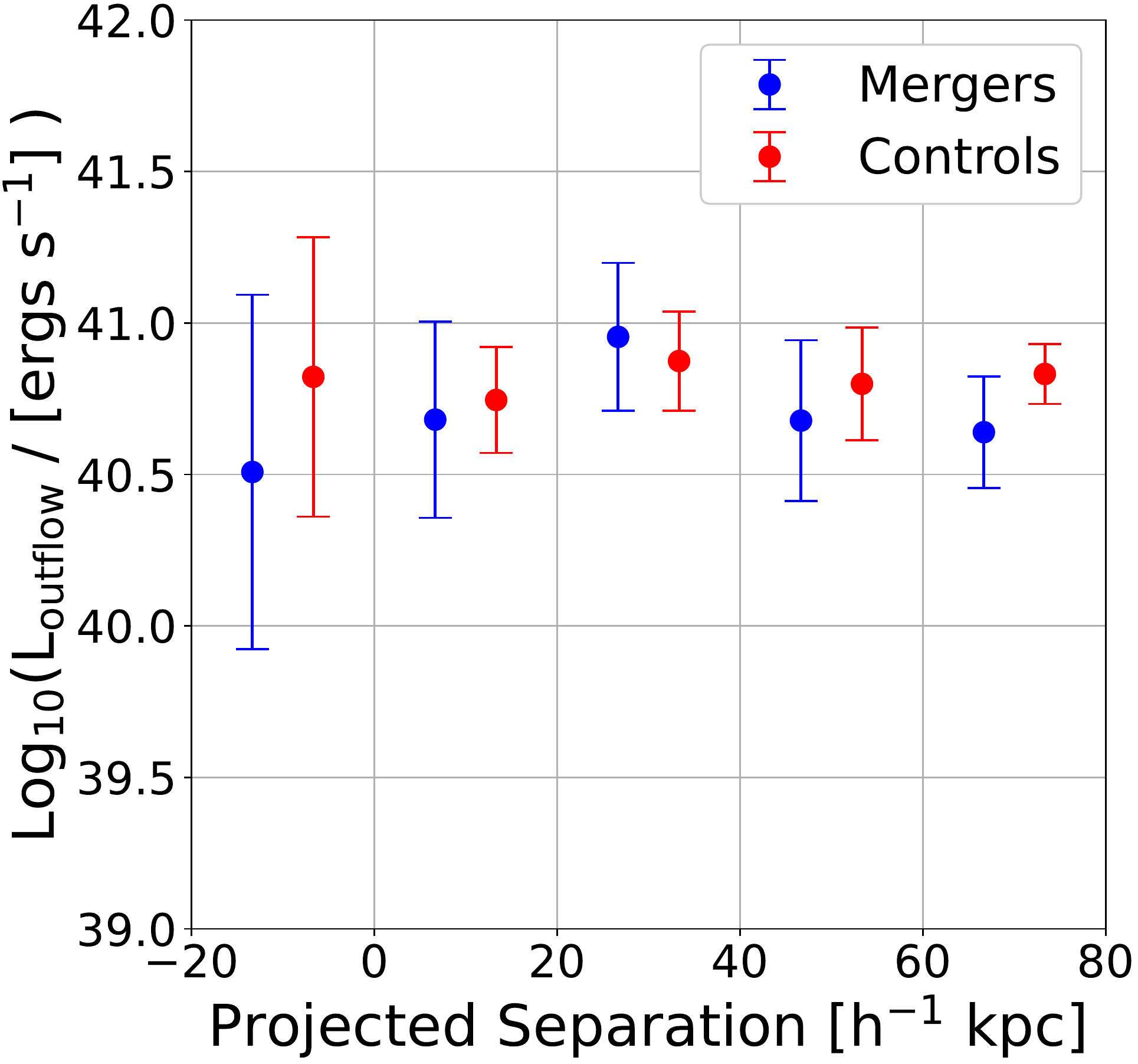}

\end{subfigure}
\begin{subfigure}{0.33\textwidth}
  \centering
  
  \includegraphics[width=1\linewidth]{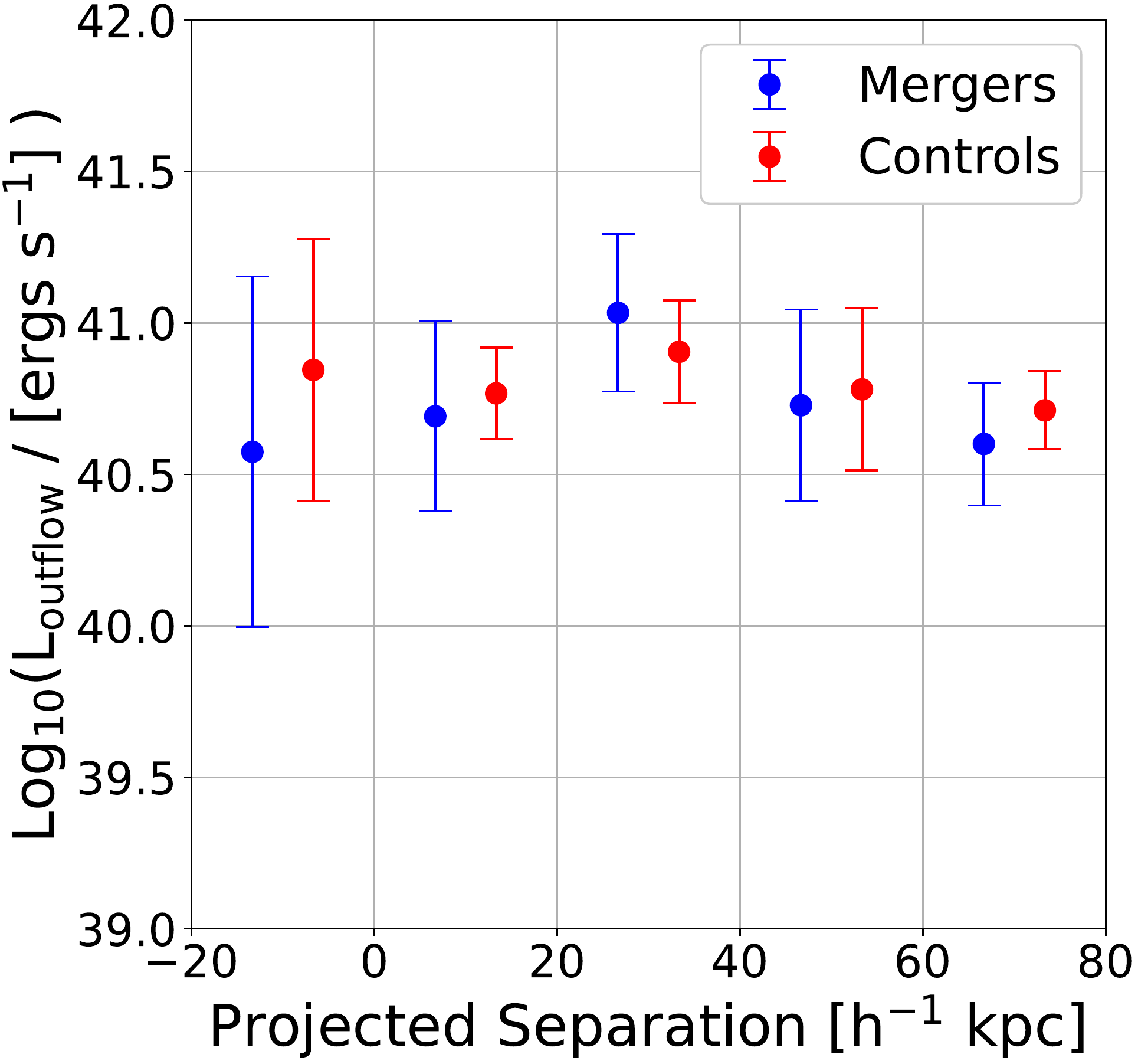}

\end{subfigure}
\caption{Average outflow Luminosity in each bin of projected physical separation $r_p$ for the WISE+BPT sample (left panel), K03 AGN sample (middle panel) and K01 AGN sample (right panel). Errors are computed using a 2$\sigma$ standard error of the mean. Each merger is only required to have a minimum of \textit{one} control with an outflow in order to be included in the plot. Horizontal spacing between data points \textit{within} each bin is arbitrary and only serves to enhance readability.}
\label{fig:Binned_AGN_OLum}
\end{figure*}

Figure \ref{fig:Binned_AGN_OLum} shows the outflow luminosity in the mergers and controls as a function of $r_p$ for the three AGN sub-samples. Again, the final bin in the WISE+BPT AGN plot does not have any objects because there is no merger with an outflow that also has a control with an outflow in that bin. Since the uncertainty on the flux and redshift is exceedingly small for each object, we compute uncertainties using the standard error of the mean here. In all three sub-samples, we again see that there is no significant difference between the mergers and controls. Again, we omit the outflow luminosity figure for the SF sample because there are not enough outflows in this sample to draw any significant conclusions. 

\begin{figure*}
\centering
\begin{subfigure}{0.33\textwidth}
  \centering
  \includegraphics[width=1\linewidth]{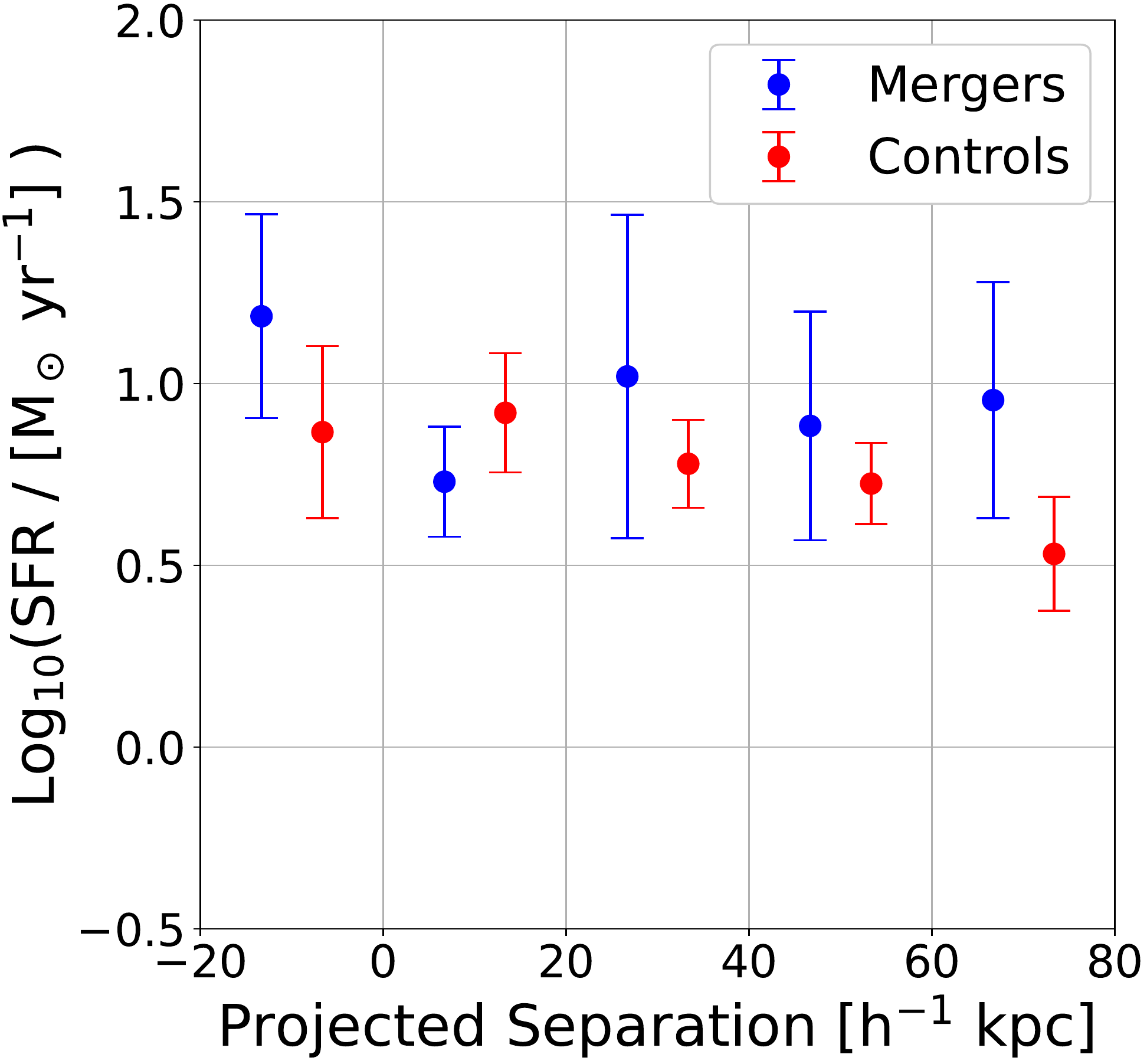}

\end{subfigure}%
\begin{subfigure}{0.33\textwidth}
  \centering
  \includegraphics[width=1\linewidth]{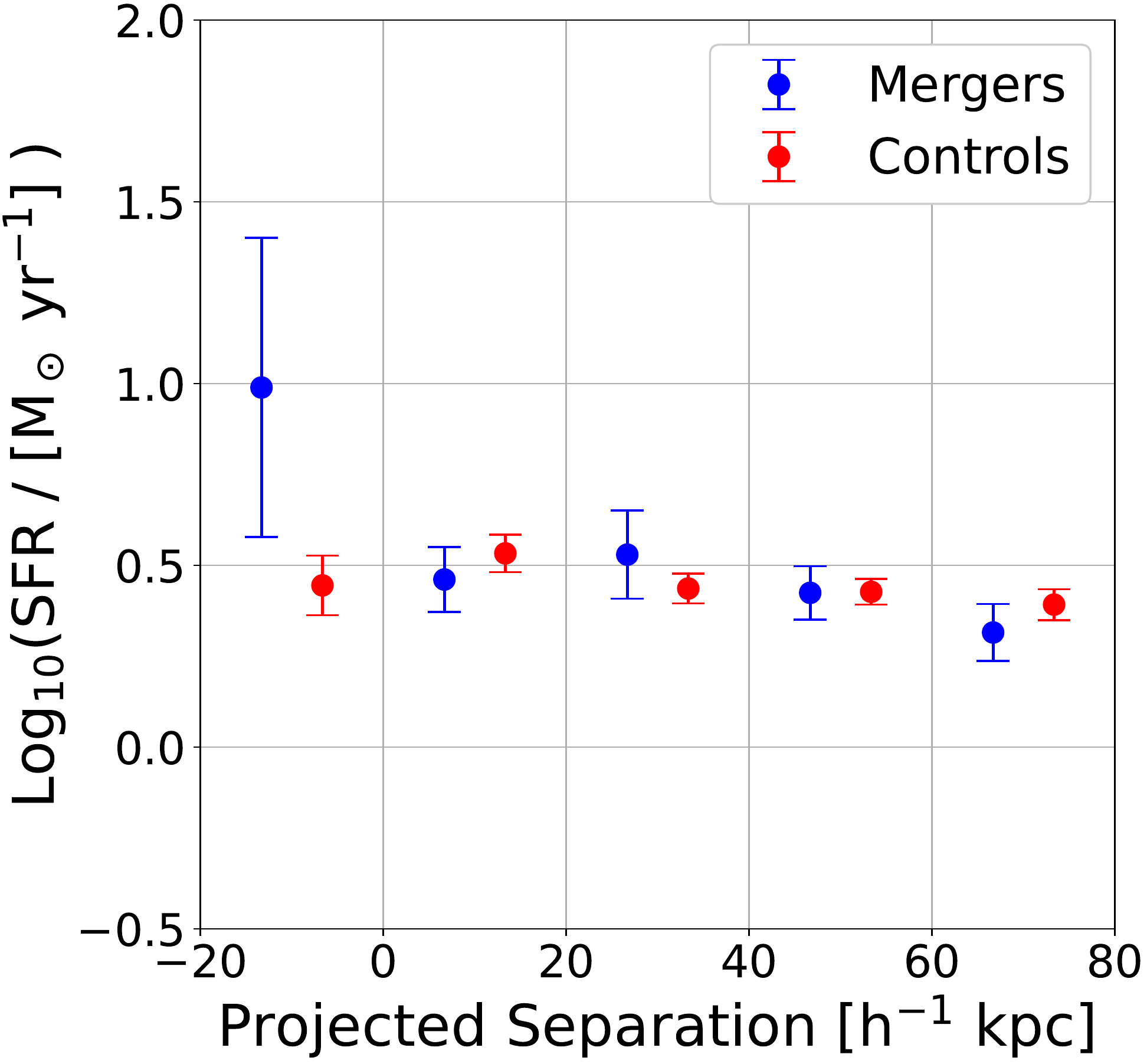}

\end{subfigure}
\begin{subfigure}{0.33\textwidth}
  \centering
  \includegraphics[width=1\linewidth]{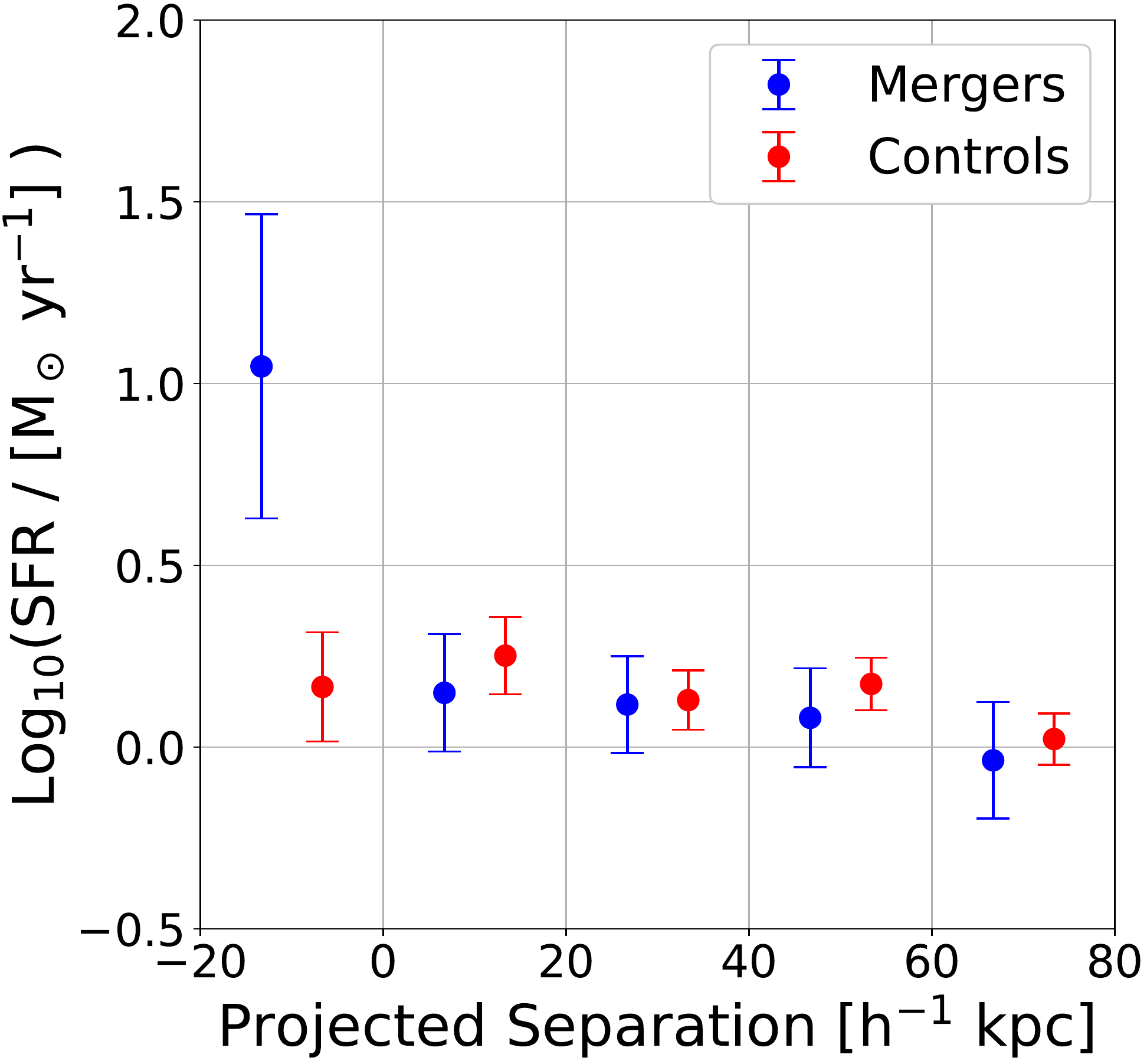}

\end{subfigure}
\caption{Average SFR in each bin of projected physical separation $r_p$ for the WISE+BPT sample (left panel), K03 AGN sample (middle panel) and K01 AGN sample (right panel). Errors computed using a 2$\sigma$ standard error of the mean. Horizontal spacing between data points \textit{within} each bin is arbitrary and only serves to enhance readability.}
\label{fig:SFR_rp}
\end{figure*}

We note that since we have not matched our controls in SFR for AGN, it is important to examine the SFR in mergers and controls to see if this could impact possible outflow trends presented here. Figure \ref{fig:SFR_rp} shows the SFR as a function of $r_p$ in the mergers and controls. For the WISE+BPT AGN sample we see that there is no appreciable difference between the merger and control SFR, though there is possible slight elevation seen in the post-merger bin and 60-80 kpc bin for the mergers. In the K01 and K03 AGN samples, we see a clear enhancement of the SFR in the post-mergers, and only the post-mergers. This is expected, since the SFR is known to naturally increase in our post-mergers sample \citep[see Figure 5 and Figure 6 in ][]{2013MNRAS.435.3627E}. However, elevations in SFR are unlikely to impact our results since the SFR enhancement in the post-merger bin is likely not sufficient to drive outflows as suggested by Figure \ref{fig:LumInc}, where it is evident that the outflow incidence remains at zero percent at all but the highest [OIII]~$\lambda$5007 luminosities. Hence, the effect of not matching our controls in SFR is insufficient to cause any significant impact on our results. Moreover, there is no enhancement in the outflow incidence, velocity, or luminosity in the post-merger bin that could be caused by an elevated SFR.

\begin{figure*}
\centering
\begin{subfigure}{0.33\textwidth}
  \centering
  
  \includegraphics[width=1\linewidth]{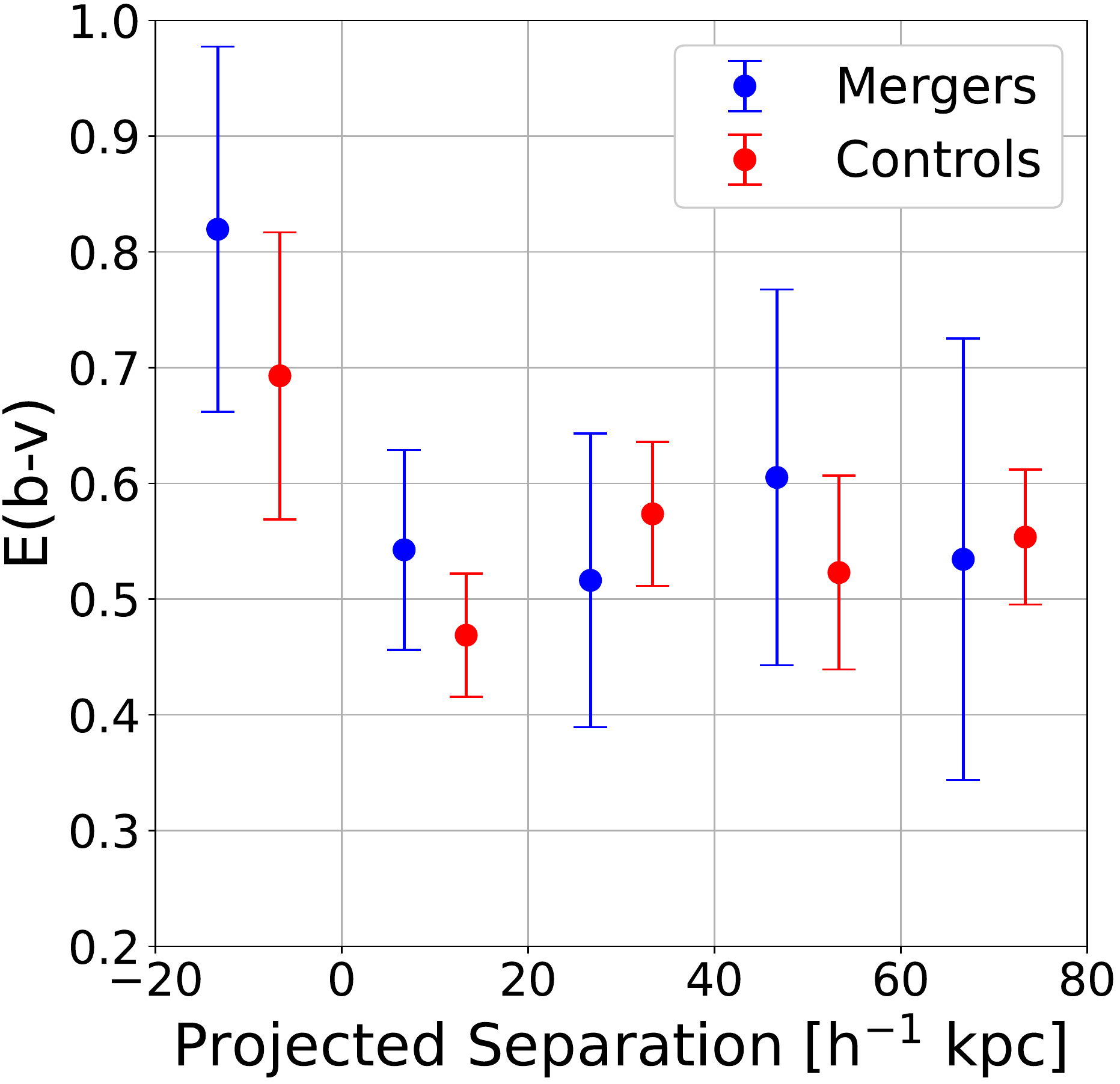}

\end{subfigure}%
\begin{subfigure}{0.33\textwidth}
  \centering
  
  \includegraphics[width=1\linewidth]{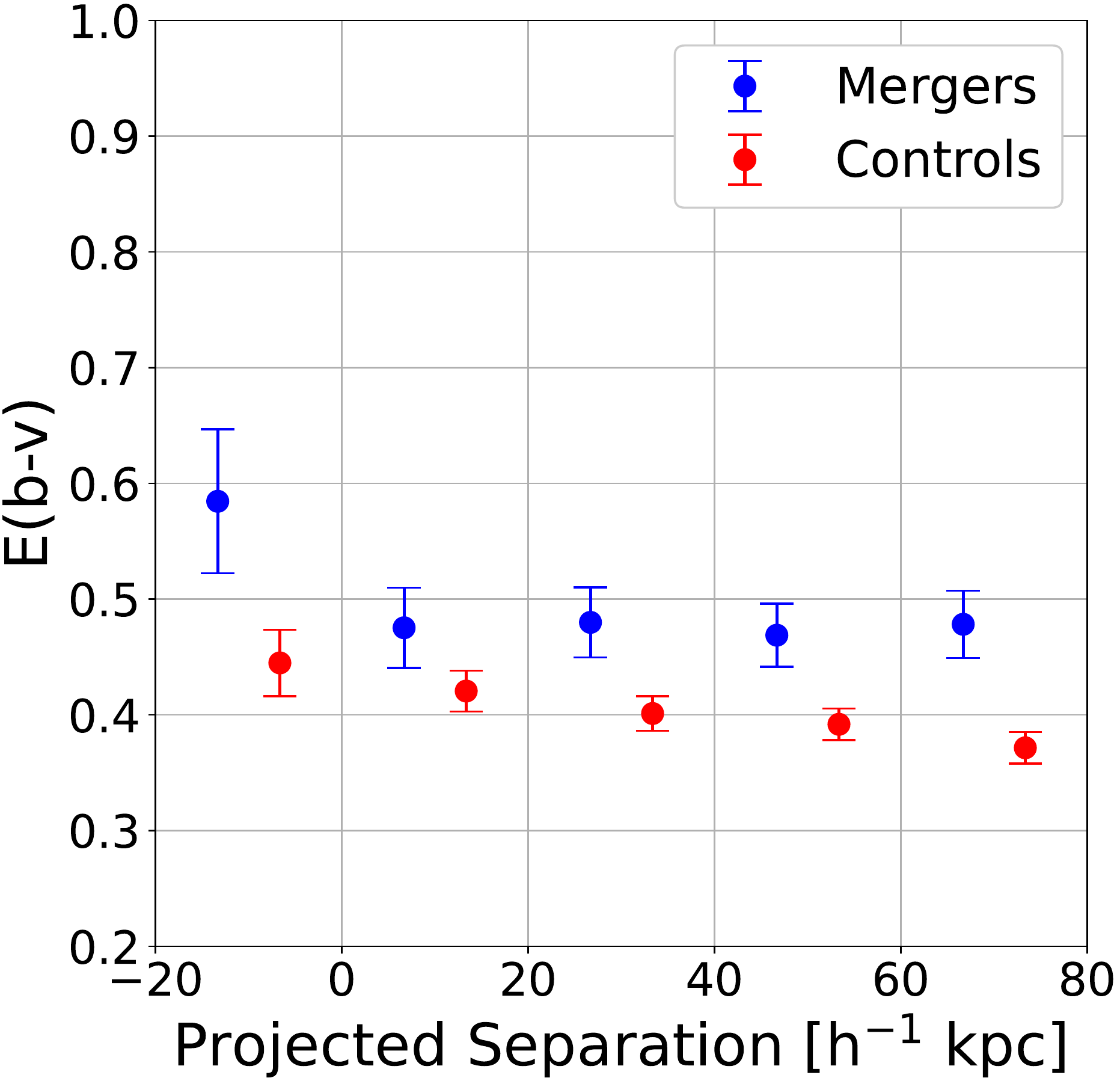}

\end{subfigure}
\begin{subfigure}{0.33\textwidth}
  \centering
  
  \includegraphics[width=1\linewidth]{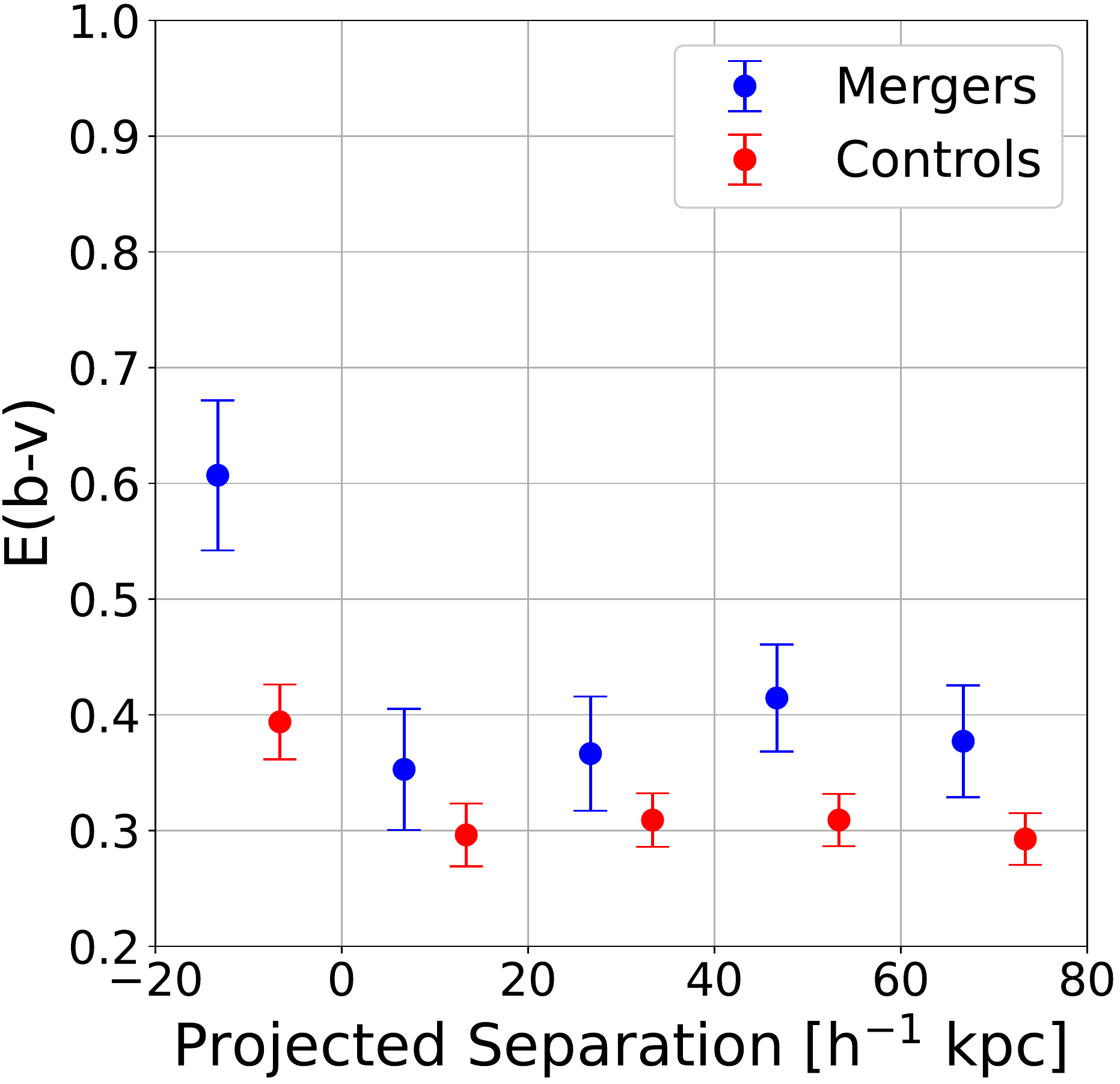}

\end{subfigure}
\caption{Average dust extinction in each bin of projected physical separation $r_p$ for the WISE+BPT sample (left panel), K03 AGN sample (middle panel) and K01 AGN sample (right panel). Errors are computed using a 2$\sigma$ standard error of the mean. Horizontal spacing between data points \textit{within} each bin is arbitrary and only serves to enhance readability.}
\label{fig:Binned_AGN_Extinction}
\end{figure*}

\begin{figure}
    \centering
    \includegraphics[width=0.70\linewidth]{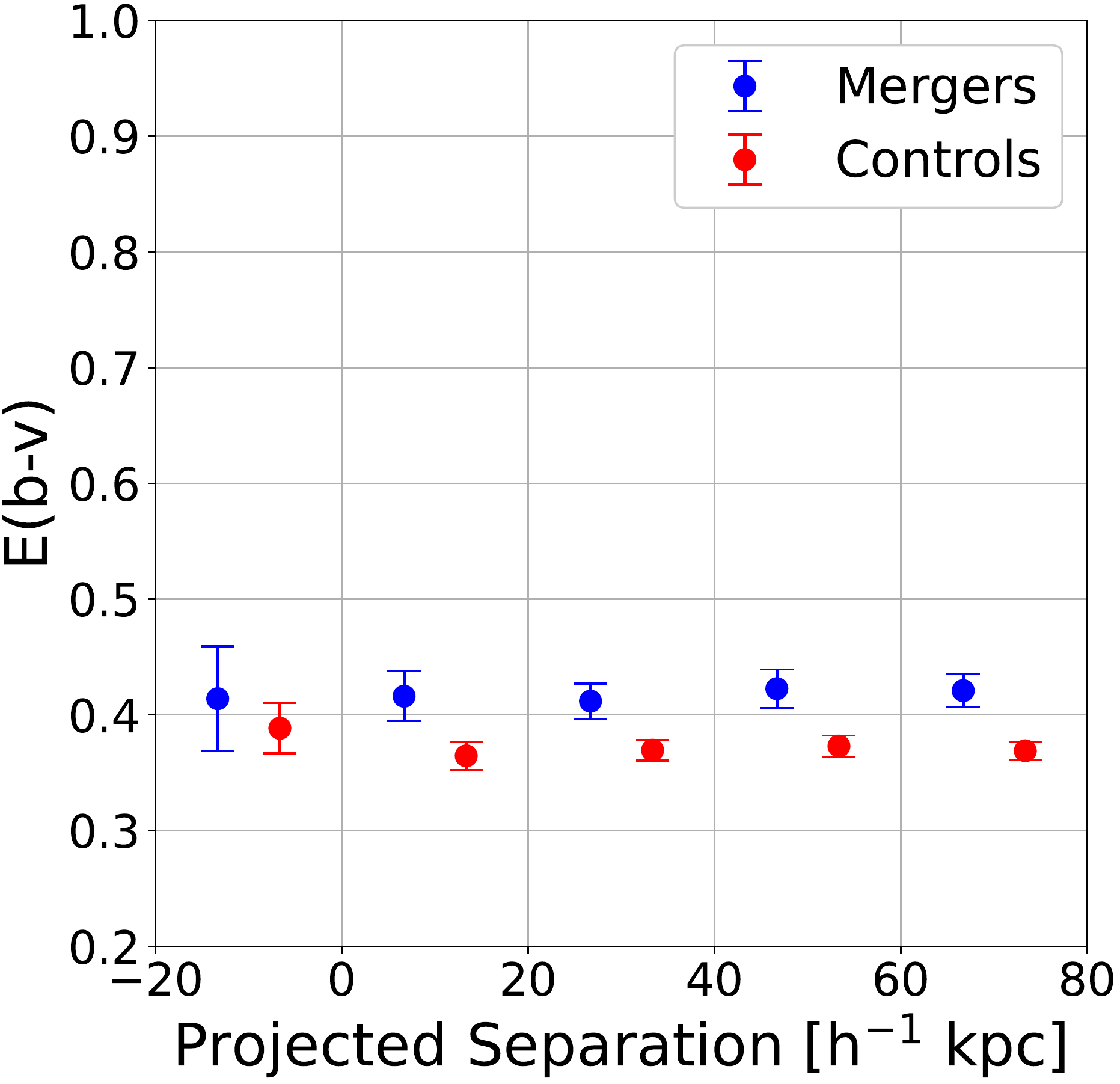}
    \caption{Average extinction in each bin of projected physical separation $r_p$ for SF galaxies. Errors computed using a 2$\sigma$ standard error of the mean. Horizontal spacing between data points \textit{within} each bin is arbitrary and only serves to enhance readability.}
    \label{fig:Binned_SF_Extinction}
\end{figure}

\section{Discussion}\label{sec: Discussion}

The above analysis yields several main results. First, we see in Figures \ref{fig:LumInc} and \ref{fig:VelLum} that the ionised outflow incidence and velocities, as traced by the [OIII]~$\lambda$5007 line, are significantly lower in SF galaxies compared with AGN even when normalizing by the luminosity of the [OIII]~$\lambda$5007 line. While the outflow incidence in both AGN and SF galaxies increases with the [OIII]~$\lambda$5007 luminosity, the outflow incidence is by far the greatest in galaxies that harbour an AGN. This general result is consistent with numerous other studies that report a low outflow incidence in SF galaxies compared to AGN \citep[ e.g.][]{2014A&A...562A..21C,2016MNRAS.456.1195H,2017A&A...606A..36C,2019ApJ...886...11L,2020A&A...633A.134L,2021MNRAS.503.5134A}. Similarly, the outflow velocity in SF galaxies is significantly less than those galaxies hosting an AGN, ($\sim$300~km~s$^{-1}$ vs. $\sim$700~km~s$^{-1}$). We only note an increase in average outflow velocity in the most luminous [OIII]~$\lambda$5007 bins. While we do expect a correlation with outflow velocity and [OIII]~$\lambda$5007 luminosity, others \citep[e.g.][]{2013ApJ...776...27V,2017A&A...601A.143F} have shown a much stronger trend with (bolometric) luminosity and outflow velocity. The discrepancy can, at least in part, be attributed to different measures of outflow velocity (e.g. using purely the velocity offset, 90th percentile of the emission line, or some combination of these) and different sample luminosity ranges (often being several orders of magnitude higher than ours). However, this may not \textit{fully} explain our discrepancy with [OIII]~$\lambda$5007 luminosity and outflow velocity. Figure 1 in \citet[][]{2021MNRAS.505.5469S} shows a strong correlation between $W_{80}$ and [OIII]~$\lambda$5007 luminosity for local ($z<0.4$) AGN in SDSS in our luminosity range. While we do not find such a strong correlation, we note our results are self-consistent and are robust for the questions we are investigating.

It is important to note that outflow incidence rates vary widely throughout the literature. Our bulk outflow detection rates in our AGN sub-samples are slightly lower than the lower-end estimates of outflow incidence found in the literature. For instance, \citet[][]{2019ApJ...884...54M,2020MNRAS.493.3081R,2020MNRAS.492.4680W,2021MNRAS.507.3985S,2021MNRAS.503.5134A} find outflow incidence rates in the $\sim$10-30 per cent range. However, \citet[][]{2012ApJ...757...86S,2013ApJ...776...27V,2014MNRAS.440.3202V,2017A&A...603A..99P,2017ApJ...850..140T,2018ApJ...865....5R} find that outflows are exceedingly common, at both moderate and high ($z \lesssim 2.6$) redshifts, with incidence rates in the $\sim$40-95 per cent range. There are many reasons for this wide range, including differences in spectral resolution (higher resolution data often find more outflows than lower resolution data), sample luminosities (outflows tend to be more common in galaxies with higher bolometric luminosities), and different fitting techniques. Even though outflow incidence is highly variable in the literature, overall trends such as outflow velocity/incidence increasing with bolometric luminosity, are generally consistent. 

In SF galaxies, the outflow incidence shows a significant increase only at the highest [OIII]~$\lambda$5007 luminosities in our study, as indicated in Figure \ref{fig:LumInc}. Indeed, the outflow incidence in SF galaxies is near zero percent, except at the highest luminosities where it is about five per cent. Even then, the outflow detection rate is far below that found in the other sub-samples. This is consistent with other studies in which the outflow incidence is $\sim$ zero per cent in galaxies at the lowest luminosities \citep[e.g.][]{2017A&A...606A..36C}, but is significantly higher in more luminous SF galaxies at redshifts closer to the peak in the cosmic star formation history \citep[e.g.][]{2010ApJ...719.1503R,2012ApJ...758..135K,2014ApJ...794..156R,2019ApJ...875...21F}, perhaps suggesting that there is some threshold above which SF can drive powerful outflows. Indeed, studies have suggested a minimum SFR surface density for outflows to be launched in SF galaxies at both moderate ($z \sim 1$) \citep[][]{2012ApJ...758..135K} and low ($z \sim 0$) \citep[][]{2020MNRAS.493.3081R} redshifts. Again from Figure \ref{fig:LumInc}, it is clear that for log($L_{\text{[OIII]}}$ [ergs s$^{-1}]$)$ \gtrsim 40.5$ the outflow incidence for galaxies harbouring AGN is significantly above that for the SF galaxies with comparable luminosities -- a result that is independent of presence or absence of a galaxy merger. This suggests that the presence and energetics of the AGN is needed to drive large-scale outflows, at least at these moderate luminosities. Alternatively, the conditions in the galaxies hosting AGNs could be such that it is easier to drive the outflow. For example, recent studies of the central kiloparsec region of AGN hosts have found lower molecular gas fractions compared with SF galaxies \citep[][]{2021MNRAS.505L..46E}, a result that is consistent with recent MaNGA IFU studies \citep[][]{2018RMxAA..54..217S}. The greatest molecular depletion factors found by \citet[][]{2021MNRAS.505L..46E} are at the highest luminosities, possibly suggesting that either the radiation field or shocks produced by the AGN reduces the central molecular gas fraction, making it easier to drive the currently observed outflows. 

These results also suggest, for the first time, that the outflow incidence is dependent on the AGN selection technique, with Figure \ref{fig:LumInc} showing that optical+mid-infrared selected AGN have the highest outflow fraction compared to purely optical AGNs at a given luminosity, especially at the highest [OIII]~$\lambda$5007 luminosities. It is not immediately apparent why this might be the case. This result is not a consequence of differences in AGN luminosity between AGN classes, and the mid-infrared selected AGN in our sample are also required to be optically BPT AGNs. One possibility is that mid-infrared selected AGN are found in galaxies with elevated star formation rates compared to optical AGN with the same luminosity, at least at higher luminosities. Hence, it is possible that both the AGN and enhanced star formation activity in these galaxies contribute to driving powerful outflows. Alternatively, the nuclear environment of our mid-infrared selected AGN may be denser than our optical AGN, thereby providing more raw material to power the outflow. It is worth noting that the outflow incidence in mid-infrared selected AGN is over 60 per cent at the highest luminosities, and may be slightly elevated compared to optical AGN in other higher luminosity bins. This 60 per cent outflow incidence is similar to \citet[][]{2022arXiv220413238B}, where they found ionised outflows in $\sim$60 per cent of their mid-infrared selected AGN, with similar [OIII]~$\lambda$5007 luminosities. Note, however, the outflow incidence in WISE+BPT AGN is only significantly elevated in the highest [OIII]~$\lambda$5007 luminosity bin in the control sample. While this result is only apparent in the control sample, the control sample has about two to three times the number of objects in each sample compared to the corresponding luminosity bin in the mergers (15 vs. 36 WISE+BPT AGN, 8 vs. 29 K01 AGN, and 10 vs. 28 K03 AGN), hence our control sample provides a more statistically robust result. Further, while there is marginal overlap in the error bars at the 95 per cent confidence level, they do not overlap at the 93 per cent confidence level. To assess the true significance of the difference in outflow incidence between the WISE+BPT AGN and optical AGN in this bin, we utilize Fisher's exact test and a binomial Continuous Distribution Function (CDF) to test for homogeneity between the samples. Our null hypothesis is that the WISE+BPT AGN and K01 AGN (or K03 AGN) have the same proportion of outflows. Fisher's exact test returns a p-value of only $\sim$0.003 in both sample comparisons. The binomial CDF returns a p-value of $\sim$0.03 in both sample comparisons. Therefore, we reject the null hypothesis and conclude that outflows are more common in the WISE+BPT AGN compared to optical AGN in the highest luminosity bin ($ 41.8 \leq L_{\mathrm{[OIII]~\lambda5007}} \leq 42.5$) to a confidence level greater than $\sim$97 per cent. Therefore, our work demonstrates that at the highest [OIII]~$\lambda$5007 luminosities, optical+mid-infrared selected AGNs do have a statistically significant higher outflow rate compared to AGNs that are only selected optically. Thus, at least at sufficiently high [OIII]~$\lambda$5007 luminosities, mid-infrared colour selection might be an effective strategy for finding outflows in large samples of galaxies.  

The main result of this work is, while the galaxy sub-type and luminosity have a clear impact on the incidence and velocity of outflows, Figures \ref{fig:Binned_AGN_Rate} and \ref{fig:Binned_SF_Rate} show that there is no statistically significant difference between these properties in mergers relative to a matched control sample, at any merger stage. We note that while the AGN incidence and SFR in mergers are generally expected to increase with decreasing pair separation, because we are requiring our controls to be matched in [OIII]~$\lambda$5007 luminosity, in addition to mass, redshift, and local galaxy density, we can isolate the role of the merging environment on the outflow properties. We note that the goal of this work is to determine the effect of the merger alone on the presence and properties of outflows. In order to perform the test we have designed, we must match in [OIII]~$\lambda$5007 luminosity when constructing our control sample because we know outflow incidence and velocity is correlated with luminosity. It is also clear that matching in stellar mass and redshift is critical since outflow incidence is found to be a function of stellar mass \citep[e.g.][]{2021MNRAS.503.5134A} and IFU studies have shown that outflows are centrally concentrated \citep[][]{2020ApJ...905..166L, 2021MNRAS.503.5134A,2021MNRAS.500.3802H}, indicating that aperture size will strongly impact outflow detection and properties. After these matching criteria are met, it is clear that there is no effect on the presence, velocity, and luminosity of outflows that results from the merger environment, at least as traced by the [OIII]~$\lambda$5007 line. 

If galaxy mergers cause gas inflows induced by gravitational torques, one might suppose that the outflows might be more difficult to drive in galaxy mergers compared to their matched controls. However our study demonstrates that no such impact is seen in the large scale outflows. We note that in this work, we only explore the presence of ionised outflows, their velocities, and their luminosities. The derived outflow properties, such as the mass outflow rate and mass loading factor, require knowledge of additional parameters such as the radial extent of the outflow and the electron density in the outflowing gas, which are difficult or impossible to constrain given the single aperture data presented here. As discussed in \citet[][]{2018NatAs...2..198H}, the electron density alone is a large source of uncertainty in these energetics calculations, with resulting calculations having up to three orders of magnitude uncertainty. We are also limited to the spatial regions covered by the SDSS fibres, so we cannot explore the spatial extent of the outflows. A robust determination of the outflow energetics will require high signal-to-noise and high spatial resolution IFU data in order to constrain the extent and mass of the outflow. Thus, while there is no clear difference in the outflow incidence and velocities in mergers relative to isolated controls, the mass outflow rates could be different, particularly if there are differences in the electron density within the entrained wind or in the spatial extent of the warm ionised gas in mergers relative to their matched controls. 

While there have not been any large-scale systematic studies that compare outflow properties in various merger stages relative to a robust control sample, \citet[][]{2014MNRAS.439.2701H} examined 115 ULIRGs for excess H$_2$ emission, which they interpreted as evidence of an outflow. Their findings are mixed. They find a weak positive correlation between H$_2$ and tidal tails of a galaxy, whose length can be used to estimate the stage of a galaxy merger. This correlation suggests that the outflows they found are more likely to be in advanced mergers. However, they do not find evidence of H$_2$ being correlated with nuclear separation. It should be stressed that, unlike this work, their study did not compare their mergers to a robust control sample. Nonetheless, their findings also suggest that there is no correlation between outflows and mergers, and no dependence on merger stage. 

It is important to note several caveats to our work. First, we do not explicitly match our AGN sub-sample in SFR. As a result, the SFR in our post-merger sample is enhanced relative to their controls. However, this enhancement in SFR does not produce an excess of outflows in our post-merger sample, implying that the SFR is not high enough to influence the results of our study. Second, the ionised outflows probed by the [OIII]~$\lambda$5007 line are based on single-fibre observations. There could be differences in mergers relative to their controls seen in the centrally concentrated gas -- the outflow signatures could be diluted in the larger aperture observations presented here. Indeed, \citet[][]{2010MNRAS.406L..55D} suggest that major mergers can enhance outflows in the nuclear region of galaxies, but cannot generate large-scale outflows. Higher spatial resolution observations would be required to explore differences on scales closer to those at which the outflows are physically launched. The SDSS utilizes a three arcsec fibre, which corresponds to a physical distance of $\sim5\pm2$~kpc at a redshift of $0.0835\pm0.0355$ (the average and standard deviation of the redshift in our pair sample), which is comparable to the effective radius of the galaxies in our sample. While aperture effects may play a role, recent MaNGA results \citep[e.g.][]{2021MNRAS.503.5134A} have shown that the outflows traced by the [OIII]~$\lambda$5007 line are centrally concentrated, with $\sim$67 per cent of outflowing galaxies showing the bulk of their broad-component wind signature encompassed within the effective radius of the galaxy. Further, there could be systematic differences in the attenuation of the optical emission lines in mergers relative to the controls, either as a result of foreground material from the inflowing gas from the merger, or changes in the dust attenuation in the entrained wind as a result of metallicity changes in the outflowing gas from the merger. There could also be differences in the metal enrichment of the outflowing gas from nuclear star formation, a phenomenon that has been previously suggested in low redshift galaxies in the MaNGA survey by \citet[][]{2021MNRAS.503.5134A}. Further, it is possible that the inflowing gas in mergers could cause significant obscuration of the optical line emission, preferentially causing outflows in mergers to be more elusive than in the control when using optical lines. We examine this possibility by plotting the extinction, as measured by E(B-V), as a function of $r_p$. Figure \ref{fig:Binned_AGN_Extinction} shows the extinction in mergers and controls in each bin for our AGN sub-samples. Note that in the WISE+BPT AGN extinction we see an elevated extinction in the post-mergers for \textit{both} the mergers and controls. In the K03 and K01 AGN, we see that there is a significantly heightened extinction in the mergers compared to the controls in nearly every $r_p$ bin. Figure \ref{fig:Binned_SF_Extinction} shows a similar extinction plot for SF galaxies. In any case, we see that there is generally a higher level of obscuration in the mergers compared to the controls in all sub-samples, except in the WISE+BPT sample. Finally, we note that the [OIII]~$\lambda$5007 luminosity may not be reliable for tracing AGN luminosity in the WISE+BPT AGN sample. As shown in Figure \ref{fig:Binned_AGN_Extinction}, the WISE+BPT AGN tend to be more heavily obscured than the optical AGN. If the central AGN in the WISE+BPT sample additionally heats the dust near the central engine, the [OIII]~$\lambda$5007 line might not be as accurate of a tracer for AGN bolometric luminosity as it is in the less obscured optically identified AGN.

\section{Conclusions}\label{sec: Conclusion}

In this paper, we have conducted the first large-scale systematic study of ionised outflows in galaxy pairs and post-mergers matched to a carefully constructed, robust control sample in order to determine the effect of mergers on the prevalence and properties of these outflows, and to explore the dependence on merger stage. Our sample consists of $\sim$4,000 galaxies at various merger stages drawn from the SDSS DR7, comprised of SF galaxies, optical AGN, and mid-infrared colour-selected AGN, classified based on the K01 and K03 AGN classification schemes. Each merger is matched in stellar mass, redshift, local density of galaxies and [OIII]~$\lambda$5007 luminosity (SF galaxies are additionally matched in SFR) to three isolated control galaxies. We examined the kinematics of the [OIII]~$\lambda$5007  emission line to search for blue-shifted, asymmetrical emission line profiles, which are interpreted as ionised outflows. Our results are summarized as follows:
\begin{enumerate}
    \item{The merging environment has no statistically significant impact on the incidence or properties of the ionised outflows, and there is no dependence seen on merger stage. }
    \item{Ionised outflows are significantly less common in SF galaxies compared to galaxies that host an AGN in the sample as a whole, even when normalizing by the luminosity of the [OIII]~$\lambda$5007 line. This result is consistent with numerous other studies exploring ionised outflows in galaxies. Further, outflows are much less common in LINERs compared to Seyferts.}
    \item{The outflow incidence and velocity increase with [OIII]~$\lambda$5007 luminosity for both SF galaxies and galaxies harbouring AGN. In contrast to AGN, there is a threshold in the SF galaxies of log($L_{\text{[OIII]}}$ (ergs s$^{-1})$)$ \gtrsim 41.5$ below which the outflow incidence is near zero percent. }
    \item{Outflow velocities are lowest in the SF galaxies, with an average of $\sim$300~km~s$^{-1}$. The average outflow velocity in all AGN sub-samples is $\sim$700~km~s$^{-1}$.}
    \item{The outflow incidence depends on the presence of AGN and AGN selection technique, with optical+mid-infrared selected AGN showing the largest outflow incidence. SF galaxies have an outflow incidence of only $\sim$0.5 per cent, while K03 AGN have an incidence of $\sim$7 per cent and K01 AGN have an incidence of $\sim$14 per cent. Optical+mid-infrared selected AGN have an average incidence of $\sim$40 per cent, with an incidence rate of $\sim60$ per cent at the highest [OIII]~$\lambda$5007 luminosities. Mid-infrared colour selection may therefore be an effective pre-selection strategy for finding outflows in large samples of galaxies}.
    \item{The outflow luminosity is lowest in SF galaxies, with an average of $\log_{\text{10}}(L_{\text{outflow}} [\text{ergs~s}^{-1}])$ $\sim$40.0. AGNs have a slightly elevated average outflow luminosity of $\log_{\text{10}}(L_{\text{outflow}} [\text{ergs~s}^{-1}]$) $\sim$40.7.}
\end{enumerate}

While these results suggest that the merging environment does not have a significant impact on the presence or properties of ionised outflows, our observations are based on large aperture, single-fibre observations. If outflows are more concentrated in the centres of galaxies, the signatures of the outflowing gas could be diluted in the spectra analysed here. Further, optical emission lines in mergers could be more heavily attenuated compared to the controls. Near-infrared spectroscopic observations are necessary to fully understand the effects of extinction and the resulting impact on the outflow incidence and properties. In order to obtain a complete understanding of how ionised outflows behave in mergers, higher spatial resolution near-infrared IFU data are needed.

\section{Acknowledgements}
The authors thank the anonymous referee for providing insightful feedback that improved the quality and clarity of the manuscript. W. M. is grateful to R. Weigel for discussions on the statistical analysis conducted in this work, T. Bohn for his helpful plotting routines, and S. Silayi for advice on fitting spectra efficiently. G. C. acknowledges support by the National Science Foundation, under grant No. AST 1817233. L. B. acknowledges support from NASA award \#80NSSC20K0502. This publication makes use of data products from the {\it Wide-field Infrared Survey Explorer}, which is a joint project of the University of California, Los Angeles, and the Jet Propulsion Laboratory/California Institute of Technology, funded by the National Aeronautics and Space Administration. We are grateful to the MPA/JHU group for access to their data products and catalogues (maintained by Jarle Brinchmann at \url{http://www.mpa-garching.mpg.de/SDSS/}). Funding for the SDSS and SDSS--II has been provided by the Alfred P. Sloan Foundation, the Participating Institutions, the National Science Foundation, the U.S. Department of Energy, the National Aeronautics and Space Administration, the Japanese Monbukagakusho, the Max Planck Society, and the Higher Education Funding Council for England. The SDSS Web Site is \url{http://www.sdss.org/}. The SDSS is managed by the Astrophysical Research Consortium for the Participating Institutions. The Participating Institutions are the American Museum of Natural History, Astrophysical Institute Potsdam, University of Basel, University of Cambridge, Case Western Reserve University, University of Chicago, Drexel University, Fermilab, the Institute for Advanced Study, the Japan Participation Group, Johns Hopkins University, the Joint Institute for Nuclear Astrophysics, the Kavli Institute for Particle Astrophysics and Cosmology, the Korean Scientist Group, and the Chinese Academy.

The spectral fits carried out in this work were run on ARGO, a research computing cluster provided by the Office of Research Computing at George Mason University, VA. (\url{ http://orc.gmu.edu}). This research made use of Astropy,\footnote{\url{http://www.astropy.org}} a community-developed core Python package for Astronomy \citep{2018AJ....156..123A}, as well as \textsc{topcat} \citep{2005ASPC..347...29T}.

\section{Data Availability}
The data sets used in this publication were derived from the open-source code Bayesian AGN Decomposition Analysis for SDSS Spectra available at \url{https://github.com/remingtonsexton/BADASS3}, the Wide-field Infrared Survey Explorer (\url{https://wise2.ipac.caltech.edu/docs/release/allsky/}), the MPA/JHU catalogue (\url{http://www.mpa-garching.mpg.de/SDSS/}), and the Sloan Digital Sky Survey (SDSS; \url{http://www.sdss.org/}). Our sample was assembled from the SDSS Data Release 7 Main Galaxy Sample \citep[][]{2009ApJS..182..543A}. Corresponding spectra were queried from the SDSS Data Release 12 \citep[][]{2015ApJS..219...12A}, which can be accessed from the traditional Science Archive Server \url{https://dr12.sdss.org/sas}. The processed data sets underlying this article will be shared on reasonable request to the corresponding author.


\bibliographystyle{mnras}
\bibliography{my.bib} 

\begin{thebibliography}{}
\makeatletter
\relax
\def\mn@urlcharsother{\let\do\@makeother \do\$\do\&\do\#\do\^\do\_\do\%\do\~}
\def\mn@doi{\begingroup\mn@urlcharsother \@ifnextchar [ {\mn@doi@}
  {\mn@doi@[]}}
\def\mn@doi@[#1]#2{\def\@tempa{#1}\ifx\@tempa\@empty \href
  {http://dx.doi.org/#2} {doi:#2}\else \href {http://dx.doi.org/#2} {#1}\fi
  \endgroup}
\def\mn@eprint#1#2{\mn@eprint@#1:#2::\@nil}
\def\mn@eprint@arXiv#1{\href {http://arxiv.org/abs/#1} {{\tt arXiv:#1}}}
\def\mn@eprint@dblp#1{\href {http://dblp.uni-trier.de/rec/bibtex/#1.xml}
  {dblp:#1}}
\def\mn@eprint@#1:#2:#3:#4\@nil{\def\@tempa {#1}\def\@tempb {#2}\def\@tempc
  {#3}\ifx \@tempc \@empty \let \@tempc \@tempb \let \@tempb \@tempa \fi \ifx
  \@tempb \@empty \def\@tempb {arXiv}\fi \@ifundefined
  {mn@eprint@\@tempb}{\@tempb:\@tempc}{\expandafter \expandafter \csname
  mn@eprint@\@tempb\endcsname \expandafter{\@tempc}}}

\bibitem[\protect\citeauthoryear{{Abazajian} et~al.,}{{Abazajian}
  et~al.}{2009}]{2009ApJS..182..543A}
{Abazajian} K.~N.,  et~al., 2009, \mn@doi [\apjs]
  {10.1088/0067-0049/182/2/543}, \href
  {https://ui.adsabs.harvard.edu/abs/2009ApJS..182..543A} {182, 543}

\bibitem[\protect\citeauthoryear{{Agostino} et~al.,}{{Agostino}
  et~al.}{2021}]{2021ApJ...922..156A}
{Agostino} C.~J.,  et~al., 2021, \mn@doi [\apj] {10.3847/1538-4357/ac1e8d},
  \href {https://ui.adsabs.harvard.edu/abs/2021ApJ...922..156A} {922, 156}

\bibitem[\protect\citeauthoryear{{Alam} et~al.,}{{Alam}
  et~al.}{2015}]{2015ApJS..219...12A}
{Alam} S.,  et~al., 2015, \mn@doi [\apjs] {10.1088/0067-0049/219/1/12}, \href
  {https://ui.adsabs.harvard.edu/abs/2015ApJS..219...12A} {219, 12}

\bibitem[\protect\citeauthoryear{{Astropy Collaboration} et~al.,}{{Astropy
  Collaboration} et~al.}{2018}]{2018AJ....156..123A}
{Astropy Collaboration} et~al., 2018, \mn@doi [\aj] {10.3847/1538-3881/aabc4f},
  \href {https://ui.adsabs.harvard.edu/abs/2018AJ....156..123A} {156, 123}

\bibitem[\protect\citeauthoryear{{Avery} et~al.,}{{Avery}
  et~al.}{2021}]{2021MNRAS.503.5134A}
{Avery} C.~R.,  et~al., 2021, \mn@doi [\mnras] {10.1093/mnras/stab780}, \href
  {https://ui.adsabs.harvard.edu/abs/2021MNRAS.503.5134A} {503, 5134}

\bibitem[\protect\citeauthoryear{{Bae} \& {Woo}}{{Bae} \&
  {Woo}}{2016}]{2016ApJ...828...97B}
{Bae} H.-J.,  {Woo} J.-H.,  2016, \mn@doi [\apj] {10.3847/0004-637X/828/2/97},
  \href {https://ui.adsabs.harvard.edu/abs/2016ApJ...828...97B} {828, 97}

\bibitem[\protect\citeauthoryear{{Baldwin}, {Phillips}  \&
  {Terlevich}}{{Baldwin} et~al.}{1981}]{1981PASP...93....5B}
{Baldwin} J.~A.,  {Phillips} M.~M.,   {Terlevich} R.,  1981, \mn@doi [\pasp]
  {10.1086/130766}, \href
  {https://ui.adsabs.harvard.edu/abs/1981PASP...93....5B} {93, 5}

\bibitem[\protect\citeauthoryear{{Balmaverde} et~al.,}{{Balmaverde}
  et~al.}{2016}]{2016A&A...585A.148B}
{Balmaverde} B.,  et~al., 2016, \mn@doi [\aap] {10.1051/0004-6361/201526694},
  \href {https://ui.adsabs.harvard.edu/abs/2016A&A...585A.148B} {585, A148}

\bibitem[\protect\citeauthoryear{{Barnes} \& {Hernquist}}{{Barnes} \&
  {Hernquist}}{1991}]{1991ApJ...370L..65B}
{Barnes} J.~E.,  {Hernquist} L.~E.,  1991, \mn@doi [\apjl] {10.1086/185978},
  \href {https://ui.adsabs.harvard.edu/abs/1991ApJ...370L..65B} {370, L65}

\bibitem[\protect\citeauthoryear{{Baron} et~al.,}{{Baron}
  et~al.}{2018}]{2018MNRAS.480.3993B}
{Baron} D.,  et~al., 2018, \mn@doi [\mnras] {10.1093/mnras/sty2113}, \href
  {https://ui.adsabs.harvard.edu/abs/2018MNRAS.480.3993B} {480, 3993}

\bibitem[\protect\citeauthoryear{{Barrows}, {Sandberg Lacy}, {Kennefick},
  {Comerford}, {Kennefick}  \& {Berrier}}{{Barrows}
  et~al.}{2013}]{2013ApJ...769...95B}
{Barrows} R.~S.,  {Sandberg Lacy} C.~H.,  {Kennefick} J.,  {Comerford} J.~M.,
  {Kennefick} D.,   {Berrier} J.~C.,  2013, \mn@doi [\apj]
  {10.1088/0004-637X/769/2/95}, \href
  {https://ui.adsabs.harvard.edu/abs/2013ApJ...769...95B} {769, 95}

\bibitem[\protect\citeauthoryear{{Blecha}, {Snyder}, {Satyapal}  \&
  {Ellison}}{{Blecha} et~al.}{2018}]{2018MNRAS.478.3056B}
{Blecha} L.,  {Snyder} G.~F.,  {Satyapal} S.,   {Ellison} S.~L.,  2018, \mn@doi
  [\mnras] {10.1093/mnras/sty1274}, \href
  {https://ui.adsabs.harvard.edu/abs/2018MNRAS.478.3056B} {478, 3056}

\bibitem[\protect\citeauthoryear{{Blumenthal} \& {Barnes}}{{Blumenthal} \&
  {Barnes}}{2018}]{2018MNRAS.479.3952B}
{Blumenthal} K.~A.,  {Barnes} J.~E.,  2018, \mn@doi [\mnras]
  {10.1093/mnras/sty1605}, \href
  {https://ui.adsabs.harvard.edu/abs/2018MNRAS.479.3952B} {479, 3952}

\bibitem[\protect\citeauthoryear{{Bohn}, {Canalizo}, {Satyapal}  \&
  {Sales}}{{Bohn} et~al.}{2022}]{2022arXiv220413238B}
{Bohn} T.,  {Canalizo} G.,  {Satyapal} S.,   {Sales} L.,  2022, arXiv e-prints,
  \href {https://ui.adsabs.harvard.edu/abs/2022arXiv220413238B} {p.
  arXiv:2204.13238}

\bibitem[\protect\citeauthoryear{{Brinchmann}, {Charlot}, {White}, {Tremonti},
  {Kauffmann}, {Heckman}  \& {Brinkmann}}{{Brinchmann}
  et~al.}{2004}]{2004MNRAS.351.1151B}
{Brinchmann} J.,  {Charlot} S.,  {White} S.~D.~M.,  {Tremonti} C.,  {Kauffmann}
  G.,  {Heckman} T.,   {Brinkmann} J.,  2004, \mn@doi [\mnras]
  {10.1111/j.1365-2966.2004.07881.x}, \href
  {https://ui.adsabs.harvard.edu/abs/2004MNRAS.351.1151B} {351, 1151}

\bibitem[\protect\citeauthoryear{{Bruce}, {Dunlop}, {Mortlock}, {Kocevski},
  {McGrath}  \& {Rosario}}{{Bruce} et~al.}{2016}]{2016MNRAS.458.2391B}
{Bruce} V.~A.,  {Dunlop} J.~S.,  {Mortlock} A.,  {Kocevski} D.~D.,  {McGrath}
  E.~J.,   {Rosario} D.~J.,  2016, \mn@doi [\mnras] {10.1093/mnras/stw467},
  \href {https://ui.adsabs.harvard.edu/abs/2016MNRAS.458.2391B} {458, 2391}

\bibitem[\protect\citeauthoryear{{Canalizo} \& {Stockton}}{{Canalizo} \&
  {Stockton}}{2001}]{2001ApJ...555..719C}
{Canalizo} G.,  {Stockton} A.,  2001, \mn@doi [\apj] {10.1086/321520}, \href
  {https://ui.adsabs.harvard.edu/abs/2001ApJ...555..719C} {555, 719}

\bibitem[\protect\citeauthoryear{{Carniani} et~al.,}{{Carniani}
  et~al.}{2015}]{2015A&A...580A.102C}
{Carniani} S.,  et~al., 2015, \mn@doi [\aap] {10.1051/0004-6361/201526557},
  \href {https://ui.adsabs.harvard.edu/abs/2015A&A...580A.102C} {580, A102}

\bibitem[\protect\citeauthoryear{{Cicone} et~al.,}{{Cicone}
  et~al.}{2014}]{2014A&A...562A..21C}
{Cicone} C.,  et~al., 2014, \mn@doi [\aap] {10.1051/0004-6361/201322464}, \href
  {https://ui.adsabs.harvard.edu/abs/2014A&A...562A..21C} {562, A21}

\bibitem[\protect\citeauthoryear{{Cicone}, {Maiolino}  \& {Marconi}}{{Cicone}
  et~al.}{2016}]{2016A&A...588A..41C}
{Cicone} C.,  {Maiolino} R.,   {Marconi} A.,  2016, \mn@doi [\aap]
  {10.1051/0004-6361/201424514}, \href
  {https://ui.adsabs.harvard.edu/abs/2016A&A...588A..41C} {588, A41}

\bibitem[\protect\citeauthoryear{{Cisternas} et~al.,}{{Cisternas}
  et~al.}{2011}]{2011ApJ...726...57C}
{Cisternas} M.,  et~al., 2011, \mn@doi [\apj] {10.1088/0004-637X/726/2/57},
  \href {https://ui.adsabs.harvard.edu/abs/2011ApJ...726...57C} {726, 57}

\bibitem[\protect\citeauthoryear{{Comerford}, {Nevin}, {Stemo},
  {M{\"u}ller-S{\'a}nchez}, {Barrows}, {Cooper}  \& {Newman}}{{Comerford}
  et~al.}{2018}]{2018ApJ...867...66C}
{Comerford} J.~M.,  {Nevin} R.,  {Stemo} A.,  {M{\"u}ller-S{\'a}nchez} F.,
  {Barrows} R.~S.,  {Cooper} M.~C.,   {Newman} J.~A.,  2018, \mn@doi [\apj]
  {10.3847/1538-4357/aae2b4}, \href
  {https://ui.adsabs.harvard.edu/abs/2018ApJ...867...66C} {867, 66}

\bibitem[\protect\citeauthoryear{{Concas}, {Popesso}, {Brusa}, {Mainieri},
  {Erfanianfar}  \& {Morselli}}{{Concas} et~al.}{2017}]{2017A&A...606A..36C}
{Concas} A.,  {Popesso} P.,  {Brusa} M.,  {Mainieri} V.,  {Erfanianfar} G.,
  {Morselli} L.,  2017, \mn@doi [\aap] {10.1051/0004-6361/201629519}, \href
  {https://ui.adsabs.harvard.edu/abs/2017A&A...606A..36C} {606, A36}

\bibitem[\protect\citeauthoryear{{Cresci} et~al.,}{{Cresci}
  et~al.}{2015}]{2015A&A...582A..63C}
{Cresci} G.,  et~al., 2015, \mn@doi [\aap] {10.1051/0004-6361/201526581}, \href
  {https://ui.adsabs.harvard.edu/abs/2015A&A...582A..63C} {582, A63}

\bibitem[\protect\citeauthoryear{{Darg} et~al.,}{{Darg}
  et~al.}{2010}]{2010MNRAS.401.1043D}
{Darg} D.~W.,  et~al., 2010, \mn@doi [\mnras]
  {10.1111/j.1365-2966.2009.15686.x}, \href
  {https://ui.adsabs.harvard.edu/abs/2010MNRAS.401.1043D} {401, 1043}

\bibitem[\protect\citeauthoryear{{Davies} et~al.,}{{Davies}
  et~al.}{2019}]{2019ApJ...873..122D}
{Davies} R.~L.,  et~al., 2019, \mn@doi [\apj] {10.3847/1538-4357/ab06f1}, \href
  {https://ui.adsabs.harvard.edu/abs/2019ApJ...873..122D} {873, 122}

\bibitem[\protect\citeauthoryear{{Debuhr}, {Quataert}, {Ma}  \&
  {Hopkins}}{{Debuhr} et~al.}{2010}]{2010MNRAS.406L..55D}
{Debuhr} J.,  {Quataert} E.,  {Ma} C.-P.,   {Hopkins} P.,  2010, \mn@doi
  [\mnras] {10.1111/j.1745-3933.2010.00881.x}, \href
  {https://ui.adsabs.harvard.edu/abs/2010MNRAS.406L..55D} {406, L55}

\bibitem[\protect\citeauthoryear{{Ellison}, {Patton}, {Simard}  \&
  {McConnachie}}{{Ellison} et~al.}{2008}]{2008AJ....135.1877E}
{Ellison} S.~L.,  {Patton} D.~R.,  {Simard} L.,   {McConnachie} A.~W.,  2008,
  \mn@doi [\aj] {10.1088/0004-6256/135/5/1877}, \href
  {https://ui.adsabs.harvard.edu/abs/2008AJ....135.1877E} {135, 1877}

\bibitem[\protect\citeauthoryear{{Ellison}, {Patton}, {Simard}, {McConnachie},
  {Baldry}  \& {Mendel}}{{Ellison} et~al.}{2010}]{2010MNRAS.407.1514E}
{Ellison} S.~L.,  {Patton} D.~R.,  {Simard} L.,  {McConnachie} A.~W.,  {Baldry}
  I.~K.,   {Mendel} J.~T.,  2010, \mn@doi [\mnras]
  {10.1111/j.1365-2966.2010.17076.x}, \href
  {https://ui.adsabs.harvard.edu/abs/2010MNRAS.407.1514E} {407, 1514}

\bibitem[\protect\citeauthoryear{{Ellison}, {Patton}, {Mendel}  \&
  {Scudder}}{{Ellison} et~al.}{2011}]{2011MNRAS.418.2043E}
{Ellison} S.~L.,  {Patton} D.~R.,  {Mendel} J.~T.,   {Scudder} J.~M.,  2011,
  \mn@doi [\mnras] {10.1111/j.1365-2966.2011.19624.x}, \href
  {https://ui.adsabs.harvard.edu/abs/2011MNRAS.418.2043E} {418, 2043}

\bibitem[\protect\citeauthoryear{{Ellison}, {Mendel}, {Scudder}, {Patton}  \&
  {Palmer}}{{Ellison} et~al.}{2013a}]{2013MNRAS.430.3128E}
{Ellison} S.~L.,  {Mendel} J.~T.,  {Scudder} J.~M.,  {Patton} D.~R.,   {Palmer}
  M. J.~D.,  2013a, \mn@doi [\mnras] {10.1093/mnras/sts546}, \href
  {https://ui.adsabs.harvard.edu/abs/2013MNRAS.430.3128E} {430, 3128}

\bibitem[\protect\citeauthoryear{{Ellison}, {Mendel}, {Patton}  \&
  {Scudder}}{{Ellison} et~al.}{2013b}]{2013MNRAS.435.3627E}
{Ellison} S.~L.,  {Mendel} J.~T.,  {Patton} D.~R.,   {Scudder} J.~M.,  2013b,
  \mn@doi [\mnras] {10.1093/mnras/stt1562}, \href
  {https://ui.adsabs.harvard.edu/abs/2013MNRAS.435.3627E} {435, 3627}

\bibitem[\protect\citeauthoryear{{Ellison}, {Patton}  \& {Hickox}}{{Ellison}
  et~al.}{2015}]{2015MNRAS.451L..35E}
{Ellison} S.~L.,  {Patton} D.~R.,   {Hickox} R.~C.,  2015, \mn@doi [\mnras]
  {10.1093/mnrasl/slv061}, \href
  {https://ui.adsabs.harvard.edu/abs/2015MNRAS.451L..35E} {451, L35}

\bibitem[\protect\citeauthoryear{{Ellison}, {Viswanathan}, {Patton},
  {Bottrell}, {McConnachie}, {Gwyn}  \& {Cuillandre}}{{Ellison}
  et~al.}{2019}]{2019MNRAS.487.2491E}
{Ellison} S.~L.,  {Viswanathan} A.,  {Patton} D.~R.,  {Bottrell} C.,
  {McConnachie} A.~W.,  {Gwyn} S.,   {Cuillandre} J.-C.,  2019, \mn@doi
  [\mnras] {10.1093/mnras/stz1431}, \href
  {https://ui.adsabs.harvard.edu/abs/2019MNRAS.487.2491E} {487, 2491}

\bibitem[\protect\citeauthoryear{{Ellison} et~al.,}{{Ellison}
  et~al.}{2021}]{2021MNRAS.505L..46E}
{Ellison} S.~L.,  et~al., 2021, \mn@doi [\mnras] {10.1093/mnrasl/slab047},
  \href {https://ui.adsabs.harvard.edu/abs/2021MNRAS.505L..46E} {505, L46}

\bibitem[\protect\citeauthoryear{{Fensch} et~al.,}{{Fensch}
  et~al.}{2017}]{2017MNRAS.465.1934F}
{Fensch} J.,  et~al., 2017, \mn@doi [\mnras] {10.1093/mnras/stw2920}, \href
  {https://ui.adsabs.harvard.edu/abs/2017MNRAS.465.1934F} {465, 1934}

\bibitem[\protect\citeauthoryear{{Fiore} et~al.,}{{Fiore}
  et~al.}{2017}]{2017A&A...601A.143F}
{Fiore} F.,  et~al., 2017, \mn@doi [\aap] {10.1051/0004-6361/201629478}, \href
  {https://ui.adsabs.harvard.edu/abs/2017A&A...601A.143F} {601, A143}

\bibitem[\protect\citeauthoryear{{Fluetsch} et~al.,}{{Fluetsch}
  et~al.}{2019}]{2019MNRAS.483.4586F}
{Fluetsch} A.,  et~al., 2019, \mn@doi [\mnras] {10.1093/mnras/sty3449}, \href
  {https://ui.adsabs.harvard.edu/abs/2019MNRAS.483.4586F} {483, 4586}

\bibitem[\protect\citeauthoryear{{Fluetsch} et~al.,}{{Fluetsch}
  et~al.}{2021}]{2021MNRAS.505.5753F}
{Fluetsch} A.,  et~al., 2021, \mn@doi [\mnras] {10.1093/mnras/stab1666}, \href
  {https://ui.adsabs.harvard.edu/abs/2021MNRAS.505.5753F} {505, 5753}

\bibitem[\protect\citeauthoryear{{Foreman-Mackey}, {Hogg}, {Lang}  \&
  {Goodman}}{{Foreman-Mackey} et~al.}{2013}]{2013PASP..125..306F}
{Foreman-Mackey} D.,  {Hogg} D.~W.,  {Lang} D.,   {Goodman} J.,  2013, \mn@doi
  [\pasp] {10.1086/670067}, \href
  {https://ui.adsabs.harvard.edu/abs/2013PASP..125..306F} {125, 306}

\bibitem[\protect\citeauthoryear{{F{\"o}rster Schreiber} et~al.,}{{F{\"o}rster
  Schreiber} et~al.}{2019}]{2019ApJ...875...21F}
{F{\"o}rster Schreiber} N.~M.,  et~al., 2019, \mn@doi [\apj]
  {10.3847/1538-4357/ab0ca2}, \href
  {https://ui.adsabs.harvard.edu/abs/2019ApJ...875...21F} {875, 21}

\bibitem[\protect\citeauthoryear{{Geach} et~al.,}{{Geach}
  et~al.}{2018}]{2018ApJ...864L...1G}
{Geach} J.~E.,  et~al., 2018, \mn@doi [\apjl] {10.3847/2041-8213/aad8b6}, \href
  {https://ui.adsabs.harvard.edu/abs/2018ApJ...864L...1G} {864, L1}

\bibitem[\protect\citeauthoryear{{Gehrels}}{{Gehrels}}{1986}]{1986ApJ...303..336G}
{Gehrels} N.,  1986, \mn@doi [\apj] {10.1086/164079}, \href
  {https://ui.adsabs.harvard.edu/abs/1986ApJ...303..336G} {303, 336}

\bibitem[\protect\citeauthoryear{{Genzel} et~al.,}{{Genzel}
  et~al.}{2011}]{2011ApJ...733..101G}
{Genzel} R.,  et~al., 2011, \mn@doi [\apj] {10.1088/0004-637X/733/2/101}, \href
  {https://ui.adsabs.harvard.edu/abs/2011ApJ...733..101G} {733, 101}

\bibitem[\protect\citeauthoryear{{Goulding} et~al.,}{{Goulding}
  et~al.}{2018}]{2018PASJ...70S..37G}
{Goulding} A.~D.,  et~al., 2018, \mn@doi [\pasj] {10.1093/pasj/psx135}, \href
  {https://ui.adsabs.harvard.edu/abs/2018PASJ...70S..37G} {70, S37}

\bibitem[\protect\citeauthoryear{{Guolo-Pereira}, {Ruschel-Dutra},
  {Storchi-Bergmann}, {Schnorr-M{\"u}ller}, {Cid Fernandes}, {Couto}, {Dametto}
   \& {Hernandez-Jimenez}}{{Guolo-Pereira} et~al.}{2021}]{2021MNRAS.502.3618G}
{Guolo-Pereira} M.,  {Ruschel-Dutra} D.,  {Storchi-Bergmann} T.,
  {Schnorr-M{\"u}ller} A.,  {Cid Fernandes} R.,  {Couto} G.,  {Dametto} N.,
  {Hernandez-Jimenez} J.~A.,  2021, \mn@doi [\mnras] {10.1093/mnras/stab245},
  \href {https://ui.adsabs.harvard.edu/abs/2021MNRAS.502.3618G} {502, 3618}

\bibitem[\protect\citeauthoryear{{Harrison}, {Alexander}, {Mullaney}  \&
  {Swinbank}}{{Harrison} et~al.}{2014}]{2014MNRAS.441.3306H}
{Harrison} C.~M.,  {Alexander} D.~M.,  {Mullaney} J.~R.,   {Swinbank} A.~M.,
  2014, \mn@doi [\mnras] {10.1093/mnras/stu515}, \href
  {https://ui.adsabs.harvard.edu/abs/2014MNRAS.441.3306H} {441, 3306}

\bibitem[\protect\citeauthoryear{{Harrison} et~al.,}{{Harrison}
  et~al.}{2016}]{2016MNRAS.456.1195H}
{Harrison} C.~M.,  et~al., 2016, \mn@doi [\mnras] {10.1093/mnras/stv2727},
  \href {https://ui.adsabs.harvard.edu/abs/2016MNRAS.456.1195H} {456, 1195}

\bibitem[\protect\citeauthoryear{{Harrison}, {Costa}, {Tadhunter},
  {Fl{\"u}tsch}, {Kakkad}, {Perna}  \& {Vietri}}{{Harrison}
  et~al.}{2018}]{2018NatAs...2..198H}
{Harrison} C.~M.,  {Costa} T.,  {Tadhunter} C.~N.,  {Fl{\"u}tsch} A.,  {Kakkad}
  D.,  {Perna} M.,   {Vietri} G.,  2018, \mn@doi [Nature Astronomy]
  {10.1038/s41550-018-0403-6}, \href
  {https://ui.adsabs.harvard.edu/abs/2018NatAs...2..198H} {2, 198}

\bibitem[\protect\citeauthoryear{{Heckman}, {Armus}  \& {Miley}}{{Heckman}
  et~al.}{1990}]{1990ApJS...74..833H}
{Heckman} T.~M.,  {Armus} L.,   {Miley} G.~K.,  1990, \mn@doi [\apjs]
  {10.1086/191522}, \href
  {https://ui.adsabs.harvard.edu/abs/1990ApJS...74..833H} {74, 833}

\bibitem[\protect\citeauthoryear{{Hermosa Mu{\~n}oz}, {M{\'a}rquez}, {Cazzoli},
  {Masegosa}  \& {Ag{\'\i}s-Gonz{\'a}lez}}{{Hermosa Mu{\~n}oz}
  et~al.}{2022}]{2022arXiv220105080H}
{Hermosa Mu{\~n}oz} L.,  {M{\'a}rquez} I.,  {Cazzoli} S.,  {Masegosa} J.,
  {Ag{\'\i}s-Gonz{\'a}lez} B.,  2022, arXiv e-prints, \href
  {https://ui.adsabs.harvard.edu/abs/2022arXiv220105080H} {p. arXiv:2201.05080}

\bibitem[\protect\citeauthoryear{{Herrera-Camus} et~al.,}{{Herrera-Camus}
  et~al.}{2020}]{2020A&A...635A..47H}
{Herrera-Camus} R.,  et~al., 2020, \mn@doi [\aap]
  {10.1051/0004-6361/201936434}, \href
  {https://ui.adsabs.harvard.edu/abs/2020A&A...635A..47H} {635, A47}

\bibitem[\protect\citeauthoryear{{Hill} \& {Zakamska}}{{Hill} \&
  {Zakamska}}{2014}]{2014MNRAS.439.2701H}
{Hill} M.~J.,  {Zakamska} N.~L.,  2014, \mn@doi [\mnras]
  {10.1093/mnras/stu123}, \href
  {https://ui.adsabs.harvard.edu/abs/2014MNRAS.439.2701H} {439, 2701}

\bibitem[\protect\citeauthoryear{{Hogarth} et~al.,}{{Hogarth}
  et~al.}{2021}]{2021MNRAS.500.3802H}
{Hogarth} L.~M.,  et~al., 2021, \mn@doi [\mnras] {10.1093/mnras/staa3512},
  \href {https://ui.adsabs.harvard.edu/abs/2021MNRAS.500.3802H} {500, 3802}

\bibitem[\protect\citeauthoryear{{Holt}, {Tadhunter}  \& {Morganti}}{{Holt}
  et~al.}{2008}]{2008MNRAS.387..639H}
{Holt} J.,  {Tadhunter} C.~N.,   {Morganti} R.,  2008, \mn@doi [\mnras]
  {10.1111/j.1365-2966.2008.13089.x}, \href
  {https://ui.adsabs.harvard.edu/abs/2008MNRAS.387..639H} {387, 639}

\bibitem[\protect\citeauthoryear{{Hopkins}, {Hernquist}, {Cox}, {Di Matteo},
  {Robertson}  \& {Springel}}{{Hopkins} et~al.}{2006}]{2006ApJS..163....1H}
{Hopkins} P.~F.,  {Hernquist} L.,  {Cox} T.~J.,  {Di Matteo} T.,  {Robertson}
  B.,   {Springel} V.,  2006, \mn@doi [\apjs] {10.1086/499298}, \href
  {https://ui.adsabs.harvard.edu/abs/2006ApJS..163....1H} {163, 1}

\bibitem[\protect\citeauthoryear{{Hopkins}, {Cox}, {Kere{\v{s}}}  \&
  {Hernquist}}{{Hopkins} et~al.}{2008}]{2008ApJS..175..390H}
{Hopkins} P.~F.,  {Cox} T.~J.,  {Kere{\v{s}}} D.,   {Hernquist} L.,  2008,
  \mn@doi [\apjs] {10.1086/524363}, \href
  {https://ui.adsabs.harvard.edu/abs/2008ApJS..175..390H} {175, 390}

\bibitem[\protect\citeauthoryear{{Kakkad} et~al.,}{{Kakkad}
  et~al.}{2016}]{2016A&A...592A.148K}
{Kakkad} D.,  et~al., 2016, \mn@doi [\aap] {10.1051/0004-6361/201527968}, \href
  {https://ui.adsabs.harvard.edu/abs/2016A&A...592A.148K} {592, A148}

\bibitem[\protect\citeauthoryear{{Kauffmann} et~al.,}{{Kauffmann}
  et~al.}{2003}]{2003MNRAS.346.1055K}
{Kauffmann} G.,  et~al., 2003, \mn@doi [\mnras]
  {10.1111/j.1365-2966.2003.07154.x}, \href
  {https://ui.adsabs.harvard.edu/abs/2003MNRAS.346.1055K} {346, 1055}

\bibitem[\protect\citeauthoryear{{Kaviraj} et~al.,}{{Kaviraj}
  et~al.}{2013}]{2013MNRAS.429L..40K}
{Kaviraj} S.,  et~al., 2013, \mn@doi [\mnras] {10.1093/mnrasl/sls019}, \href
  {https://ui.adsabs.harvard.edu/abs/2013MNRAS.429L..40K} {429, L40}

\bibitem[\protect\citeauthoryear{{Kewley}, {Heisler}, {Dopita}  \&
  {Lumsden}}{{Kewley} et~al.}{2001}]{2001ApJS..132...37K}
{Kewley} L.~J.,  {Heisler} C.~A.,  {Dopita} M.~A.,   {Lumsden} S.,  2001,
  \mn@doi [\apjs] {10.1086/318944}, \href
  {https://ui.adsabs.harvard.edu/abs/2001ApJS..132...37K} {132, 37}

\bibitem[\protect\citeauthoryear{{Kewley}, {Rupke}, {Zahid}, {Geller}  \&
  {Barton}}{{Kewley} et~al.}{2010}]{2010ApJ...721L..48K}
{Kewley} L.~J.,  {Rupke} D.,  {Zahid} H.~J.,  {Geller} M.~J.,   {Barton} E.~J.,
   2010, \mn@doi [\apjl] {10.1088/2041-8205/721/1/L48}, \href
  {https://ui.adsabs.harvard.edu/abs/2010ApJ...721L..48K} {721, L48}

\bibitem[\protect\citeauthoryear{{Khochfar} \& {Silk}}{{Khochfar} \&
  {Silk}}{2011}]{2011MNRAS.410L..42K}
{Khochfar} S.,  {Silk} J.,  2011, \mn@doi [\mnras]
  {10.1111/j.1745-3933.2010.00976.x}, \href
  {https://ui.adsabs.harvard.edu/abs/2011MNRAS.410L..42K} {410, L42}

\bibitem[\protect\citeauthoryear{{Kocevski} et~al.,}{{Kocevski}
  et~al.}{2012}]{2012ApJ...744..148K}
{Kocevski} D.~D.,  et~al., 2012, \mn@doi [\apj] {10.1088/0004-637X/744/2/148},
  \href {https://ui.adsabs.harvard.edu/abs/2012ApJ...744..148K} {744, 148}

\bibitem[\protect\citeauthoryear{{Kornei}, {Shapley}, {Martin}, {Coil}, {Lotz},
  {Schiminovich}, {Bundy}  \& {Noeske}}{{Kornei}
  et~al.}{2012}]{2012ApJ...758..135K}
{Kornei} K.~A.,  {Shapley} A.~E.,  {Martin} C.~L.,  {Coil} A.~L.,  {Lotz}
  J.~M.,  {Schiminovich} D.,  {Bundy} K.,   {Noeske} K.~G.,  2012, \mn@doi
  [\apj] {10.1088/0004-637X/758/2/135}, \href
  {https://ui.adsabs.harvard.edu/abs/2012ApJ...758..135K} {758, 135}

\bibitem[\protect\citeauthoryear{{Lackner} et~al.,}{{Lackner}
  et~al.}{2014}]{2014AJ....148..137L}
{Lackner} C.~N.,  et~al., 2014, \mn@doi [\aj] {10.1088/0004-6256/148/6/137},
  \href {https://ui.adsabs.harvard.edu/abs/2014AJ....148..137L} {148, 137}

\bibitem[\protect\citeauthoryear{{Leung} et~al.,}{{Leung}
  et~al.}{2019}]{2019ApJ...886...11L}
{Leung} G. C.~K.,  et~al., 2019, \mn@doi [\apj] {10.3847/1538-4357/ab4a7c},
  \href {https://ui.adsabs.harvard.edu/abs/2019ApJ...886...11L} {886, 11}

\bibitem[\protect\citeauthoryear{{Lintott} et~al.,}{{Lintott}
  et~al.}{2008}]{2008MNRAS.389.1179L}
{Lintott} C.~J.,  et~al., 2008, \mn@doi [\mnras]
  {10.1111/j.1365-2966.2008.13689.x}, \href
  {https://ui.adsabs.harvard.edu/abs/2008MNRAS.389.1179L} {389, 1179}

\bibitem[\protect\citeauthoryear{{Liu}, {Shen}  \& {Strauss}}{{Liu}
  et~al.}{2012}]{2012ApJ...745...94L}
{Liu} X.,  {Shen} Y.,   {Strauss} M.~A.,  2012, \mn@doi [\apj]
  {10.1088/0004-637X/745/1/94}, \href
  {https://ui.adsabs.harvard.edu/abs/2012ApJ...745...94L} {745, 94}

\bibitem[\protect\citeauthoryear{{Liu}, {Veilleux}, {Canalizo}, {Rupke},
  {Manzano-King}, {Bohn}  \& {U}}{{Liu} et~al.}{2020}]{2020ApJ...905..166L}
{Liu} W.,  {Veilleux} S.,  {Canalizo} G.,  {Rupke} D. S.~N.,  {Manzano-King}
  C.~M.,  {Bohn} T.,   {U} V.,  2020, \mn@doi [\apj]
  {10.3847/1538-4357/abc269}, \href
  {https://ui.adsabs.harvard.edu/abs/2020ApJ...905..166L} {905, 166}

\bibitem[\protect\citeauthoryear{{Lutz} et~al.,}{{Lutz}
  et~al.}{2020}]{2020A&A...633A.134L}
{Lutz} D.,  et~al., 2020, \mn@doi [\aap] {10.1051/0004-6361/201936803}, \href
  {https://ui.adsabs.harvard.edu/abs/2020A&A...633A.134L} {633, A134}

\bibitem[\protect\citeauthoryear{{Manzano-King}, {Canalizo}  \&
  {Sales}}{{Manzano-King} et~al.}{2019}]{2019ApJ...884...54M}
{Manzano-King} C.~M.,  {Canalizo} G.,   {Sales} L.~V.,  2019, \mn@doi [\apj]
  {10.3847/1538-4357/ab4197}, \href
  {https://ui.adsabs.harvard.edu/abs/2019ApJ...884...54M} {884, 54}

\bibitem[\protect\citeauthoryear{{Mechtley} et~al.,}{{Mechtley}
  et~al.}{2016}]{2016ApJ...830..156M}
{Mechtley} M.,  et~al., 2016, \mn@doi [\apj] {10.3847/0004-637X/830/2/156},
  \href {https://ui.adsabs.harvard.edu/abs/2016ApJ...830..156M} {830, 156}

\bibitem[\protect\citeauthoryear{{Mendel}, {Simard}, {Palmer}, {Ellison}  \&
  {Patton}}{{Mendel} et~al.}{2014}]{2014ApJS..210....3M}
{Mendel} J.~T.,  {Simard} L.,  {Palmer} M.,  {Ellison} S.~L.,   {Patton} D.~R.,
   2014, \mn@doi [\apjs] {10.1088/0067-0049/210/1/3}, \href
  {https://ui.adsabs.harvard.edu/abs/2014ApJS..210....3M} {210, 3}

\bibitem[\protect\citeauthoryear{{Molyneux}, {Harrison}  \&
  {Jarvis}}{{Molyneux} et~al.}{2019}]{2019A&A...631A.132M}
{Molyneux} S.~J.,  {Harrison} C.~M.,   {Jarvis} M.~E.,  2019, \mn@doi [\aap]
  {10.1051/0004-6361/201936408}, \href
  {https://ui.adsabs.harvard.edu/abs/2019A&A...631A.132M} {631, A132}

\bibitem[\protect\citeauthoryear{{Mullaney}, {Alexander}, {Fine}, {Goulding},
  {Harrison}  \& {Hickox}}{{Mullaney} et~al.}{2013}]{2013MNRAS.433..622M}
{Mullaney} J.~R.,  {Alexander} D.~M.,  {Fine} S.,  {Goulding} A.~D.,
  {Harrison} C.~M.,   {Hickox} R.~C.,  2013, \mn@doi [\mnras]
  {10.1093/mnras/stt751}, \href
  {https://ui.adsabs.harvard.edu/abs/2013MNRAS.433..622M} {433, 622}

\bibitem[\protect\citeauthoryear{{M{\"u}ller-S{\'a}nchez}, {Comerford},
  {Nevin}, {Barrows}, {Cooper}  \& {Greene}}{{M{\"u}ller-S{\'a}nchez}
  et~al.}{2015}]{2015ApJ...813..103M}
{M{\"u}ller-S{\'a}nchez} F.,  {Comerford} J.~M.,  {Nevin} R.,  {Barrows} R.~S.,
   {Cooper} M.~C.,   {Greene} J.~E.,  2015, \mn@doi [\apj]
  {10.1088/0004-637X/813/2/103}, \href
  {https://ui.adsabs.harvard.edu/abs/2015ApJ...813..103M} {813, 103}

\bibitem[\protect\citeauthoryear{{Nestor}, {Johnson}, {Wild}, {M{\'e}nard},
  {Turnshek}, {Rao}  \& {Pettini}}{{Nestor} et~al.}{2011}]{2011MNRAS.412.1559N}
{Nestor} D.~B.,  {Johnson} B.~D.,  {Wild} V.,  {M{\'e}nard} B.,  {Turnshek}
  D.~A.,  {Rao} S.,   {Pettini} M.,  2011, \mn@doi [\mnras]
  {10.1111/j.1365-2966.2010.17865.x}, \href
  {https://ui.adsabs.harvard.edu/abs/2011MNRAS.412.1559N} {412, 1559}

\bibitem[\protect\citeauthoryear{{Nevin}, {Comerford},
  {M{\"u}ller-S{\'a}nchez}, {Barrows}  \& {Cooper}}{{Nevin}
  et~al.}{2018}]{2018MNRAS.473.2160N}
{Nevin} R.,  {Comerford} J.~M.,  {M{\"u}ller-S{\'a}nchez} F.,  {Barrows} R.,
  {Cooper} M.~C.,  2018, \mn@doi [\mnras] {10.1093/mnras/stx2433}, \href
  {https://ui.adsabs.harvard.edu/abs/2018MNRAS.473.2160N} {473, 2160}

\bibitem[\protect\citeauthoryear{{Patton}, {Ellison}, {Simard}, {McConnachie}
  \& {Mendel}}{{Patton} et~al.}{2011}]{2011MNRAS.412..591P}
{Patton} D.~R.,  {Ellison} S.~L.,  {Simard} L.,  {McConnachie} A.~W.,
  {Mendel} J.~T.,  2011, \mn@doi [\mnras] {10.1111/j.1365-2966.2010.17932.x},
  \href {https://ui.adsabs.harvard.edu/abs/2011MNRAS.412..591P} {412, 591}

\bibitem[\protect\citeauthoryear{{Patton}, {Torrey}, {Ellison}, {Mendel}  \&
  {Scudder}}{{Patton} et~al.}{2013}]{2013MNRAS.433L..59P}
{Patton} D.~R.,  {Torrey} P.,  {Ellison} S.~L.,  {Mendel} J.~T.,   {Scudder}
  J.~M.,  2013, \mn@doi [\mnras] {10.1093/mnrasl/slt058}, \href
  {https://ui.adsabs.harvard.edu/abs/2013MNRAS.433L..59P} {433, L59}

\bibitem[\protect\citeauthoryear{{Patton}, {Qamar}, {Ellison}, {Bluck},
  {Simard}, {Mendel}, {Moreno}  \& {Torrey}}{{Patton}
  et~al.}{2016}]{2016MNRAS.461.2589P}
{Patton} D.~R.,  {Qamar} F.~D.,  {Ellison} S.~L.,  {Bluck} A. F.~L.,  {Simard}
  L.,  {Mendel} J.~T.,  {Moreno} J.,   {Torrey} P.,  2016, \mn@doi [\mnras]
  {10.1093/mnras/stw1494}, \href
  {https://ui.adsabs.harvard.edu/abs/2016MNRAS.461.2589P} {461, 2589}

\bibitem[\protect\citeauthoryear{{Pearson} et~al.,}{{Pearson}
  et~al.}{2019}]{2019A&A...631A..51P}
{Pearson} W.~J.,  et~al., 2019, \mn@doi [\aap] {10.1051/0004-6361/201936337},
  \href {https://ui.adsabs.harvard.edu/abs/2019A&A...631A..51P} {631, A51}

\bibitem[\protect\citeauthoryear{{Pereira-Santaella}
  et~al.,}{{Pereira-Santaella} et~al.}{2018}]{2018A&A...616A.171P}
{Pereira-Santaella} M.,  et~al., 2018, \mn@doi [\aap]
  {10.1051/0004-6361/201833089}, \href
  {https://ui.adsabs.harvard.edu/abs/2018A&A...616A.171P} {616, A171}

\bibitem[\protect\citeauthoryear{{Perez}, {Tissera}  \& {Blaizot}}{{Perez}
  et~al.}{2009}]{2009MNRAS.397..748P}
{Perez} J.,  {Tissera} P.,   {Blaizot} J.,  2009, \mn@doi [\mnras]
  {10.1111/j.1365-2966.2009.15033.x}, \href
  {https://ui.adsabs.harvard.edu/abs/2009MNRAS.397..748P} {397, 748}

\bibitem[\protect\citeauthoryear{{Perna}, {Lanzuisi}, {Brusa}, {Mignoli}  \&
  {Cresci}}{{Perna} et~al.}{2017}]{2017A&A...603A..99P}
{Perna} M.,  {Lanzuisi} G.,  {Brusa} M.,  {Mignoli} M.,   {Cresci} G.,  2017,
  \mn@doi [\aap] {10.1051/0004-6361/201630369}, \href
  {https://ui.adsabs.harvard.edu/abs/2017A&A...603A..99P} {603, A99}

\bibitem[\protect\citeauthoryear{{Perret}, {Renaud}, {Epinat}, {Amram},
  {Bournaud}, {Contini}, {Teyssier}  \& {Lambert}}{{Perret}
  et~al.}{2014}]{2014A&A...562A...1P}
{Perret} V.,  {Renaud} F.,  {Epinat} B.,  {Amram} P.,  {Bournaud} F.,
  {Contini} T.,  {Teyssier} R.,   {Lambert} J.~C.,  2014, \mn@doi [\aap]
  {10.1051/0004-6361/201322395}, \href
  {https://ui.adsabs.harvard.edu/abs/2014A&A...562A...1P} {562, A1}

\bibitem[\protect\citeauthoryear{{Rakshit} \& {Woo}}{{Rakshit} \&
  {Woo}}{2018}]{2018ApJ...865....5R}
{Rakshit} S.,  {Woo} J.-H.,  2018, \mn@doi [\apj] {10.3847/1538-4357/aad9f8},
  \href {https://ui.adsabs.harvard.edu/abs/2018ApJ...865....5R} {865, 5}

\bibitem[\protect\citeauthoryear{{Roberts-Borsani}, {Saintonge}, {Masters}  \&
  {Stark}}{{Roberts-Borsani} et~al.}{2020}]{2020MNRAS.493.3081R}
{Roberts-Borsani} G.~W.,  {Saintonge} A.,  {Masters} K.~L.,   {Stark} D.~V.,
  2020, \mn@doi [\mnras] {10.1093/mnras/staa464}, \href
  {https://ui.adsabs.harvard.edu/abs/2020MNRAS.493.3081R} {493, 3081}

\bibitem[\protect\citeauthoryear{{Rosario}, {McGurk}, {Max}, {Shields}, {Smith}
   \& {Ammons}}{{Rosario} et~al.}{2011}]{2011ApJ...739...44R}
{Rosario} D.~J.,  {McGurk} R.~C.,  {Max} C.~E.,  {Shields} G.~A.,  {Smith}
  K.~L.,   {Ammons} S.~M.,  2011, \mn@doi [\apj] {10.1088/0004-637X/739/1/44},
  \href {https://ui.adsabs.harvard.edu/abs/2011ApJ...739...44R} {739, 44}

\bibitem[\protect\citeauthoryear{{Rosario} et~al.,}{{Rosario}
  et~al.}{2015}]{2015A&A...573A..85R}
{Rosario} D.~J.,  et~al., 2015, \mn@doi [\aap] {10.1051/0004-6361/201423782},
  \href {https://ui.adsabs.harvard.edu/abs/2015A&A...573A..85R} {573, A85}

\bibitem[\protect\citeauthoryear{{Rothberg} \& {Joseph}}{{Rothberg} \&
  {Joseph}}{2004}]{2004AJ....128.2098R}
{Rothberg} B.,  {Joseph} R.~D.,  2004, \mn@doi [\aj] {10.1086/425049}, \href
  {https://ui.adsabs.harvard.edu/abs/2004AJ....128.2098R} {128, 2098}

\bibitem[\protect\citeauthoryear{{Rubin}, {Weiner}, {Koo}, {Martin},
  {Prochaska}, {Coil}  \& {Newman}}{{Rubin} et~al.}{2010}]{2010ApJ...719.1503R}
{Rubin} K. H.~R.,  {Weiner} B.~J.,  {Koo} D.~C.,  {Martin} C.~L.,  {Prochaska}
  J.~X.,  {Coil} A.~L.,   {Newman} J.~A.,  2010, \mn@doi [\apj]
  {10.1088/0004-637X/719/2/1503}, \href
  {https://ui.adsabs.harvard.edu/abs/2010ApJ...719.1503R} {719, 1503}

\bibitem[\protect\citeauthoryear{{Rubin}, {Prochaska}, {Koo}, {Phillips},
  {Martin}  \& {Winstrom}}{{Rubin} et~al.}{2014}]{2014ApJ...794..156R}
{Rubin} K. H.~R.,  {Prochaska} J.~X.,  {Koo} D.~C.,  {Phillips} A.~C.,
  {Martin} C.~L.,   {Winstrom} L.~O.,  2014, \mn@doi [\apj]
  {10.1088/0004-637X/794/2/156}, \href
  {https://ui.adsabs.harvard.edu/abs/2014ApJ...794..156R} {794, 156}

\bibitem[\protect\citeauthoryear{{Rupke}}{{Rupke}}{2018}]{2018Galax...6..138R}
{Rupke} D.,  2018, \mn@doi [Galaxies] {10.3390/galaxies6040138}, \href
  {https://ui.adsabs.harvard.edu/abs/2018Galax...6..138R} {6, 138}

\bibitem[\protect\citeauthoryear{{Rupke}, {Veilleux}  \& {Sanders}}{{Rupke}
  et~al.}{2002}]{2002ApJ...570..588R}
{Rupke} D.~S.,  {Veilleux} S.,   {Sanders} D.~B.,  2002, \mn@doi [\apj]
  {10.1086/339789}, \href
  {https://ui.adsabs.harvard.edu/abs/2002ApJ...570..588R} {570, 588}

\bibitem[\protect\citeauthoryear{{Rupke}, {Veilleux}  \& {Sanders}}{{Rupke}
  et~al.}{2005a}]{2005ApJS..160...87R}
{Rupke} D.~S.,  {Veilleux} S.,   {Sanders} D.~B.,  2005a, \mn@doi [\apjs]
  {10.1086/432886}, \href
  {https://ui.adsabs.harvard.edu/abs/2005ApJS..160...87R} {160, 87}

\bibitem[\protect\citeauthoryear{{Rupke}, {Veilleux}  \& {Sanders}}{{Rupke}
  et~al.}{2005b}]{2005ApJS..160..115R}
{Rupke} D.~S.,  {Veilleux} S.,   {Sanders} D.~B.,  2005b, \mn@doi [\apjs]
  {10.1086/432889}, \href
  {https://ui.adsabs.harvard.edu/abs/2005ApJS..160..115R} {160, 115}

\bibitem[\protect\citeauthoryear{{Rupke}, {Veilleux}  \& {Sanders}}{{Rupke}
  et~al.}{2005c}]{2005ApJ...632..751R}
{Rupke} D.~S.,  {Veilleux} S.,   {Sanders} D.~B.,  2005c, \mn@doi [\apj]
  {10.1086/444451}, \href
  {https://ui.adsabs.harvard.edu/abs/2005ApJ...632..751R} {632, 751}

\bibitem[\protect\citeauthoryear{{Rupke}, {G{\"u}ltekin}  \&
  {Veilleux}}{{Rupke} et~al.}{2017}]{2017ApJ...850...40R}
{Rupke} D. S.~N.,  {G{\"u}ltekin} K.,   {Veilleux} S.,  2017, \mn@doi [\apj]
  {10.3847/1538-4357/aa94d1}, \href
  {https://ui.adsabs.harvard.edu/abs/2017ApJ...850...40R} {850, 40}

\bibitem[\protect\citeauthoryear{{S{\'a}nchez} et~al.,}{{S{\'a}nchez}
  et~al.}{2018}]{2018RMxAA..54..217S}
{S{\'a}nchez} S.~F.,  et~al., 2018, \rmxaa, \href
  {https://ui.adsabs.harvard.edu/abs/2018RMxAA..54..217S} {54, 217}

\bibitem[\protect\citeauthoryear{{Santoro}, {Tadhunter}, {Baron}, {Morganti}
  \& {Holt}}{{Santoro} et~al.}{2020}]{2020A&A...644A..54S}
{Santoro} F.,  {Tadhunter} C.,  {Baron} D.,  {Morganti} R.,   {Holt} J.,  2020,
  \mn@doi [\aap] {10.1051/0004-6361/202039077}, \href
  {https://ui.adsabs.harvard.edu/abs/2020A&A...644A..54S} {644, A54}

\bibitem[\protect\citeauthoryear{{Satyapal}, {Ellison}, {McAlpine}, {Hickox},
  {Patton}  \& {Mendel}}{{Satyapal} et~al.}{2014}]{2014MNRAS.441.1297S}
{Satyapal} S.,  {Ellison} S.~L.,  {McAlpine} W.,  {Hickox} R.~C.,  {Patton}
  D.~R.,   {Mendel} J.~T.,  2014, \mn@doi [\mnras] {10.1093/mnras/stu650},
  \href {https://ui.adsabs.harvard.edu/abs/2014MNRAS.441.1297S} {441, 1297}

\bibitem[\protect\citeauthoryear{{Schawinski}, {Koss}, {Berney}  \&
  {Sartori}}{{Schawinski} et~al.}{2015}]{2015MNRAS.451.2517S}
{Schawinski} K.,  {Koss} M.,  {Berney} S.,   {Sartori} L.~F.,  2015, \mn@doi
  [\mnras] {10.1093/mnras/stv1136}, \href
  {https://ui.adsabs.harvard.edu/abs/2015MNRAS.451.2517S} {451, 2517}

\bibitem[\protect\citeauthoryear{{Scholtz} et~al.,}{{Scholtz}
  et~al.}{2021}]{2021MNRAS.505.5469S}
{Scholtz} J.,  et~al., 2021, \mn@doi [\mnras] {10.1093/mnras/stab1631}, \href
  {https://ui.adsabs.harvard.edu/abs/2021MNRAS.505.5469S} {505, 5469}

\bibitem[\protect\citeauthoryear{{Schweizer}}{{Schweizer}}{1982}]{1982ApJ...252..455S}
{Schweizer} F.,  1982, \mn@doi [\apj] {10.1086/159573}, \href
  {https://ui.adsabs.harvard.edu/abs/1982ApJ...252..455S} {252, 455}

\bibitem[\protect\citeauthoryear{{Scudder}, {Ellison}  \& {Mendel}}{{Scudder}
  et~al.}{2012a}]{2012MNRAS.423.2690S}
{Scudder} J.~M.,  {Ellison} S.~L.,   {Mendel} J.~T.,  2012a, \mn@doi [\mnras]
  {10.1111/j.1365-2966.2012.21080.x}, \href
  {https://ui.adsabs.harvard.edu/abs/2012MNRAS.423.2690S} {423, 2690}

\bibitem[\protect\citeauthoryear{{Scudder}, {Ellison}, {Torrey}, {Patton}  \&
  {Mendel}}{{Scudder} et~al.}{2012b}]{2012MNRAS.426..549S}
{Scudder} J.~M.,  {Ellison} S.~L.,  {Torrey} P.,  {Patton} D.~R.,   {Mendel}
  J.~T.,  2012b, \mn@doi [\mnras] {10.1111/j.1365-2966.2012.21749.x}, \href
  {https://ui.adsabs.harvard.edu/abs/2012MNRAS.426..549S} {426, 549}

\bibitem[\protect\citeauthoryear{{Scudder}, {Ellison}, {Momjian}, {Rosenberg},
  {Torrey}, {Patton}, {Fertig}  \& {Mendel}}{{Scudder}
  et~al.}{2015}]{2015MNRAS.449.3719S}
{Scudder} J.~M.,  {Ellison} S.~L.,  {Momjian} E.,  {Rosenberg} J.~L.,  {Torrey}
  P.,  {Patton} D.~R.,  {Fertig} D.,   {Mendel} J.~T.,  2015, \mn@doi [\mnras]
  {10.1093/mnras/stv588}, \href
  {https://ui.adsabs.harvard.edu/abs/2015MNRAS.449.3719S} {449, 3719}

\bibitem[\protect\citeauthoryear{{Sexton}, {Matzko}, {Darden}, {Canalizo}  \&
  {Gorjian}}{{Sexton} et~al.}{2021}]{2021MNRAS.500.2871S}
{Sexton} R.~O.,  {Matzko} W.,  {Darden} N.,  {Canalizo} G.,   {Gorjian} V.,
  2021, \mn@doi [\mnras] {10.1093/mnras/staa3278}, \href
  {https://ui.adsabs.harvard.edu/abs/2021MNRAS.500.2871S} {500, 2871}

\bibitem[\protect\citeauthoryear{{Shah} et~al.,}{{Shah}
  et~al.}{2020}]{2020ApJ...904..107S}
{Shah} E.~A.,  et~al., 2020, \mn@doi [\apj] {10.3847/1538-4357/abbf59}, \href
  {https://ui.adsabs.harvard.edu/abs/2020ApJ...904..107S} {904, 107}

\bibitem[\protect\citeauthoryear{{Shen}, {Liu}, {Greene}  \& {Strauss}}{{Shen}
  et~al.}{2011}]{2011ApJ...735...48S}
{Shen} Y.,  {Liu} X.,  {Greene} J.~E.,   {Strauss} M.~A.,  2011, \mn@doi [\apj]
  {10.1088/0004-637X/735/1/48}, \href
  {https://ui.adsabs.harvard.edu/abs/2011ApJ...735...48S} {735, 48}

\bibitem[\protect\citeauthoryear{{Silk}}{{Silk}}{2013}]{2013ApJ...772..112S}
{Silk} J.,  2013, \mn@doi [\apj] {10.1088/0004-637X/772/2/112}, \href
  {https://ui.adsabs.harvard.edu/abs/2013ApJ...772..112S} {772, 112}

\bibitem[\protect\citeauthoryear{{Simmons}, {Urry}, {Schawinski}, {Cardamone}
  \& {Glikman}}{{Simmons} et~al.}{2012}]{2012ApJ...761...75S}
{Simmons} B.~D.,  {Urry} C.~M.,  {Schawinski} K.,  {Cardamone} C.,   {Glikman}
  E.,  2012, \mn@doi [\apj] {10.1088/0004-637X/761/1/75}, \href
  {https://ui.adsabs.harvard.edu/abs/2012ApJ...761...75S} {761, 75}

\bibitem[\protect\citeauthoryear{{Smethurst}, {Simmons}, {Lintott}  \&
  {Shanahan}}{{Smethurst} et~al.}{2019}]{2019MNRAS.489.4016S}
{Smethurst} R.~J.,  {Simmons} B.~D.,  {Lintott} C.~J.,   {Shanahan} J.,  2019,
  \mn@doi [\mnras] {10.1093/mnras/stz2443}, \href
  {https://ui.adsabs.harvard.edu/abs/2019MNRAS.489.4016S} {489, 4016}

\bibitem[\protect\citeauthoryear{{Smethurst} et~al.,}{{Smethurst}
  et~al.}{2021}]{2021MNRAS.507.3985S}
{Smethurst} R.~J.,  et~al., 2021, \mn@doi [\mnras] {10.1093/mnras/stab2340},
  \href {https://ui.adsabs.harvard.edu/abs/2021MNRAS.507.3985S} {507, 3985}

\bibitem[\protect\citeauthoryear{{Soto} \& {Martin}}{{Soto} \&
  {Martin}}{2012}]{2012ApJS..203....3S}
{Soto} K.~T.,  {Martin} C.~L.,  2012, \mn@doi [\apjs]
  {10.1088/0067-0049/203/1/3}, \href
  {https://ui.adsabs.harvard.edu/abs/2012ApJS..203....3S} {203, 3}

\bibitem[\protect\citeauthoryear{{Soto}, {Martin}, {Prescott}  \&
  {Armus}}{{Soto} et~al.}{2012}]{2012ApJ...757...86S}
{Soto} K.~T.,  {Martin} C.~L.,  {Prescott} M.~K.~M.,   {Armus} L.,  2012,
  \mn@doi [\apj] {10.1088/0004-637X/757/1/86}, \href
  {https://ui.adsabs.harvard.edu/abs/2012ApJ...757...86S} {757, 86}

\bibitem[\protect\citeauthoryear{{Springel}, {Di Matteo}  \&
  {Hernquist}}{{Springel} et~al.}{2005}]{2005MNRAS.361..776S}
{Springel} V.,  {Di Matteo} T.,   {Hernquist} L.,  2005, \mn@doi [\mnras]
  {10.1111/j.1365-2966.2005.09238.x}, \href
  {https://ui.adsabs.harvard.edu/abs/2005MNRAS.361..776S} {361, 776}

\bibitem[\protect\citeauthoryear{{Stasi{\'n}ska}, {Cid Fernandes}, {Mateus},
  {Sodr{\'e}}  \& {Asari}}{{Stasi{\'n}ska} et~al.}{2006}]{2006MNRAS.371..972S}
{Stasi{\'n}ska} G.,  {Cid Fernandes} R.,  {Mateus} A.,  {Sodr{\'e}} L.,
  {Asari} N.~V.,  2006, \mn@doi [\mnras] {10.1111/j.1365-2966.2006.10732.x},
  \href {https://ui.adsabs.harvard.edu/abs/2006MNRAS.371..972S} {371, 972}

\bibitem[\protect\citeauthoryear{{Sturm} et~al.,}{{Sturm}
  et~al.}{2011}]{2011ApJ...733L..16S}
{Sturm} E.,  et~al., 2011, \mn@doi [\apjl] {10.1088/2041-8205/733/1/L16}, \href
  {https://ui.adsabs.harvard.edu/abs/2011ApJ...733L..16S} {733, L16}

\bibitem[\protect\citeauthoryear{{Swinbank} et~al.,}{{Swinbank}
  et~al.}{2019}]{2019MNRAS.487..381S}
{Swinbank} A.~M.,  et~al., 2019, \mn@doi [\mnras] {10.1093/mnras/stz1275},
  \href {https://ui.adsabs.harvard.edu/abs/2019MNRAS.487..381S} {487, 381}

\bibitem[\protect\citeauthoryear{{Taylor}}{{Taylor}}{2005}]{2005ASPC..347...29T}
{Taylor} M.~B.,  2005, in {Shopbell} P.,  {Britton} M.,   {Ebert} R.,  eds,
  Astronomical Society of the Pacific Conference Series Vol. 347, Astronomical
  Data Analysis Software and Systems XIV. p.~29

\bibitem[\protect\citeauthoryear{{Thorp}, {Ellison}, {Simard}, {S{\'a}nchez}
  \& {Antonio}}{{Thorp} et~al.}{2019}]{2019MNRAS.482L..55T}
{Thorp} M.~D.,  {Ellison} S.~L.,  {Simard} L.,  {S{\'a}nchez} S.~F.,
  {Antonio} B.,  2019, \mn@doi [\mnras] {10.1093/mnrasl/sly185}, \href
  {https://ui.adsabs.harvard.edu/abs/2019MNRAS.482L..55T} {482, L55}

\bibitem[\protect\citeauthoryear{{Toba}, {Bae}, {Nagao}, {Woo}, {Wang},
  {Wagner}, {Sun}  \& {Chang}}{{Toba} et~al.}{2017}]{2017ApJ...850..140T}
{Toba} Y.,  {Bae} H.-J.,  {Nagao} T.,  {Woo} J.-H.,  {Wang} W.-H.,  {Wagner}
  A.~Y.,  {Sun} A.-L.,   {Chang} Y.-Y.,  2017, \mn@doi [\apj]
  {10.3847/1538-4357/aa918a}, \href
  {https://ui.adsabs.harvard.edu/abs/2017ApJ...850..140T} {850, 140}

\bibitem[\protect\citeauthoryear{{Toomre} \& {Toomre}}{{Toomre} \&
  {Toomre}}{1972}]{1972ApJ...178..623T}
{Toomre} A.,  {Toomre} J.,  1972, \mn@doi [\apj] {10.1086/151823}, \href
  {https://ui.adsabs.harvard.edu/abs/1972ApJ...178..623T} {178, 623}

\bibitem[\protect\citeauthoryear{{Veilleux} \& {Rupke}}{{Veilleux} \&
  {Rupke}}{2005}]{2005ASPC..331..313V}
{Veilleux} S.,  {Rupke} D.~S.,  2005, in {Braun} R.,  ed.,  Astronomical
  Society of the Pacific Conference Series Vol. 331, Extra-Planar Gas. p.~313

\bibitem[\protect\citeauthoryear{{Veilleux} et~al.,}{{Veilleux}
  et~al.}{2013}]{2013ApJ...776...27V}
{Veilleux} S.,  et~al., 2013, \mn@doi [\apj] {10.1088/0004-637X/776/1/27},
  \href {https://ui.adsabs.harvard.edu/abs/2013ApJ...776...27V} {776, 27}

\bibitem[\protect\citeauthoryear{{Veilleux}, {Maiolino}, {Bolatto}  \&
  {Aalto}}{{Veilleux} et~al.}{2020}]{2020A&ARv..28....2V}
{Veilleux} S.,  {Maiolino} R.,  {Bolatto} A.~D.,   {Aalto} S.,  2020, \mn@doi
  [\aapr] {10.1007/s00159-019-0121-9}, \href
  {https://ui.adsabs.harvard.edu/abs/2020A&ARv..28....2V} {28, 2}

\bibitem[\protect\citeauthoryear{{Villar Mart{\'\i}n}, {Emonts}, {Humphrey},
  {Cabrera Lavers}  \& {Binette}}{{Villar Mart{\'\i}n}
  et~al.}{2014}]{2014MNRAS.440.3202V}
{Villar Mart{\'\i}n} M.,  {Emonts} B.,  {Humphrey} A.,  {Cabrera Lavers} A.,
  {Binette} L.,  2014, \mn@doi [\mnras] {10.1093/mnras/stu448}, \href
  {https://ui.adsabs.harvard.edu/abs/2014MNRAS.440.3202V} {440, 3202}

\bibitem[\protect\citeauthoryear{{Villforth} et~al.,}{{Villforth}
  et~al.}{2014}]{2014MNRAS.439.3342V}
{Villforth} C.,  et~al., 2014, \mn@doi [\mnras] {10.1093/mnras/stu173}, \href
  {https://ui.adsabs.harvard.edu/abs/2014MNRAS.439.3342V} {439, 3342}

\bibitem[\protect\citeauthoryear{{Villforth} et~al.,}{{Villforth}
  et~al.}{2017}]{2017MNRAS.466..812V}
{Villforth} C.,  et~al., 2017, \mn@doi [\mnras] {10.1093/mnras/stw3037}, \href
  {https://ui.adsabs.harvard.edu/abs/2017MNRAS.466..812V} {466, 812}

\bibitem[\protect\citeauthoryear{{Weiner} et~al.,}{{Weiner}
  et~al.}{2009}]{2009ApJ...692..187W}
{Weiner} B.~J.,  et~al., 2009, \mn@doi [\apj] {10.1088/0004-637X/692/1/187},
  \href {https://ui.adsabs.harvard.edu/abs/2009ApJ...692..187W} {692, 187}

\bibitem[\protect\citeauthoryear{{Westmoquette}, {Clements}, {Bendo}  \&
  {Khan}}{{Westmoquette} et~al.}{2012}]{2012MNRAS.424..416W}
{Westmoquette} M.~S.,  {Clements} D.~L.,  {Bendo} G.~J.,   {Khan} S.~A.,  2012,
  \mn@doi [\mnras] {10.1111/j.1365-2966.2012.21214.x}, \href
  {https://ui.adsabs.harvard.edu/abs/2012MNRAS.424..416W} {424, 416}

\bibitem[\protect\citeauthoryear{{Weston}, {McIntosh}, {Brodwin}, {Mann},
  {Cooper}, {McConnell}  \& {Nielsen}}{{Weston}
  et~al.}{2017}]{2017MNRAS.464.3882W}
{Weston} M.~E.,  {McIntosh} D.~H.,  {Brodwin} M.,  {Mann} J.,  {Cooper} A.,
  {McConnell} A.,   {Nielsen} J.~L.,  2017, \mn@doi [\mnras]
  {10.1093/mnras/stw2620}, \href
  {https://ui.adsabs.harvard.edu/abs/2017MNRAS.464.3882W} {464, 3882}

\bibitem[\protect\citeauthoryear{{Wong} et~al.,}{{Wong}
  et~al.}{2011}]{2011ApJ...728..119W}
{Wong} K.~C.,  et~al., 2011, \mn@doi [\apj] {10.1088/0004-637X/728/2/119},
  \href {https://ui.adsabs.harvard.edu/abs/2011ApJ...728..119W} {728, 119}

\bibitem[\protect\citeauthoryear{{Woo}, {Bae}, {Son}  \& {Karouzos}}{{Woo}
  et~al.}{2016}]{2016ApJ...817..108W}
{Woo} J.-H.,  {Bae} H.-J.,  {Son} D.,   {Karouzos} M.,  2016, \mn@doi [\apj]
  {10.3847/0004-637X/817/2/108}, \href
  {https://ui.adsabs.harvard.edu/abs/2016ApJ...817..108W} {817, 108}

\bibitem[\protect\citeauthoryear{{Woods}, {Geller}, {Kurtz}, {Westra},
  {Fabricant}  \& {Dell'Antonio}}{{Woods} et~al.}{2010}]{2010AJ....139.1857W}
{Woods} D.~F.,  {Geller} M.~J.,  {Kurtz} M.~J.,  {Westra} E.,  {Fabricant}
  D.~G.,   {Dell'Antonio} I.,  2010, \mn@doi [\aj]
  {10.1088/0004-6256/139/5/1857}, \href
  {https://ui.adsabs.harvard.edu/abs/2010AJ....139.1857W} {139, 1857}

\bibitem[\protect\citeauthoryear{{Wylezalek}, {Flores}, {Zakamska}, {Greene}
  \& {Riffel}}{{Wylezalek} et~al.}{2020}]{2020MNRAS.492.4680W}
{Wylezalek} D.,  {Flores} A.~M.,  {Zakamska} N.~L.,  {Greene} J.~E.,   {Riffel}
  R.~A.,  2020, \mn@doi [\mnras] {10.1093/mnras/staa062}, \href
  {https://ui.adsabs.harvard.edu/abs/2020MNRAS.492.4680W} {492, 4680}

\bibitem[\protect\citeauthoryear{{Xu} et~al.,}{{Xu}
  et~al.}{2012}]{2012ApJ...760...72X}
{Xu} C.~K.,  et~al., 2012, \mn@doi [\apj] {10.1088/0004-637X/760/1/72}, \href
  {https://ui.adsabs.harvard.edu/abs/2012ApJ...760...72X} {760, 72}

\bibitem[\protect\citeauthoryear{{Zakamska} \& {Greene}}{{Zakamska} \&
  {Greene}}{2014}]{2014MNRAS.442..784Z}
{Zakamska} N.~L.,  {Greene} J.~E.,  2014, \mn@doi [\mnras]
  {10.1093/mnras/stu842}, \href
  {https://ui.adsabs.harvard.edu/abs/2014MNRAS.442..784Z} {442, 784}

\bibitem[\protect\citeauthoryear{{Zakamska} et~al.,}{{Zakamska}
  et~al.}{2016}]{2016MNRAS.459.3144Z}
{Zakamska} N.~L.,  et~al., 2016, \mn@doi [\mnras] {10.1093/mnras/stw718}, \href
  {https://ui.adsabs.harvard.edu/abs/2016MNRAS.459.3144Z} {459, 3144}

\bibitem[\protect\citeauthoryear{{Zubovas}, {Nayakshin}, {King}  \&
  {Wilkinson}}{{Zubovas} et~al.}{2013}]{2013MNRAS.433.3079Z}
{Zubovas} K.,  {Nayakshin} S.,  {King} A.,   {Wilkinson} M.,  2013, \mn@doi
  [\mnras] {10.1093/mnras/stt952}, \href
  {https://ui.adsabs.harvard.edu/abs/2013MNRAS.433.3079Z} {433, 3079}

\makeatother
\end{thebibliography}







\bsp    
\label{lastpage}
\end{document}